\documentstyle[12pt,epsf]{article}
\begin{document}

\title{A modified Bragg's law of Class I Bragg resonances of linear long waves excited by an array of artificial bars}
\author{Huan-Wen Liu\\
\small School of Naval Architecture and Maritime,\\
\small Zhejiang Ocean University, Zhoushan 316022, Zhejiang, PR China}

\date{}
\maketitle
%


{\bf Abstract:} Based on the band theory of Bloch waves, a modified Bragg's law is established for Class I Bragg resonances when linear long waves are reflected by a finite periodic array of artificial bars, including rectangular, parabolic, rectified cosinoidal, isosceles trapezoidal, and isosceles triangular bars.
The modified Bragg's law is described by a function of several parameters such as the bar shape, dimensionless bar height with respect to the global water depth, and dimensionless bar width with respect to the incident wavelength.
It is found that Bragg's law is just a limiting case of the modified Bragg's law as the bar height or width approaches zero. This fact tells us that the Bragg's law that accurately describes Bragg resonances in X-ray crystallography cannot be simply applied to Bragg resonances of linear long water waves excited by any finite periodic array of artificial bars, because the height and width of any artificial bar cannot be zero.
Based on the modified Bragg's law, the phenomenon of phase downshifting can be well explained and accurately predicted. It is revealed that the phase downshifting becomes more significant with increasing the cross-sectional area of artificial bars.








\section{Introduction}






Bragg resonant reflection of X-rays excited by crystals was first observed by Bragg and Bragg (1913), who also established the resonance law, i.e., the maximal intensity of the reflected ray occurs when the periodic spacing between adjacent atoms is an integer multiple of the half-wavelength of the normally incident X-rays. Their pioneering work  ``X-rays and crystal structures" earned them the Nobel Prize in 1915.


Bragg resonant reflection of water surface waves excited by sinusoidal ripples on an otherwise horizontal, flat seabed was theoretically revealed by Davies (1982), who derived a closed-form solution of the reflection coefficient based on the regular perturbation, where the ripple amplitude was assumed to be very small with respect to the ripple wavelength, $d$, and the surface wavelength, $L$. It is shown by Davies' perturbation solution that there exists a resonant peak near the critical ratio with the ripple wavelength being half the surface wavelength, i.e., $2d/L=1$. This theoretical prediction was soon confirmed by flume experiments (Heathershaw, 1982; Davies and Heathershaw, 1984). It has been recognized that, Bragg resonant reflection of water waves not only provides an explanation to the formation of longshore sandbars on beaches, but also provides a mechanism of coastal protection. Since Davies' work, Bragg resonances excited by sinusoidal ripples have been intensively investigated, see
Mei (1985),
Kirby (1986),
Dalrymple and Kirby (1986),
Yoon and Liu (1987),
Davies et al. (1989),
Belzons et al. (1991),
Guazzelli et al. (1992),
Rey et al. (1996),
Madsen et al. (2006),
Ardhuin and Magne (2007),
Liu and Yue (1998),
Alam et al. (2010),
Couston et al. (2017),
Liu et al. (2019a),
Peng et al. (2020),
Gao et al. (2021) and so on.

Inspired by the fact that natural sinusoidal ripples can excite Bragg resonances and reflect the incident waves substantially, Mei et al. (1988) proposed the concept of artificial bars to protect drilling platforms on the oil fields in the Ekofisk of the North Sea against storm-wave attack, where the artificial bars on a flat seabed are conceived to be low in height, small in size, equidistant in spacing, and parallel to the coast. Later, the semi-circular bars and rectangular bars were considered numerically by Kirby (1987) and  Mattioli (1990), respectively, and the rectified cosinoidal bars were tested in a flume experiment by Kirby and Anton (1990) and in a field experiment by Bailard et al. (1990, 1992). Up to date, the artificial bars have drawn considerable attention, see
Hsu et al. (2003),
Cho et al. (2004),
Jeon and Cho (2006),
Wang et al. (2006),
Chang and Liou (2007),
Tsai et al. (2011),
Liu et al. (2015a, 2015b),
Liu et al. (2016),
Liu et al. (2020),
Guo et al. (2021),
Xie (2022) and so on.

For X-ray reflection by crystals, the quantitative mathematical principle of Bragg resonances is clear, which is Bragg's law, owing to the fact that X-rays are non-mechanical waves and their velocities do not depend on the propagation medium. However, for water wave reflection by sinusoidal ripples or artificial bars, the mathematical principle of Bragg resonances becomes quite complicated and is still unclear, mainly because water waves are mechanical and their phase velocities depend on water bodies, which are related to water depth and the number, amplitude, width, spacing, configuration of ripples or bars.

When the relative amplitudes of ripples or bars with respect to both the bottom wavelength and the water depth over the flat bottom are small enough, the mechanism of the nonlinear wave-wave interaction over a flat bottom derived by Phillips (1960) based on the regular perturbation was often borrowed as an approximate Bragg resonance mechanism, in which one or more of the free-surface wave components in Phillips' (1960) resonance conditions was replaced by periodic bottom components with zero frequencies since the ripples or bars are fixed in time, see p. 316-317 in Mei (1985), p. 499-500 in Madsen et al. (2006), and p. 5-6 in Xu et al. (2015). It is noted that the seabed in the problem of Bragg resonances is uneven due to the presence of artificial bars, while the seabed assumed in Phillips' (1960) mechanism is even. Obviously, the aforementioned borrowing is not reasonable in the strict sense, it is a choice without a choice. The condition for linear Class I Bragg resonances involving one bottom and two surface wave components determined in such an approximate way is in fact a throwback to Bragg's law in X-ray crystallography. That is why those Bragg resonances calculated by using perturbation methods (Davies, 1982; Miles, 1981;  Mei, 1985) still obey the traditional Bragg's law and occur nearly at $2d/L=1$.

However, when the relative amplitude of ripples or bars with respect to the global water depth and the bottom wavelength is not so small, the small parameter hypothesis in the perturbation method is no longer satisfied, so Phillips' (1960) mechanism of the nonlinear wave-wave interaction over a flat seabed cannot be employed. Equivalently, Bragg's law, which is a special case of Phillips' (1960) mechanism for Class I Bragg resonances, is no longer applicable. Indeed, in the case with the relative amplitude of ripples or bars being not so small, Class I Bragg resonances were often found to be excited at frequencies significantly lower than those predicted by the traditional Bragg's law, called phase (or frequency) downshift, see
Kirby (1986),
Mattioli (1990),
Massel (1993),
Belzons et al. (1991),
Guazzelli et al. (1992),
Rey (1992),
Liu and Yue (1998),
Madsen et al. (2006),
Liao et al. (2016),
Liu et al. (2018),
Liu et al. (2019a),
and Liang et al. (2020)
for sinusoidal ripples, and see
Kirby and Anton (1990),
Chang and Liou (2007),
Linton (2011),
Liu et al. (2015a),
Liu et al. (2016),
Liu et al. (2019b), and
Liu et al. (2020),
and Xie (2022) for artificial bars.

The phase downshift was attributed to evanescent modes (Guazzelli et al., 1992), nonlinear effects related to the seabed fluctuation (Liu and Yue, 1998; Madsen et al., 2006; Ardhuin and Magne, 2007; Peng et al., 2022), the decrease of phase velocity over bars (Chang and Liou, 2007; Liu et al., 2015a; Liu et al., 2020; Guo et al., 2021), and the bed slope (Tsai et al., 2017).
Recently, using Mathieu instability theorem, Liang et al. (2020) established a quantitative formula to elucidate the resonance mechanism of wave scattering by sinusoidal ripples, which also shows that the phase downshift is attributed to the bottom nonlinearity. However, up to date, any quantitative formula to improve Bragg's law for water wave reflection by a finite periodic array of artificial bars has never been found. Linton (2001, p.524) appealed to modify Bragg's law borrowed from X-ray crystallography for water wave reflection by artificial bars.

In this paper, based on Bloch band structures of linear long waves over five types of artificial bars (rectangular, parabolic, rectified cosinoidal, isosceles trapezoidal, and  isosceles triangular bars) given by An and Ye (2004), Liu (2017) and Liu et al. (2019b), a modified Bragg's law in the linear long wave range is established to replace the traditional Bragg's law.



\begin{figure}
\vspace*{-40mm}
\centerline{\hspace*{-3mm}\epsfxsize=3.0in \epsffile{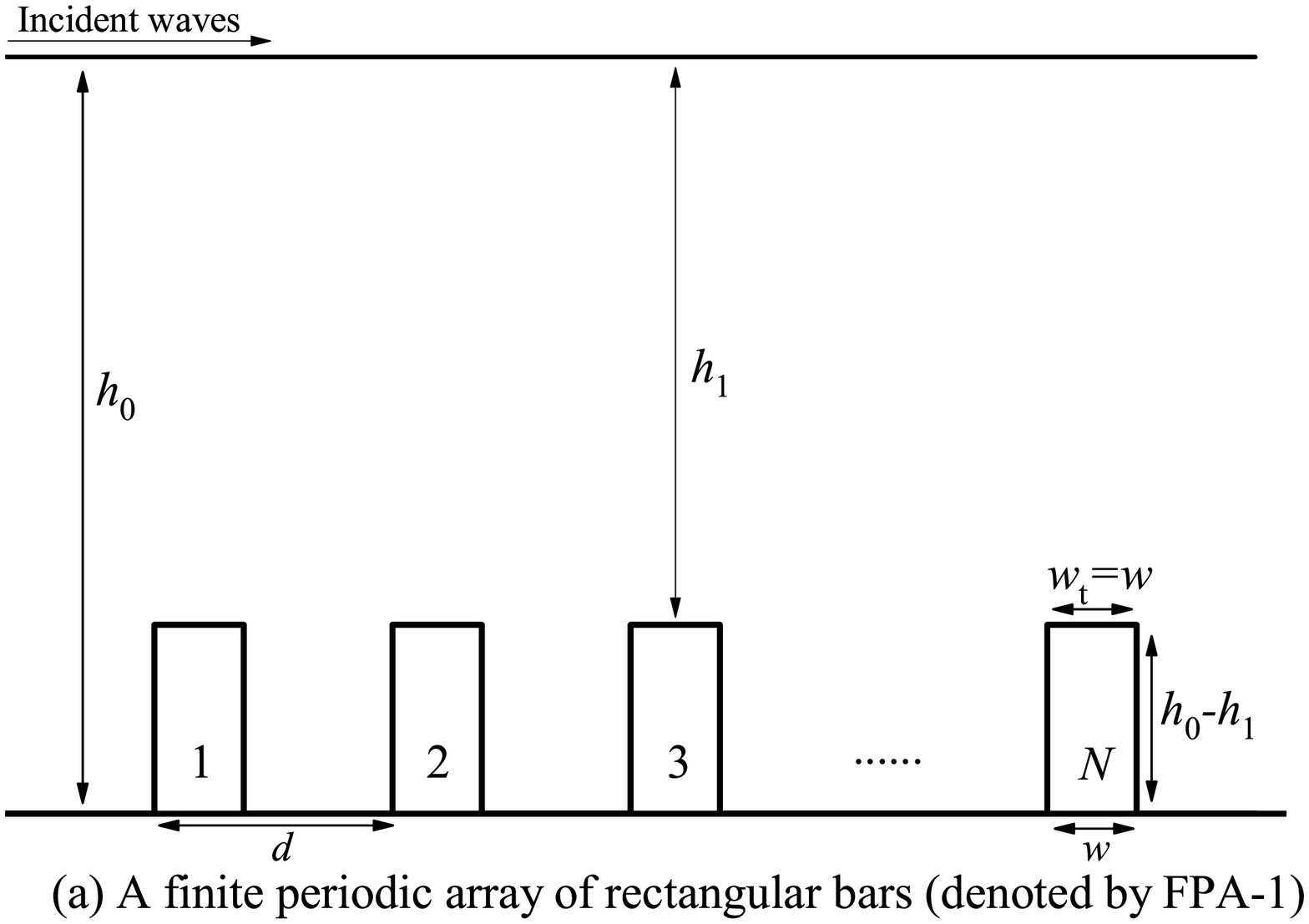}\hspace{-20mm}\epsfxsize=3.0in \epsffile{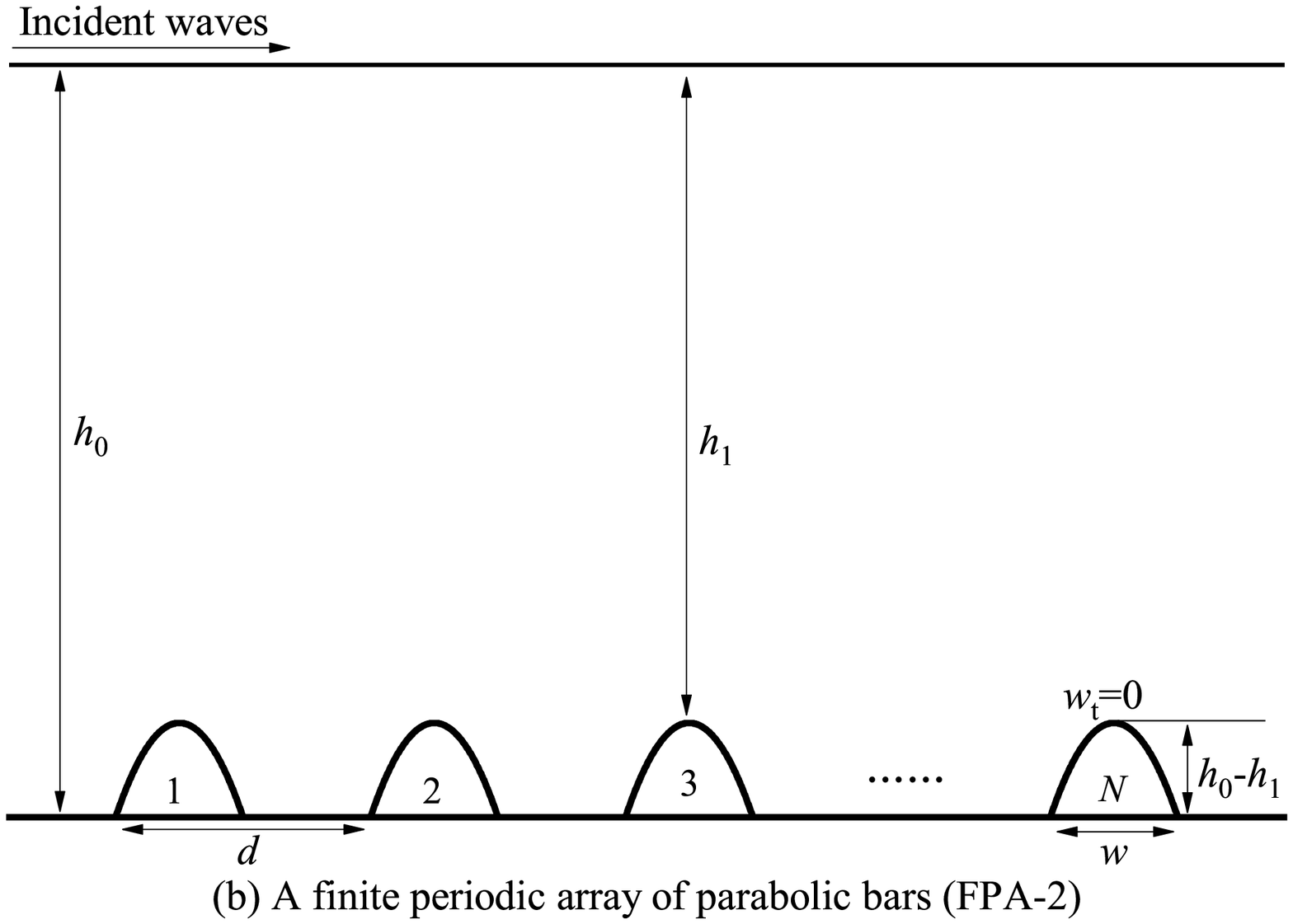}}
\vspace*{-16mm}
\centerline{\hspace*{-3mm}\epsfxsize=3.0in \epsffile{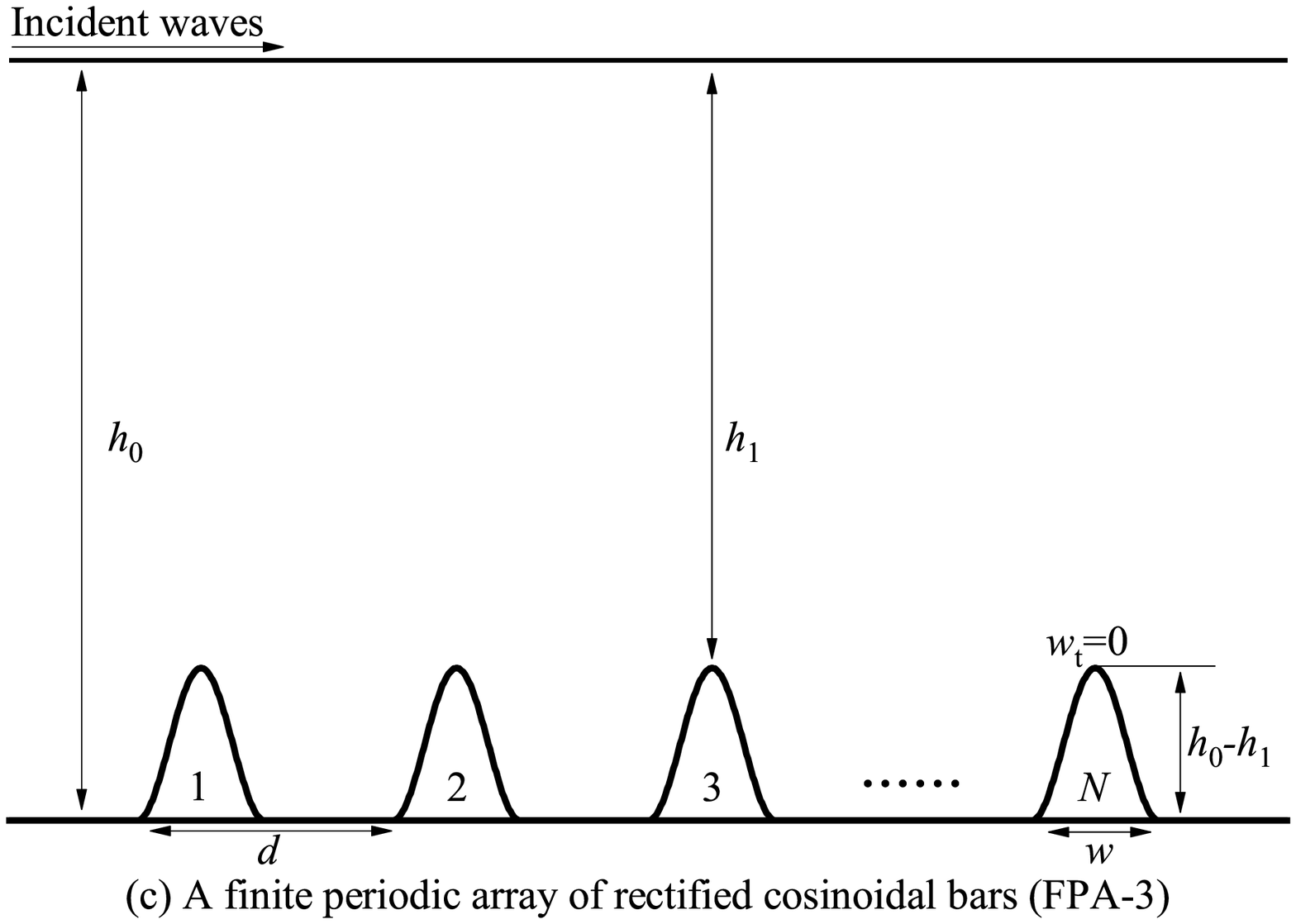}\hspace{-20mm}\epsfxsize=3.0in  \epsffile{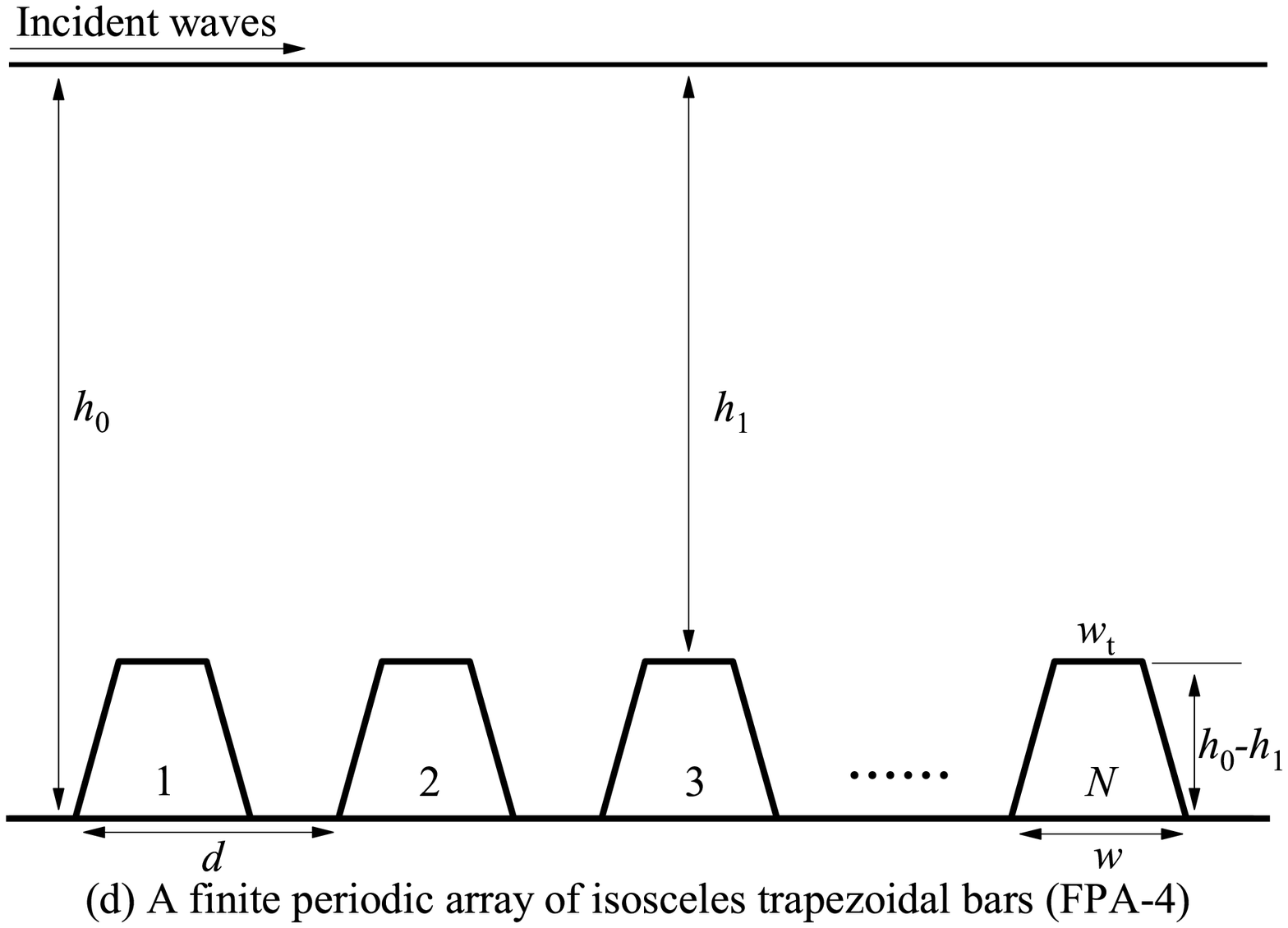}\hspace{-20mm}\epsfxsize=3.0in  \epsffile{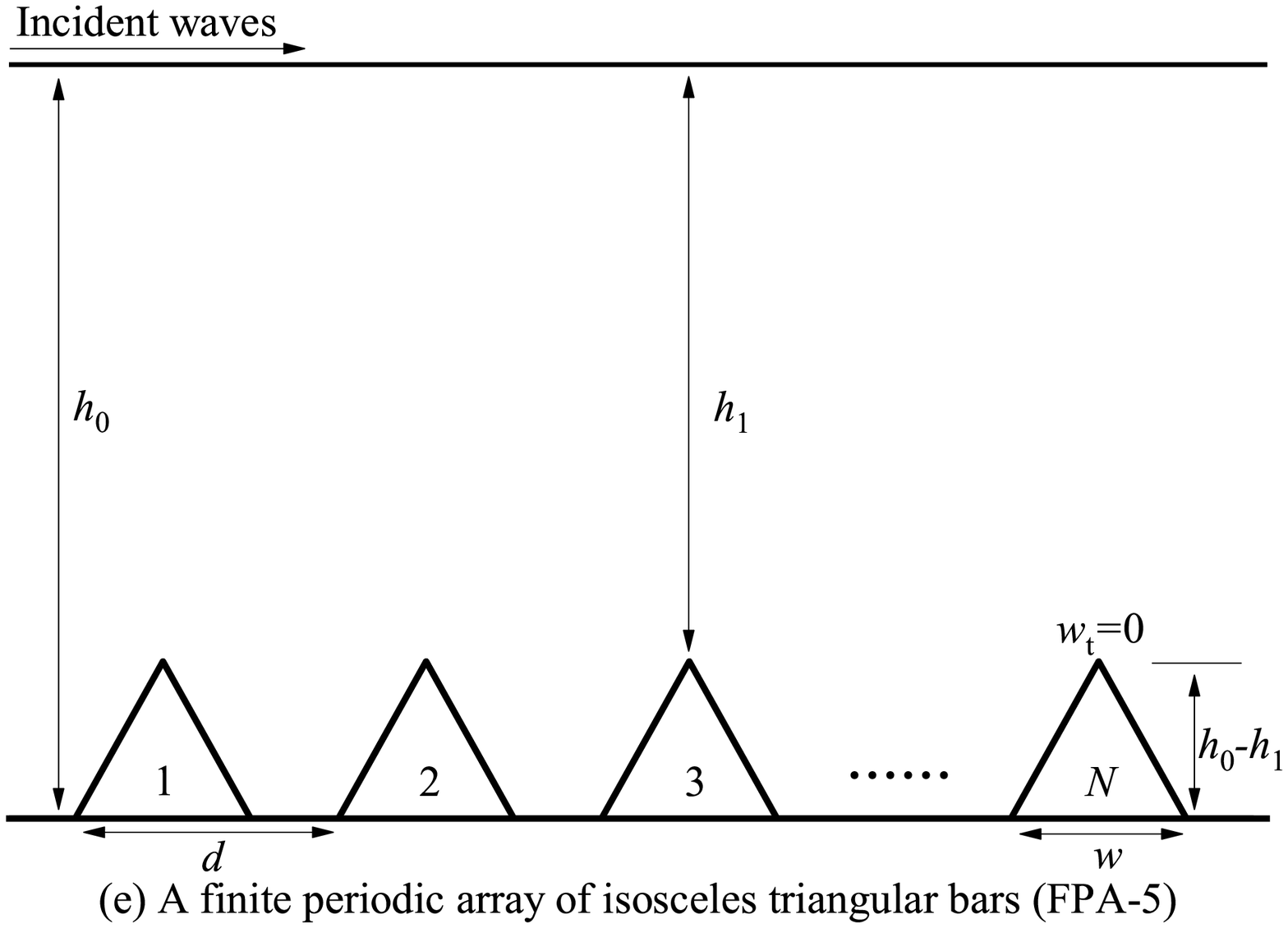}}
\vspace*{-11mm}
\caption[kdv]{\label{fig1} A finite periodic array of artificial bars: FPA-$j$, $j$=1,...,5. (a) FPA-1; (b) FPA-2; (c) FPA-3; (d) FPA-4; (e) FPA-5.}
\end{figure}

\section{Reflection of linear long waves by a finite periodic array of bars}

In this section, we consider reflection of linear long waves by a finite periodic array of artificial bars on an otherwise flat seabed. The artificial bars include the following five types: rectangular bars, parabolic bars, rectified cosinoidal bars, isosceles trapezoidal bars, and isosceles triangular bars, see Figure \ref{fig1}(a)-(e), where $N$ is the total number of bars, $h_0$ is the water depth at foot line of bars, $h_1$ is the submergence of bars, thus $h_0-h_1$ is the bar height, $d$ is the distance between any two adjacent bars. For all the five types of bars, their bottom and top widths are uniformly denoted by $w$ and $w_t$. Clearly, $w_t=w$ for rectangular bars, and $w_t=0$ for parabolic, rectified cosinoidal, and isosceles triangular bars. For convenience, the five types of finite periodic arrays are denoted by FPA-$j$, $j$=1,...,5, respectively, see Figure \ref{fig1}(a)-(e).

Suppose that the angular frequency of incident waves coming from the left side is $\omega$. According to the linear long-wave theory (Mei, 1989), the surface elevation $\eta(x)$ satisfies the linear long-wave equation as follows
\begin{equation}
h(x)\eta''(x)+h'(x)\eta'(x)+\frac{\omega^2}{g}\eta(x)=0,\label{1}
\end{equation}
where $h(x)$ is the water depth, and $g$ the gravitational acceleration.

Let $L$ denote the wavelength of the waves in the region with constant water depth $h_0$. We introduce the following dimensionless quantities:
\begin{equation}
H=\frac{h_0-h_1}{h_0},\quad W=\frac{w}{L},\quad W_t=\frac{w_t}{L},\quad D=\frac{d}{L}.\label{3}
\end{equation}

\subsection{Linear long wave reflection by FPA-1}

 Based on the the transfer matrix approach (Devillard et al., 1988), for linear long waves reflected by a finite periodic array of rectangular bars, FPA-1, with the total number of bars being $N$, it is  easy to obtain the analytical expression of the reflection coefficient as
\begin{equation}
K_R(1,H,W,W,D,N)=\frac{\left|c^{(1)}_2\right|}{\left|c^{(1)}_1\right|},
\label{Ref-Rectangle}
\end{equation}
where
\begin{eqnarray}
&&\left(\begin{array}{ll}
c^{(1)}_1\\
c^{(1)}_2
\end{array}
\right)
=\left(
\begin{array}{cc}
\frac{1}{2}  &  -\frac{\mbox{i}}{2}  \\
\frac{1}{2}  & \frac{\mbox{i}}{2} \\
\end{array}
\right)\left(AQ\right)^{N-1}A\left(
\begin{array}{cc}
1  \\
-\mbox{i}  \\
\end{array}
\right),\label{26}\\
%
%
%
&&A=
%
\left(
\begin{array}{cc}
\cos \frac{2\pi W}{\sqrt{1-H}} & \frac{1}{\sqrt{1-H}}\sin \frac{2\pi W}{\sqrt{1-H}}  \\
\sqrt{1-H}\sin \frac{2\pi W}{\sqrt{1-H}}     & -\cos \frac{2\pi W}{\sqrt{1-H}}  \\
\end{array}
\right),\label{28}\\
&&Q=\left(
\begin{array}{cc}
\cos \left[2\pi(D-W)\right]     & -\sin \left[2\pi(D-W)\right]  \\
-\sin \left[2\pi(D-W)\right]    & -\cos \left[2\pi(D-W)\right]  \\
\end{array}
\right).\label{29}
\end{eqnarray}

\subsection{Linear long wave reflection by FPA-2}

For linear long waves reflected by a finite periodic array of parabolic bars, FPA-2, with the total number of bars being $N$, Liu et al. (2015b) presented an analytical solution of the reflection coefficient as
\begin{equation}
K_R(2,H,W,0,D,N)=\frac{\left|c^{(2)}_2\right|}{\left|c^{(2)}_1\right|},
\label{Paraboloid-reflection}
\end{equation}
where
\small
\begin{eqnarray}
\hspace*{-5mm}&&\left(
\begin{array}{c}
c^{(2)}_1\\
c^{(2)}_2 \\
\end{array}
\right)=\left(
\begin{array}{cc}
\frac{1}{2}  &  -\frac{\mbox{i}}{2}  \\
\frac{1}{2}  & \frac{\mbox{i}}{2} \\
\end{array}
\right)\left(B_1B_2B_1^{-1}Q\right)^{N-1}B_1B_2B_1^{-1}\left(
\begin{array}{cc}
1  \\
-\mbox{i}  \\
\end{array}
\right),\label{3.1.14}\\
\hspace*{-5mm}&&B_1=\left(
\begin{array}{cc}
\mbox{P}^{\mu}_{-\frac{1}{2}}(\sqrt{H}) & \mbox{Q}^{\mu}_{-\frac{1}{2}}(\sqrt{H}) \\
\frac{H}{2\pi W}\mbox{P}^{\mu}_{-\frac{1}{2}}(\sqrt{H})-\frac{(1-H)\sqrt{H}}{\pi W}\mbox{P}^{\mu'}_{-\frac{1}{2}}(\sqrt{H}) & \frac{H}{2\pi W}\mbox{Q}^{\mu}_{-\frac{1}{2}}(\sqrt{H})-\frac{(1-H)\sqrt{H}}{\pi W}\mbox{Q}^{\mu'}_{-\frac{1}{2}}(\sqrt{H}) \\
\end{array}
\right), \label{3.1.15}\\
\hspace*{-5mm}&&B_2=
\left(
\begin{array}{cc}
\mbox{P}^{\mu}_{-\frac{1}{2}}\left(0\right) & \mbox{Q}^{\mu}_{-\frac{1}{2}}\left(0\right) \\
\mbox{P}^{\mu'}_{-\frac{1}{2}}\left(0\right) & \mbox{Q}^{\mu'}_{-\frac{1}{2}}\left(0\right)
\end{array}
\right)^{-1}
\left(
\begin{array}{cc}
\mbox{P}^{\mu}_{-\frac{1}{2}}\left(0\right) & \mbox{Q}^{\mu}_{-\frac{1}{2}}\left(0\right) \\
-\mbox{P}^{\mu'}_{-\frac{1}{2}}\left(0\right) & -\mbox{Q}^{\mu'}_{-\frac{1}{2}}\left(0\right)
\end{array}
\right),\label{3.1.17}
\end{eqnarray}
\normalsize
where $\mu=\sqrt{\frac{1}{4}-\frac{\pi^2W^2}{H}}$, and $\mbox{P}^{\mu}_{\nu}(t)$ and $\mbox{Q}^{\mu}_{\nu}(t)$ are the associated Legendre functions of the first kind and the second kind, respectively.

\subsection{Linear long wave reflection by FPA-3}

For linear long waves reflected by a finite periodic array of rectified cosinoidal bars, FPA-3, with the total number of bars being $N$, Liu et al. (2015a) presented an analytical solution of the reflection coefficient as
\begin{equation}
K_{R}(3,H,W,0,D,N)=\frac{\left|c^{(3)}_2\right|}{\left|c^{(3)}_1\right|},
\label{Reflection-Rectified-Cosinoidal}
\end{equation}
where
\begin{eqnarray}
&&\left(
\begin{array}{c}
c^{(3)}_1\\
c^{(3)}_2 \\
\end{array}
\right)=\left(
\begin{array}{cc}
\frac{1}{2}  &  -\frac{\mbox{i}}{2}  \\
\frac{1}{2}  & \frac{\mbox{i}}{2} \\
\end{array}
\right)
\left(E_1E_2E_1^{-1}Q\right)^{N-1}E_1E_2E_1^{-1}\left(
\begin{array}{cc}
1  \\
-\mbox{i}  \\
\end{array}
\right),\label{3.1.14}\\
&&E_1=\left(
\begin{array}{cc}
\xi_1(H) & \xi_2(H) \\
-\frac{(1-H)H}{2W}\xi'_1(H) & -\frac{(1-H)H}{2W}\xi'_2(H) \\
\end{array}
\right),\quad
E_2=\left(
\begin{array}{cc}
1 & 0 \\
0 & -1 \\
\end{array}
\right),\label{3.1.15}
\end{eqnarray}
and $\xi_1(x)$ and $\xi_2(x)$ are Heun general functions (Arscott, 1995) defined as follows
\begin{eqnarray}
&&\xi_1(x)=\mbox{HeunG}\left(\frac{2H}{1+H}, -\frac{4W^2}{1+H}; 0, 0, \frac{1}{2}, 0;x\right)=\sum_{m=0}^{\infty}a_m x^m,\label{Heun-A}\\
&&\xi_2(x)=\sqrt{x}\;\mbox{HeunG}\left(\frac{2H}{1+H}, \frac{1}{4}-\frac{4W^2}{1+H}; \frac{1}{2}, \frac{1}{2}, \frac{3}{2}, 0; x\right)
=\sqrt{x}\;\sum_{m=0}^{\infty}b_m x^m, \label{Heun-B}\qquad
\end{eqnarray}
where the coefficients $a_m$ and $b_m$ are determined by the following recursive formulae
\small
\begin{eqnarray}
&&\hspace*{-5mm} a_0=1,\qquad a_1=-\frac{4W^2}{H},\\
&&\hspace*{-5mm} a_m=\frac{(m-1)\left(3mH+m-4H-1\right)-4W^2}{m(2m-1)H}a_{m-1}-\frac{(1+H)(m-2)^2}{m(2m-1)H}a_{m-2},\; m=2,3,...\qquad
\end{eqnarray}
and
\begin{eqnarray}
&&\hspace*{-6mm}b_0=1,\qquad b_1=\frac{1+H-16W^2}{12H},\\
&&\hspace*{-6mm}b_m=\frac{\left(m-\frac{1}{2}\right)\left[m(1+3H)-\frac{5}{2}H-\frac{1}{2}\right]-4W^2}{m(2m+1)H}b_{m-1}
-\frac{(1+H)\left(m-\frac{3}{2}\right)^2}{m(2m+1)H}b_{m-2},\; m=2,3,...\qquad
\end{eqnarray}

\subsection{Linear long wave reflection by FPA-4}


For linear long waves reflected by a finite periodic array of general trapezoidal bars, Chang and Liou (2007) presented an analytical solution of the reflection coefficient via dimensional parameters. Here, for wave reflection by a finite periodic array of isosceles trapezoids, FPA-4, with the total number of bars being $N$, the reflection coefficient of Chang and Liou (2007) is simplified into the following dimensionless form
\begin{equation}
K_{R}(4,H,W,W_t,D,N)=\frac{\left|c^{(4)}_2\right|}{\left|c^{(4)}_1\right|},
\label{Ref-trapezoid}
\end{equation}
where
\begin{eqnarray}
&&\left(
\begin{array}{c}
c^{(4)}_1\\
c^{(4)}_2 \\
\end{array}
\right)=\left(
\begin{array}{cc}
\frac{1}{2}  &  -\frac{\mbox{i}}{2}  \\
\frac{1}{2}  & \frac{\mbox{i}}{2} \\
\end{array}
\right)\left(F_1F_2F_1^{-1}Q\right)^{N-1}F_1F_2F_1^{-1}\left(
\begin{array}{cc}
1  \\
-\mbox{i}  \\
\end{array}
\right),\label{3.1.14}\\
&&F_1
%
%
%
%
=-\frac{\pi\beta}{2}\left(
\begin{array}{cc}
C_{01}(\alpha,\beta) & -C_{00}(\alpha,\beta) \\
C_{11}(\alpha,\beta) & -C_{10}(\alpha,\beta) \\
\end{array}
\right), \;\;
C_{ij}(\alpha,\beta)=\left|
\begin{array}{cc}
\mbox{J}_i(\alpha)&\mbox{J}_j(\beta)\\
\mbox{Y}_i(\alpha)&\mbox{Y}_j(\beta)
\end{array}
\right|,\;i,j=0,1, \label{3.1.17}\\
&&F_2=
\left(
\begin{array}{cc}
\cos \frac{2\pi W_t}{\sqrt{1-H}} & \sin \frac{2\pi W_t}{\sqrt{1-H}} \\
\sin \frac{2\pi W_t}{\sqrt{1-H}} & -\cos \frac{2\pi W_t}{\sqrt{1-H}}
\end{array}
\right),\label{3.1.18}
\end{eqnarray}
where $\alpha=\frac{2\pi}{H}\left(W- W_t\right)$, $\beta=\sqrt{1-H}\alpha$,
and $\mbox{J}_m$ and $\mbox{Y}_m$ are the Bessel functions of the first and second kinds to order $m$, respectively.

When $W_t\rightarrow W$, the isosceles trapezoidal bar degenerates into  the rectangular bar. Since $\alpha\rightarrow 0$, $\beta\rightarrow 0$, and $\frac{\beta}{\alpha}=\sqrt{1-H}$, according to Abramowitz and Stegun (1972), we have
%
%
%
\begin{equation}
C_{00}(\alpha,\beta)\approx\frac{\ln(1-H)}{\pi},\;
C_{01}(\alpha,\beta)\approx-\frac{2}{\pi\beta},\;
C_{10}(\alpha,\beta)\approx\frac{2}{\pi\alpha},\;
C_{11}(\alpha,\beta)\approx-\frac{H}{\pi \sqrt{1-H}}.\label{substitution}
\end{equation}
Substituting Eq. (\ref{substitution}) into Eq. (\ref{3.1.17}), we have
\begin{equation}
F_1\approx\left(\begin{array}{cc}
1 & 0\\
0 & \sqrt{1-H}
\end{array}\right),\quad
F_1F_2F_1^{-1}\approx A.
\end{equation}
Hence, $K_R(4,H,W,W_t,D,N)$ defined in (\ref{Ref-trapezoid}) degenerates into $K_R(1,H,W,W,D,N)$ defined in (\ref{Ref-Rectangle}).

\subsection{Linear long wave reflection by FPA-5}

For linear long waves reflected by a finite periodic array of  isosceles triangular bars, FPA-5, with the total number of bars being $N$, Liu et al. (2015a) presented an analytical solution of the reflection coefficient as
\begin{equation}
K_R(5,H,W,0,D,N)=\frac{\left|c^{(5)}_2\right|}{\left|c^{(5)}_1\right|},
\label{Ref-triangular}
\end{equation}
where
\begin{eqnarray}
&&\left(
\begin{array}{c}
c^{(5)}_1\\
c^{(5)}_2 \\
\end{array}
\right)=\left(
\begin{array}{cc}
\frac{1}{2}  &  -\frac{\mbox{i}}{2}  \\
\frac{1}{2}  & \frac{\mbox{i}}{2} \\
\end{array}
\right)\left(G_1E_2G_1^{-1}Q\right)^{N-1}G_1E_2G_1^{-1}\left(
\begin{array}{cc}
1  \\
-\mbox{i}  \\
\end{array}
\right),\label{3.1.14}\\
&&G_1
=-\frac{\pi \delta}{2}\left(
\begin{array}{cc}
C_{01} \left(\gamma, \delta\right) &
-C_{00}\left(\gamma, \delta\right) \\
C_{11} \left(\gamma, \delta\right) &
-C_{10}\left(\gamma, \delta\right) \\
\end{array}
\right), \label{3.1.17-b}
\end{eqnarray}
where $\gamma=\frac{2\pi W}{H}$, and $\delta=\sqrt{1-H}\gamma$.

\section{Dispersion relations of Bloch long waves}

\begin{figure}
\vspace*{-40mm}
\centerline{\hspace*{-3mm}\epsfxsize=3.0in \epsffile{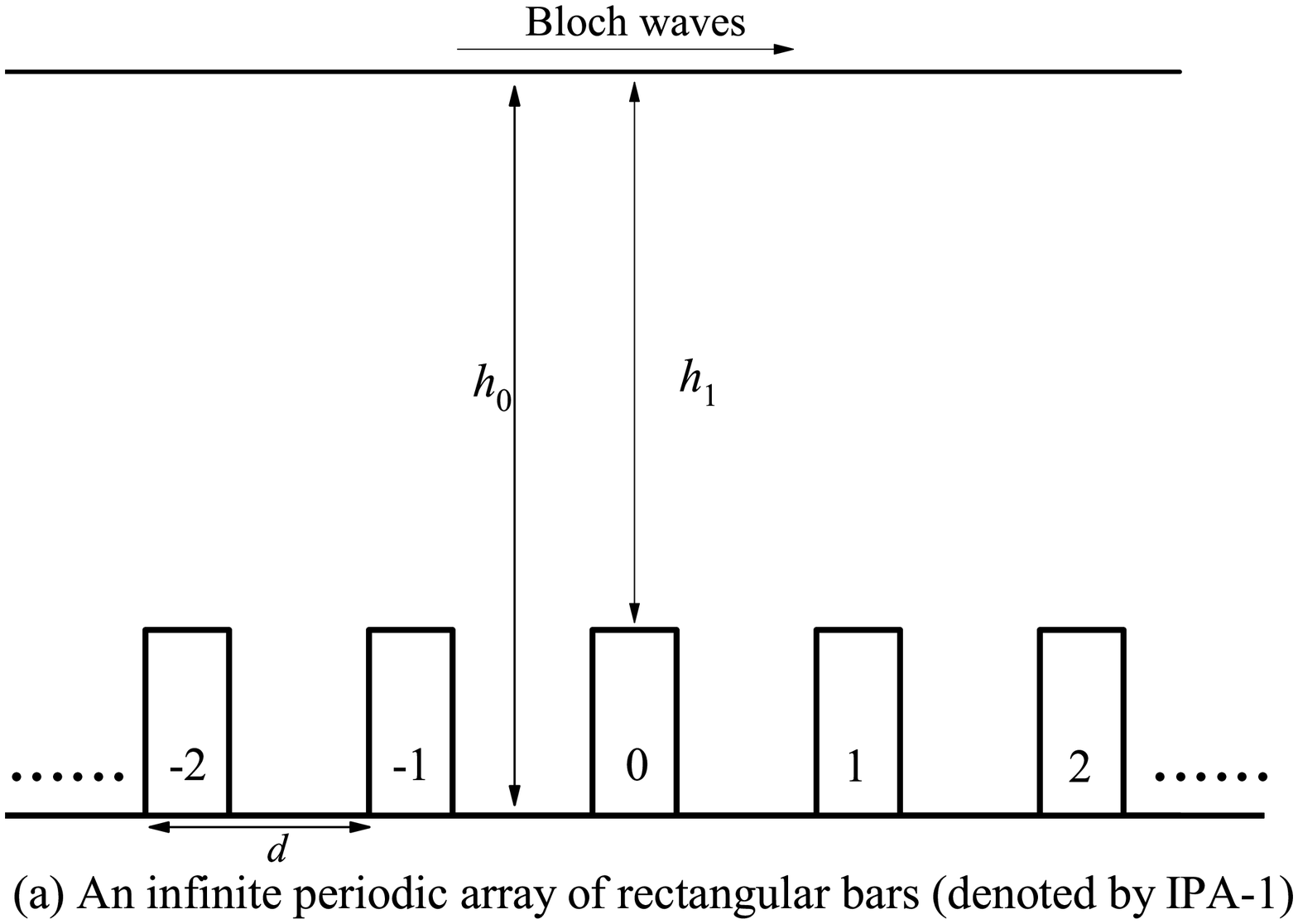}\hspace{-20mm}\epsfxsize=3.0in \epsffile{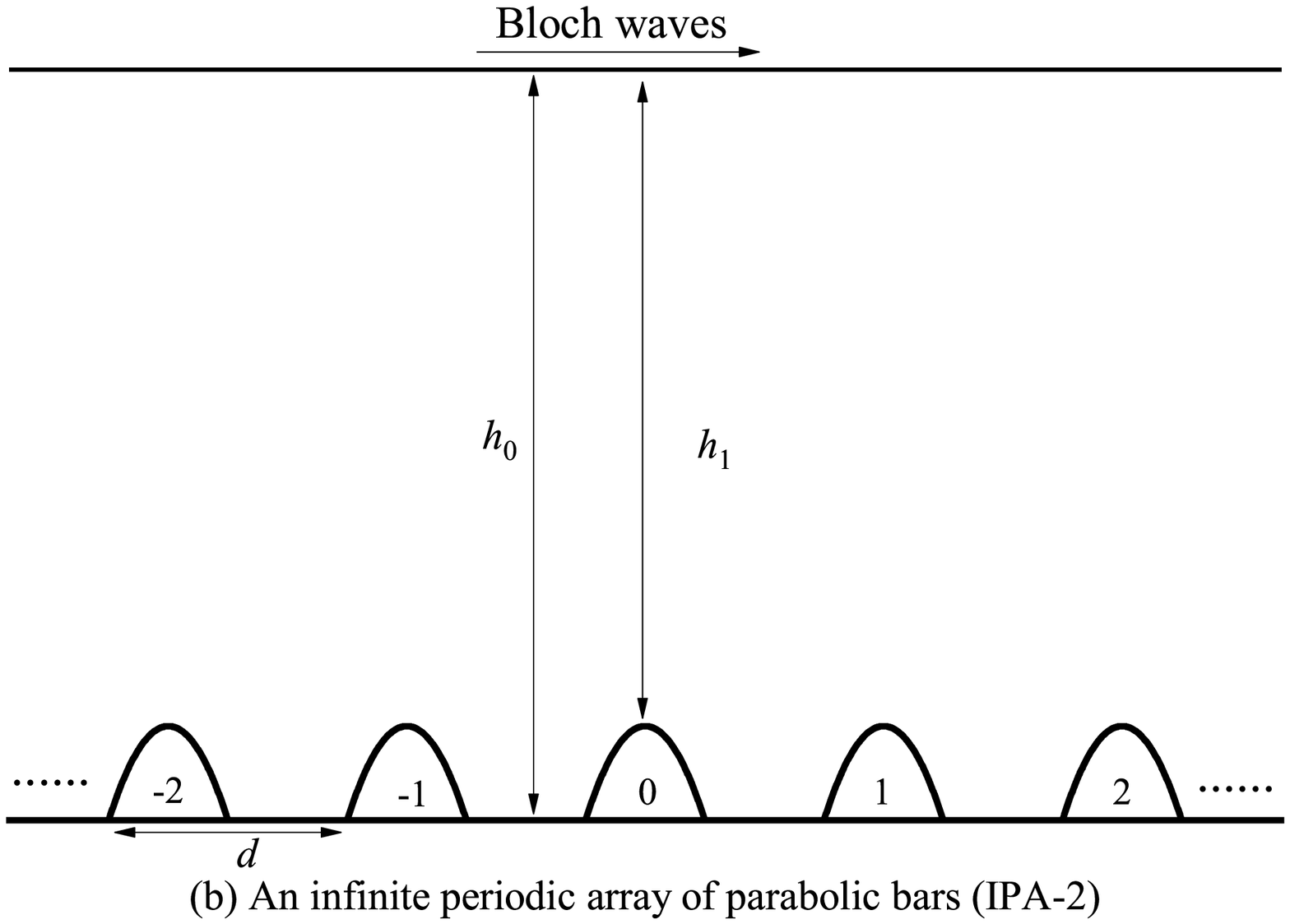}}
\vspace*{-16mm}
\centerline{\hspace*{-3mm}\epsfxsize=3.0in \epsffile{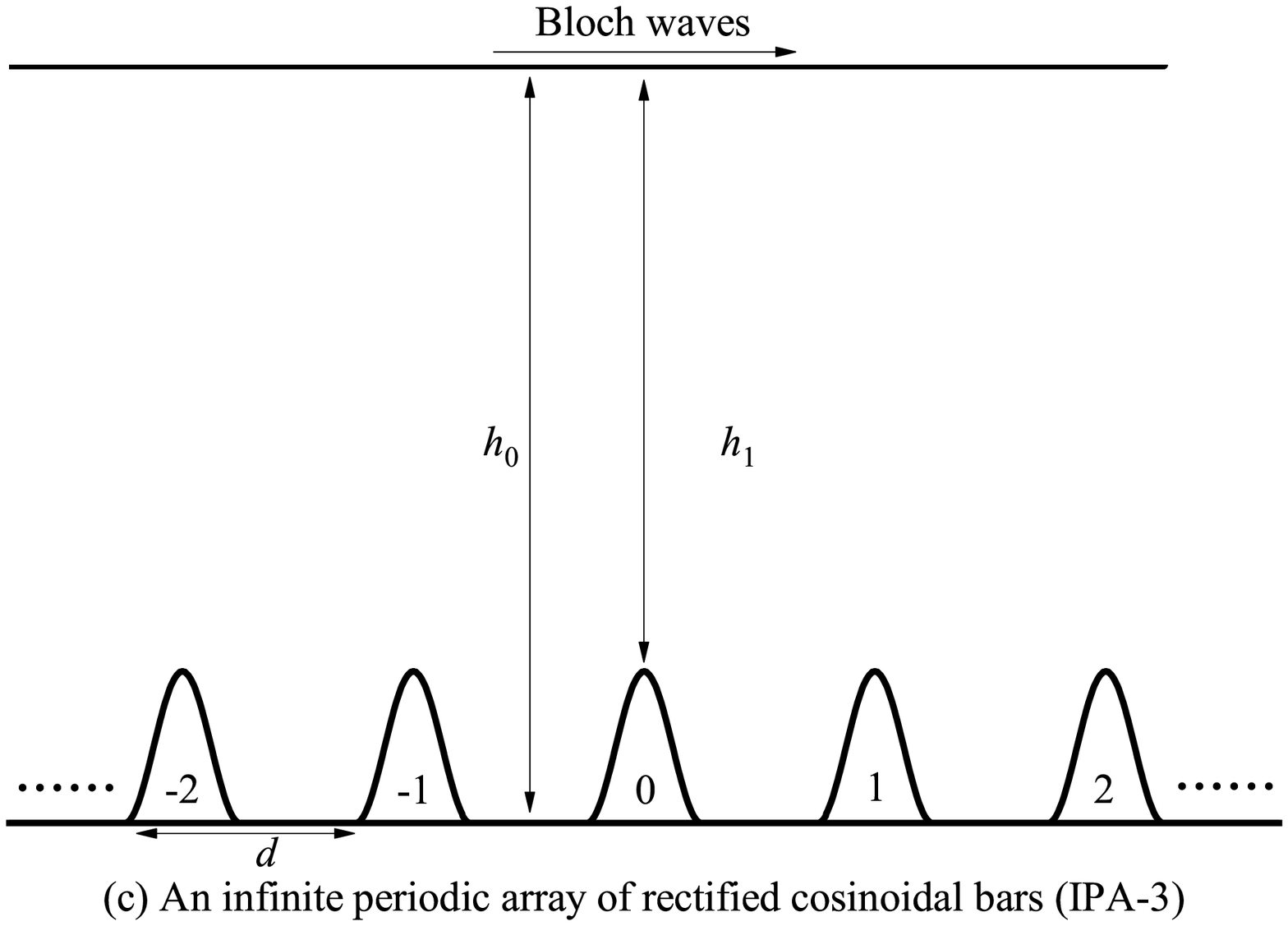}\hspace{-20mm}\epsfxsize=3.0in  \epsffile{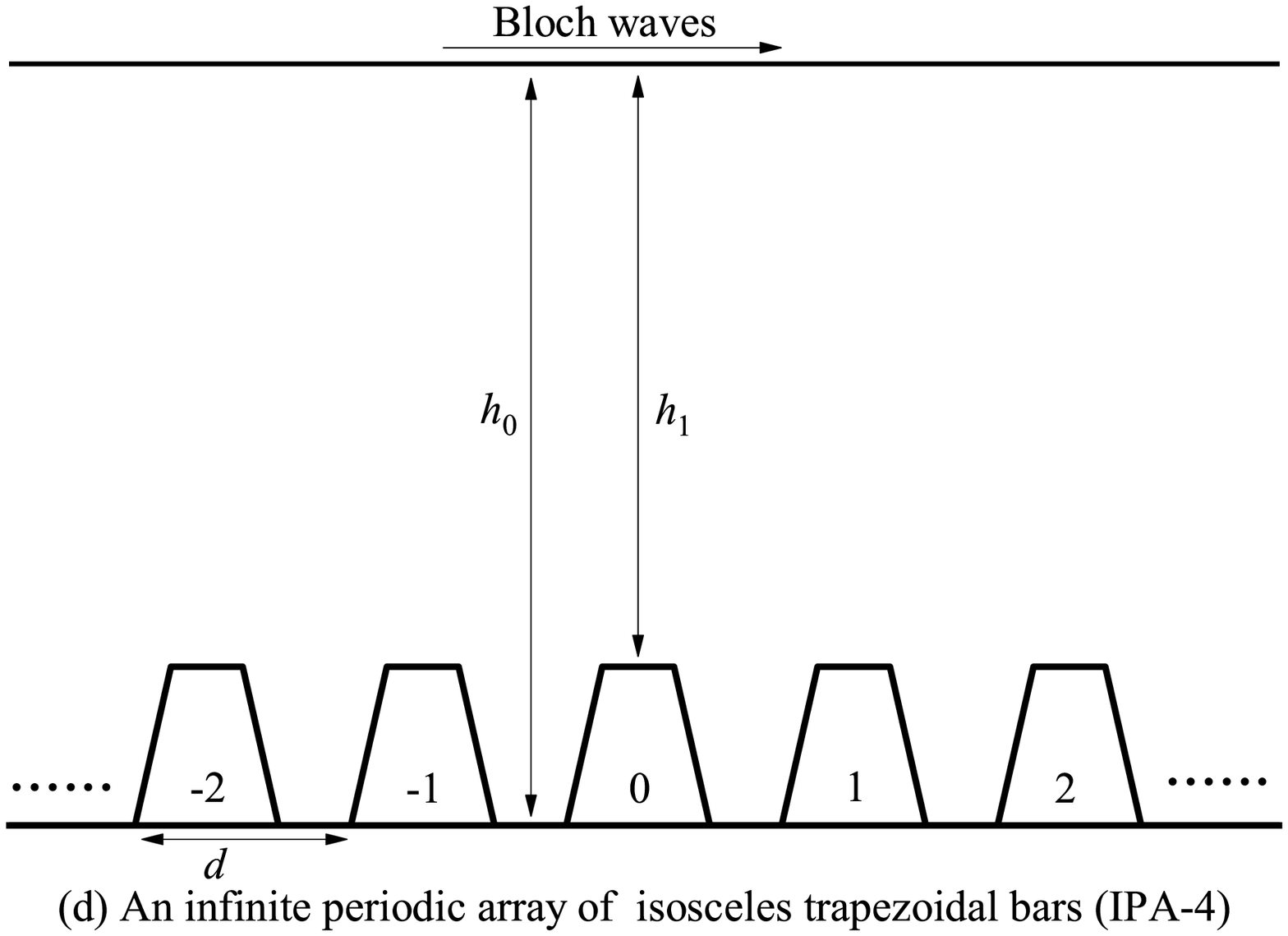}\hspace{-20mm}\epsfxsize=3.0in  \epsffile{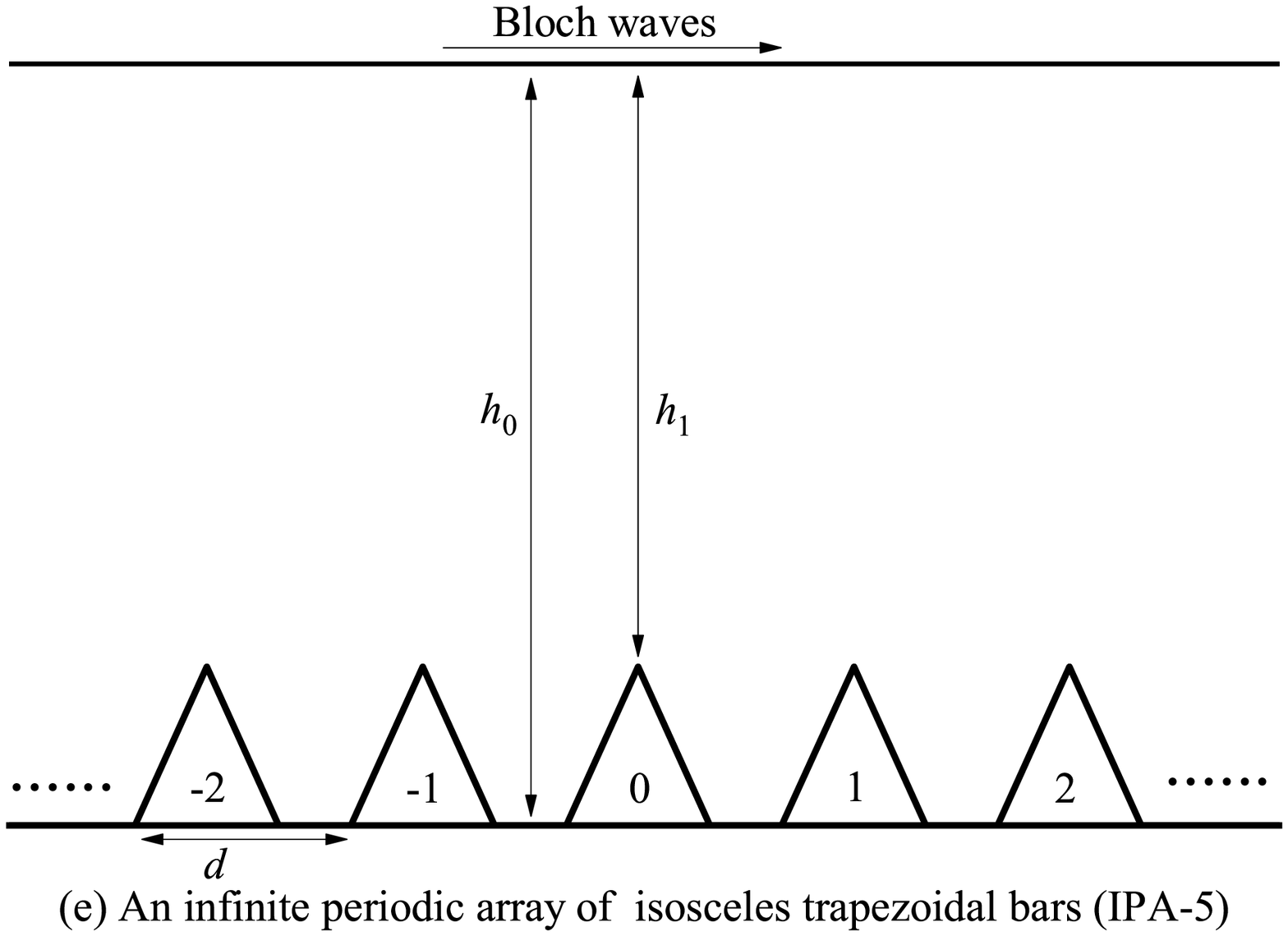}}
\vspace*{-11mm}
\caption[kdv]{\label{fig2} An infinite periodic array of artificial bars: IPA-$j$, $j$=1,2,3,4,5. (a) IPA-1; (b) IPA-2; (c) IPA-3; (d) IPA-4; (e) IPA-5.}
\end{figure}

In this section, we consider the existence of Bloch long waves over an infinite periodic array of artificial bars. Firstly, for $j$=1,...,5, we extend the finite periodic array of artificial bars, FPA-$j$, in both directions to form an infinite periodic array of artificial bars, which is denoted by IPA-$j$, see Figure \ref{fig2}(a)-(d). Then according to Bloch theorem (Ascroft and Mermin, 1976; Joannopoulos et al., 1995; Chen et al., 2005), water surface waves propagating over each IPA-$j$ ($j$=1,...,5) will exhibit Bloch state and the wave elevation can be expressed into Bloch form as
\begin{equation}
\eta_j(x) = u_j(x)\mbox{e}^{\mbox{i}K_jx}, \label{25}
\end{equation}
where $K_j$ is the wave number of Bloch waves modulated by IPA-$j$, and $u_j(x)$ is a periodic function with the period being $d$, i.e., $u_j(x+d)=u_j(x)$.

It is noted that, owing to Bloch form (\ref{25}), the representation of Bloch waves, $\eta_j(x)$, for $x$ in the entire interval $(-\infty, +\infty)$ can be obtained through the representation of $\eta_j(x)$ over a single bar. By solving Eq. (\ref{1}) analytically over a single bar and using Eq. (\ref{25}), Bloch dispersion relation can be established. The details of Bloch dispersions relations of Bloch long waves over IPA-$j$, $j$=1,...,5, are as follows.

\subsection{The dispersion relation of Bloch long waves over IPA-1}

An and Ye (2004) established a closed-form dispersion relation of Bloch waves over IPA-1 as follows
\begin{eqnarray}
\cos K_1d
=\cos \left(k_1w\right)\cos \left[k_0(d-w)\right]
-\cosh\left(\ln\frac{k_1}{k_0}\right)\sin \left(k_1w\right)\sin \left[k_0(d-w)\right], \label{rectangle0}
\end{eqnarray}
where $k_0$ and $k_1$ are Airy's wave numbers with respect to water depths $h_0$ and $h_1$, respectively. It is noted that $\omega^2\approx gk_0^2h_0$ and $\omega^2\approx gk_1^2h_1$ in the linear long-wave range. Hence, for linear long waves, Bloch dispersion relation (\ref{rectangle0}) degenerates into the following dimensionless form
\begin{equation}
\cos K_1d=U_1\cos \left[2\pi(D-W)\right]-V_1\sin \left[2\pi(D-W)\right], \label{rectangle}
\end{equation}
where
\begin{equation}
U_1(H,W)=\cos \frac{2\pi W}{\sqrt{1-H}},\quad\quad
V_1(H,W)=\frac{2-H}{2\sqrt{1-H}}\sin \frac{2\pi W}{\sqrt{1-H}}.
\end{equation}

\subsection{The dispersion relation of Bloch long waves over IPA-2}

By employing the associated Legendre functions of the first kind and the second kind, i.e., $\mbox{P}^{\mu}_{\nu}(x)$ and $\mbox{Q}^{\mu}_{\nu}(x)$, Liu et al. (2019b) derived a closed-form dispersion relation of Bloch long waves over IPA-2. However, the calculation is inconvenient since the associated Legendre functions are complex functions. Here an identical Bloch dispersion relation in terms of Frobenius series is presented as
\begin{equation}
\cos K_2d=\frac{U_2\cos \left[2\pi(D-W)\right]-V_2\sin \left[2\pi(D-W)\right]}{\eta_1(H)\eta'_2(H)-\eta'_1(H)\eta_2(H)},\label{Paraboloid}
\end{equation}
where
\begin{eqnarray}
&&U_2(H,W)=\eta_1(H)\eta'_2(H)+\eta'_1(H)\eta_2(H),\\
&&V_2(H,W)=\frac{\pi W}{2H(1-H)}{\eta_1(H)\eta_2(H)}-\frac{2H(1-H)}{\pi W} \eta'_1(H)\eta'_2(H),
\end{eqnarray}
%
and the two Frobenius series $\eta_1(x)$ and $\eta_2(x)$ are defined as follows
\begin{eqnarray}
\eta_1(x)=\sum_{m=0}^{\infty}a_{m}x^m,\quad \eta_2(x)=\sqrt{x}\sum_{m=0}^{\infty}b_{m}x^m,
\label{2.17}
\end{eqnarray}
where the coefficients $a_m$ and $b_m$ are defined by the following recursive formulae:
\begin{eqnarray}
&&a_0=1, \qquad a_1=-\frac{\pi^2W^2}{2H},\\
&&a_m=-\frac{2(m-2)^2}{m\left(2m-1\right)}a_{m-2}
-\frac{\frac{\pi^2W^2}{2H}+(m-1)(5-4m)}{m\left(2m-1\right)}a_{m-1},\;m=2,3,...
\end{eqnarray}
and
\begin{eqnarray}
&&b_0=1,\qquad b_1=\frac{H-\pi^2W^2}{6H},\\
&&b_m=-\frac{\left(2m-3\right)^2}{2m(2m+1)}b_{m-2}
-\frac{\frac{\pi^2W^2}{H}+(2m-1)(3-4m)}{2m(2m+1)}b_{m-1},\;m=2,3,...
\end{eqnarray}

\subsection{The dispersion relation of Bloch long waves over IPA-3}

Liu et al. (2019b) also derived a closed-form dispersion relation of Bloch long waves modulated by IPA-3 as follows
\begin{equation}
\cos K_3d =\frac{U_3\cos \left[2\pi(D-W)\right]-V_3\sin \left[2\pi(D-W)\right]}{\xi_1(H)\xi'_2(H)-\xi'_1(H)\xi_2(H)},\label{Rectified-band}
\end{equation}
where
\begin{eqnarray}
&&U_3(H,W)=\xi_1(H)\xi'_2(H)+\xi'_1(H)\xi_2(H),\\
&&V_3(H,W)=\frac{2W}{H(1-H)}{\xi_1(H)\xi_2(H)}-\frac{H(1-H)}{2W} \xi'_1(H)\xi'_2(H),
\end{eqnarray}
where $\xi_1(x)$ and $\xi_2(x)$ are defined in Eqs. (\ref{Heun-A}) and (\ref{Heun-B}), respectively.

\subsection{The dispersion relation of Bloch long waves over IPA-4}

By solving Eq. (\ref{1}), Liu (2017) derived a closed-form dispersion relation of Bloch long waves over IPA-4. However, in preparation of this paper, it is found that the dispersion relation given by Liu (2017) contains a mistake and it should be corrected as
\begin{equation}
\cos K_4d=-\frac{\alpha\beta\pi^2}{4}\left\{U_4\cos \left[2\pi(D-W)\right]-V_4\sin \left[2\pi(D-W)\right]\right\},\label{Bandgap-Trapezoid}
\end{equation}
where
\begin{eqnarray}
U_4(H,W,W_t)&=&\left[C_{00}(\alpha,\beta)C_{01}(\alpha,\beta)- C_{10}(\alpha,\beta)C_{11}(\alpha,\beta)\right]\sin \frac{2\pi W_t}{\sqrt{1-H}}\nonumber\\
&&+
\left[C_{00}(\alpha,\beta)C_{11}(\alpha,\beta)+ C_{01}(\alpha,\beta)C_{10}(\alpha,\beta)\right]\cos \frac{2\pi W_t}{\sqrt{1-H}},\\
V_4(H,W,W_t)&=&\frac{1}{2}\left[C^2_{00}(\alpha,\beta)+C^2_{11}(\alpha,\beta)- C^2_{10}(\alpha,\beta)-C^2_{01}(\alpha,\beta)\right]\sin \frac{2\pi W_t}{\sqrt{1-H}},\nonumber\\
&&+\left[C_{10}(\alpha,\beta)C_{00}(\alpha,\beta)
-C_{01}(\alpha,\beta)C_{11}(\alpha,\beta)\right]\cos\frac{2\pi W_t}{\sqrt{1-H}}.
\end{eqnarray}

When $W_t\rightarrow W$, the isosceles trapezoidal bar degenerates into  the rectangular bar. Substituting Eq. (\ref{substitution}) into Eq. (\ref{Bandgap-Trapezoid}) leads to the dispersion relation of Bloch long waves over IPA-1, i.e., Eq. (\ref{rectangle}).

%
%
%
%

\subsection{The dispersion relation of Bloch long waves over IPA-5}

By setting $W_t=0$ in Eq. (\ref{Bandgap-Trapezoid}), we can obtain the following closed-form dispersion relation of Bloch long waves over IPA-5:
\begin{equation}
\cos K_5d=-\frac{\gamma\delta\pi^2}{4}\left\{U_5\cos \left[2\pi(D-W)\right]-V_5\sin \left[2\pi(D-W)\right]\right\},\label{Bandgap-Triangle}
\end{equation}
where
\begin{eqnarray}
&&U_5(H,W)=C_{00}(\gamma,\delta)C_{11}(\gamma,\delta)
+C_{01}(\gamma,\delta)C_{10}(\gamma,\delta),\\
&&V_5(H,W)=C_{10}(\gamma,\delta)C_{00}(\gamma,\delta)-C_{01}(\gamma,\delta)C_{11}(\gamma,\delta).
\end{eqnarray}

\section{Correspondence between Bragg resonance band and Bloch forbidden band}

In Section 2, the reflections of linear long waves by five types of finite periodic array of artificial bars, FPA-$j$, $j$=1,...,5, were summarized. It is well known that, under certain conditions, Bragg resonances will be excited and the reflected waves may be significantly amplified.

In Section 3, the dispersion relations of Bloch long waves $\eta_j$ over IPA-$j$, $j$=1,...,5, were also summarized. According to the band theory, the Bloch wave $\eta_j$ will be characterized by the dispersion relation or the band structure corresponding to the dispersion relation. In the band structure, between two adjacent bands there may exist a forbidden band within which $|\cos Kd|>1$, i.e., the Bloch wave $\eta_j$ cannot exist, for example, see Fig. 2 in An and Ye (2003) and Fig. 2 in Liu et al. (2019b).

Bragg resonances excited by FPA-$j$ and the existence of Bloch state over IPA-$j$ are two different problems in mathematics. The  former is a boundary value problem, and the latter is an eigenvalue problem. However, they are very closely related. According to Porter and Porter (2003, p.162, lines 7-10), the frequency bands supporting the existence of Bragg resonances excited by FPA-$j$ correspond to Bloch forbidden bands caused by IPA-$j$. And according to Linton (2011, p.505, lines 31-33), if the band structure of Bloch waves over IPA-$j$ has a forbidden band, then one would expect that this would lead to a Bragg resonance band when the surface wave is propagating over FPA-$j$ at a frequency within such a forbidden band. This essential connection between Bragg resonance bands and Bloch forbidden bands was verified by Liu et al. (2019b), see their Fig. 3 for parabolic bars and Fig. 11 for rectified cosinoidal bars. Here, we further demonstrate this essential connection for rectangular bars, trapezoidal bars and triangular bars.

\begin{figure}
\vspace*{-25mm}
\centerline{\hspace*{0mm}\epsfxsize=2.8in \epsffile{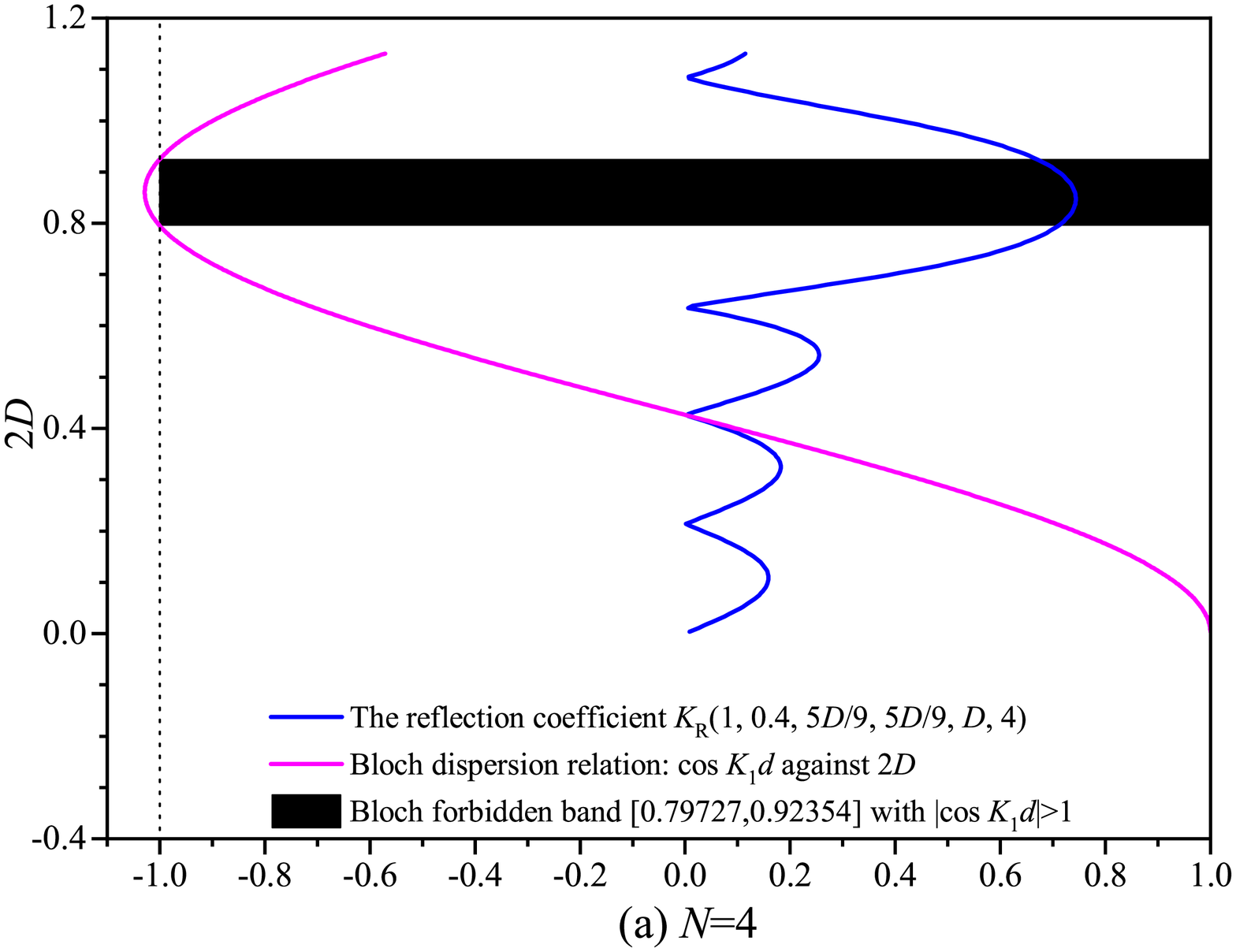}\hspace{-15mm}\epsfxsize=2.8in \epsffile{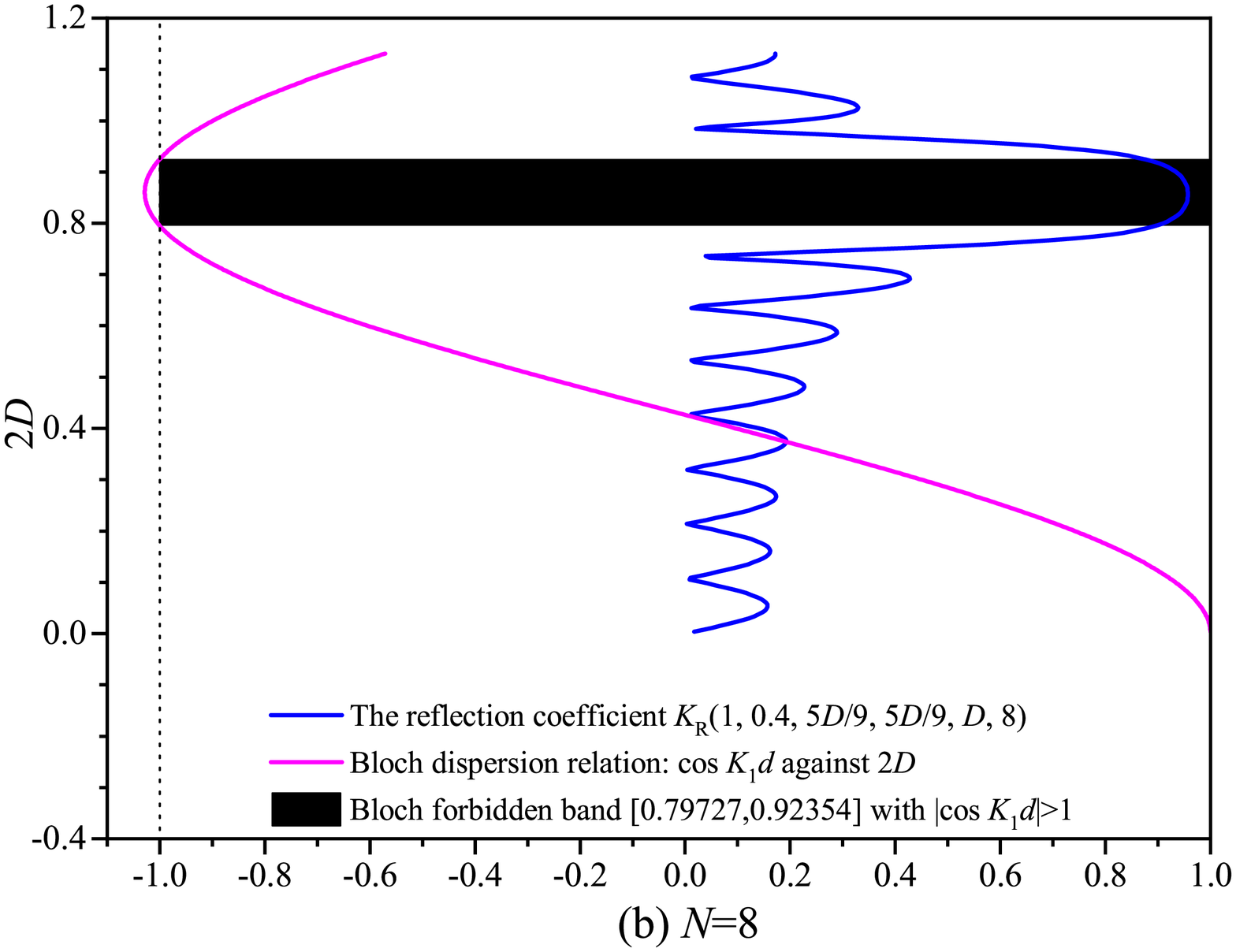}\hspace{-15mm}\epsfxsize=2.8in \epsffile{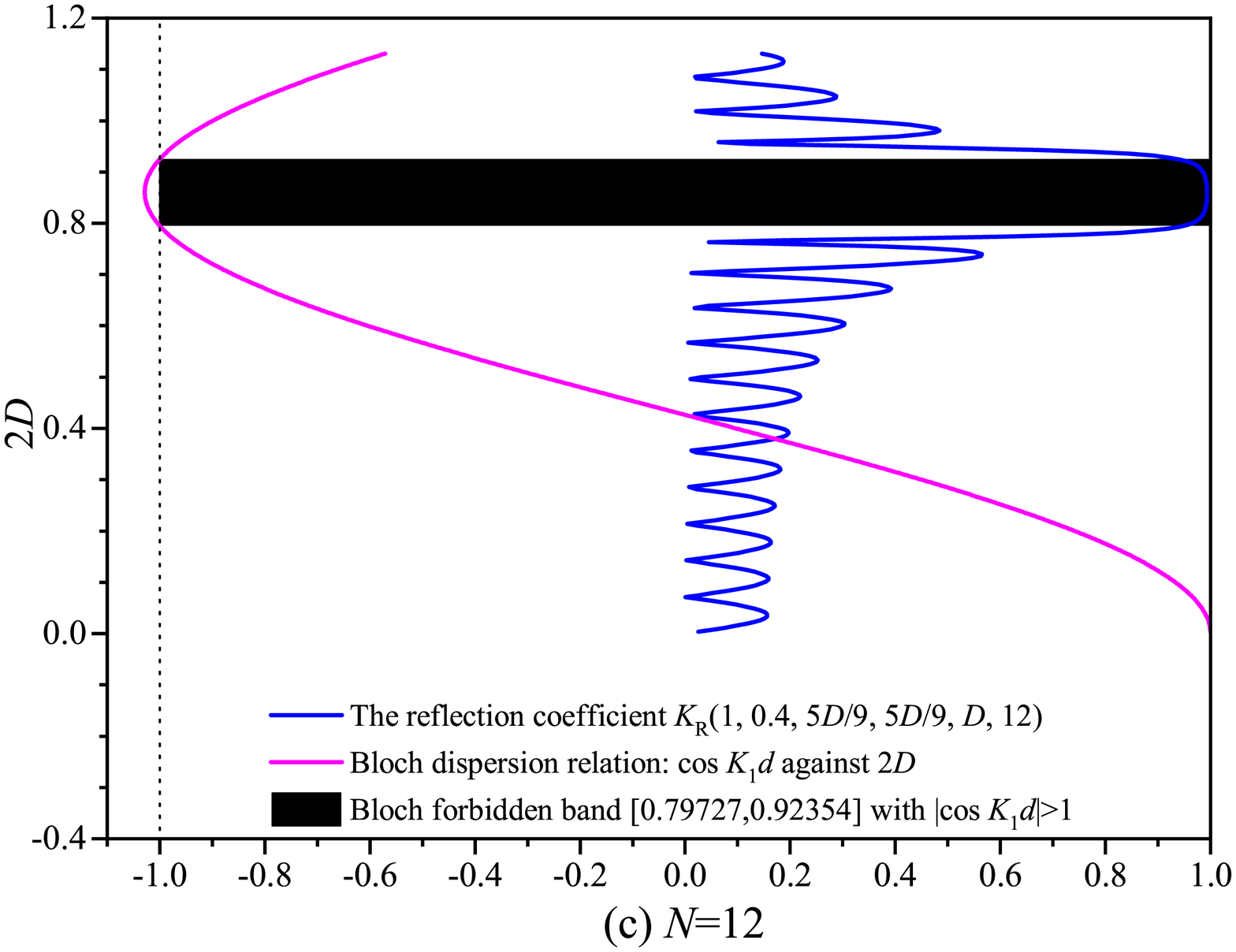}}
\vspace*{-6mm}
\centerline{\hspace*{0mm}\epsfxsize=2.8in \epsffile{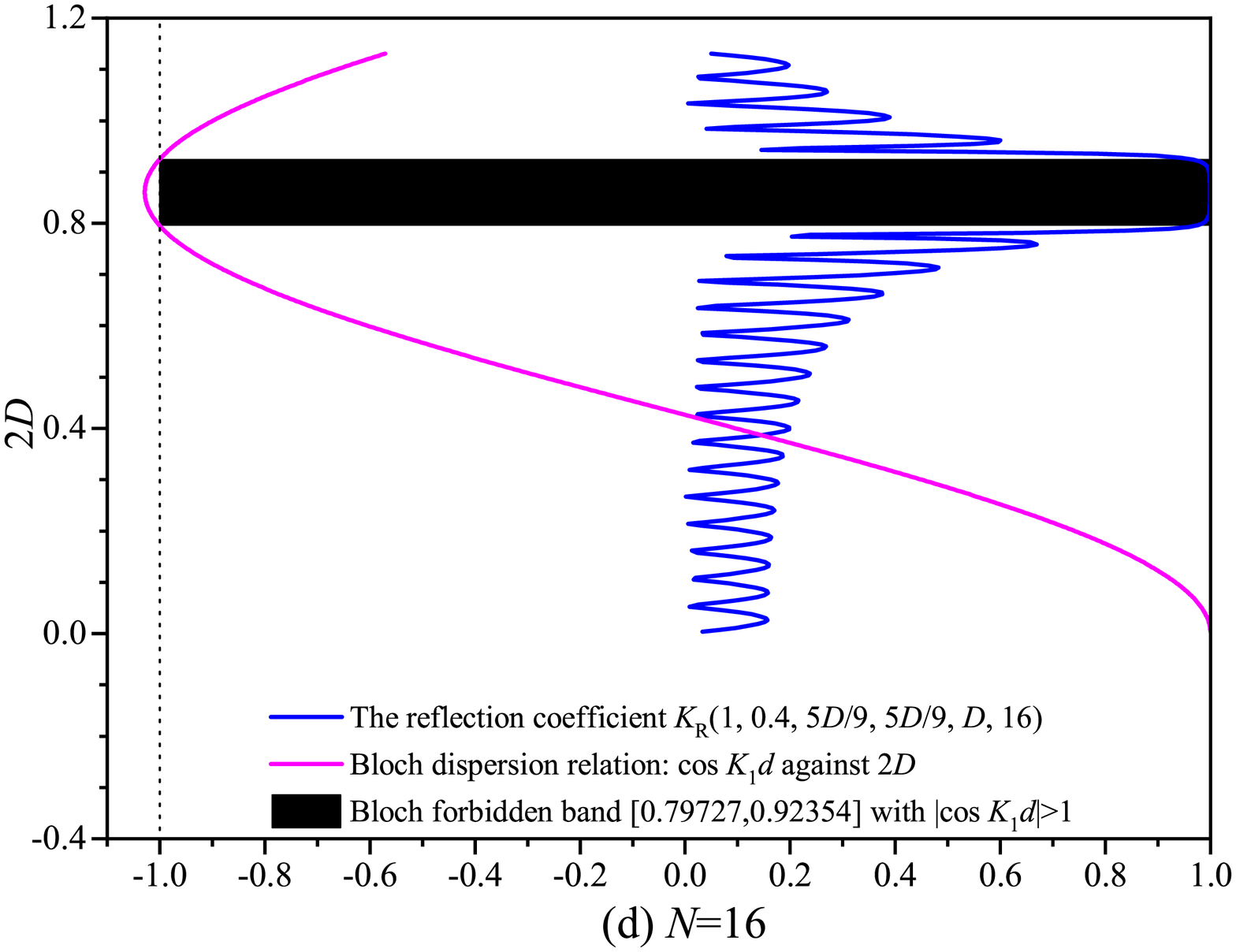}\hspace{-15mm}\epsfxsize=2.8in \epsffile{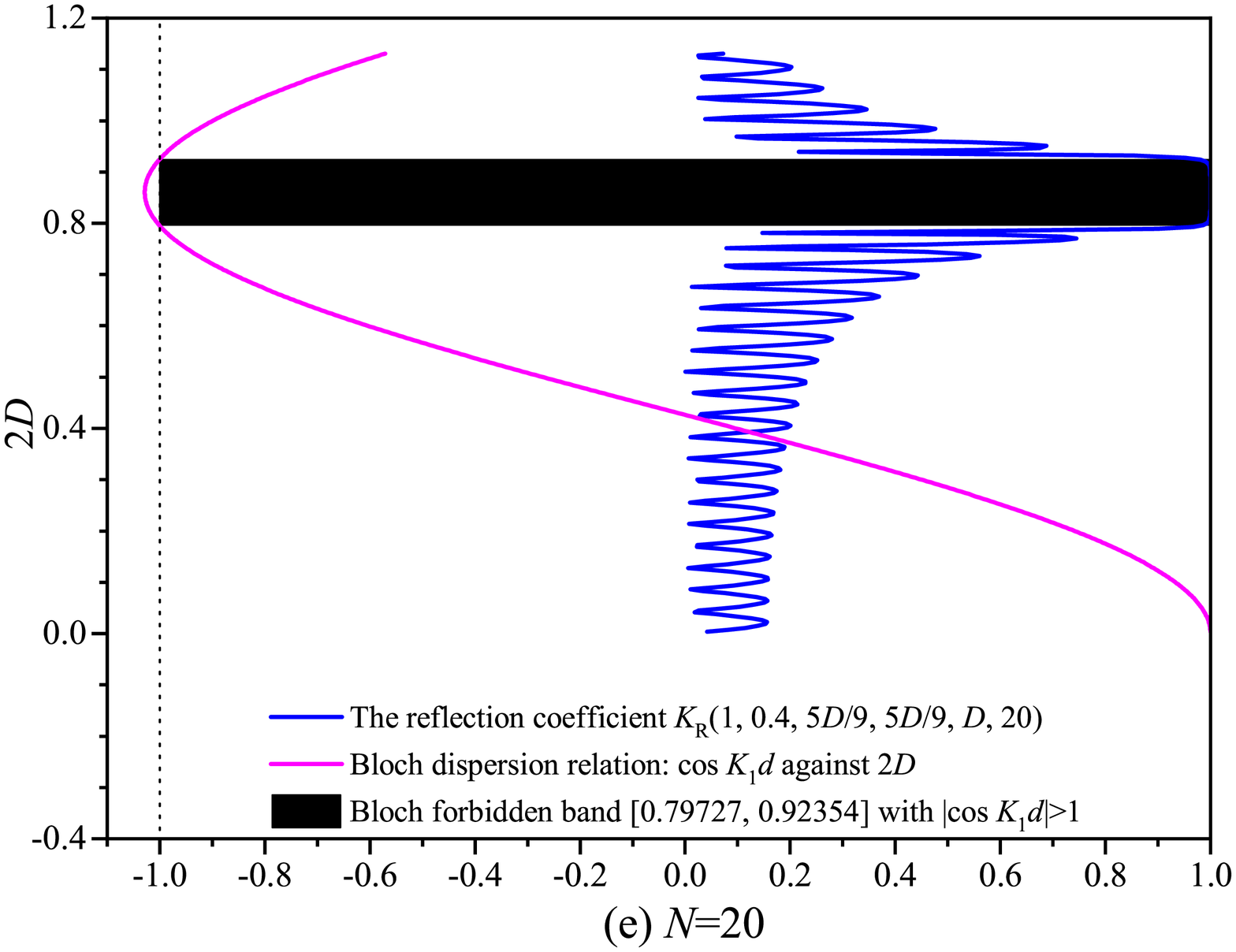}\hspace{-15mm}\epsfxsize=2.8in \epsffile{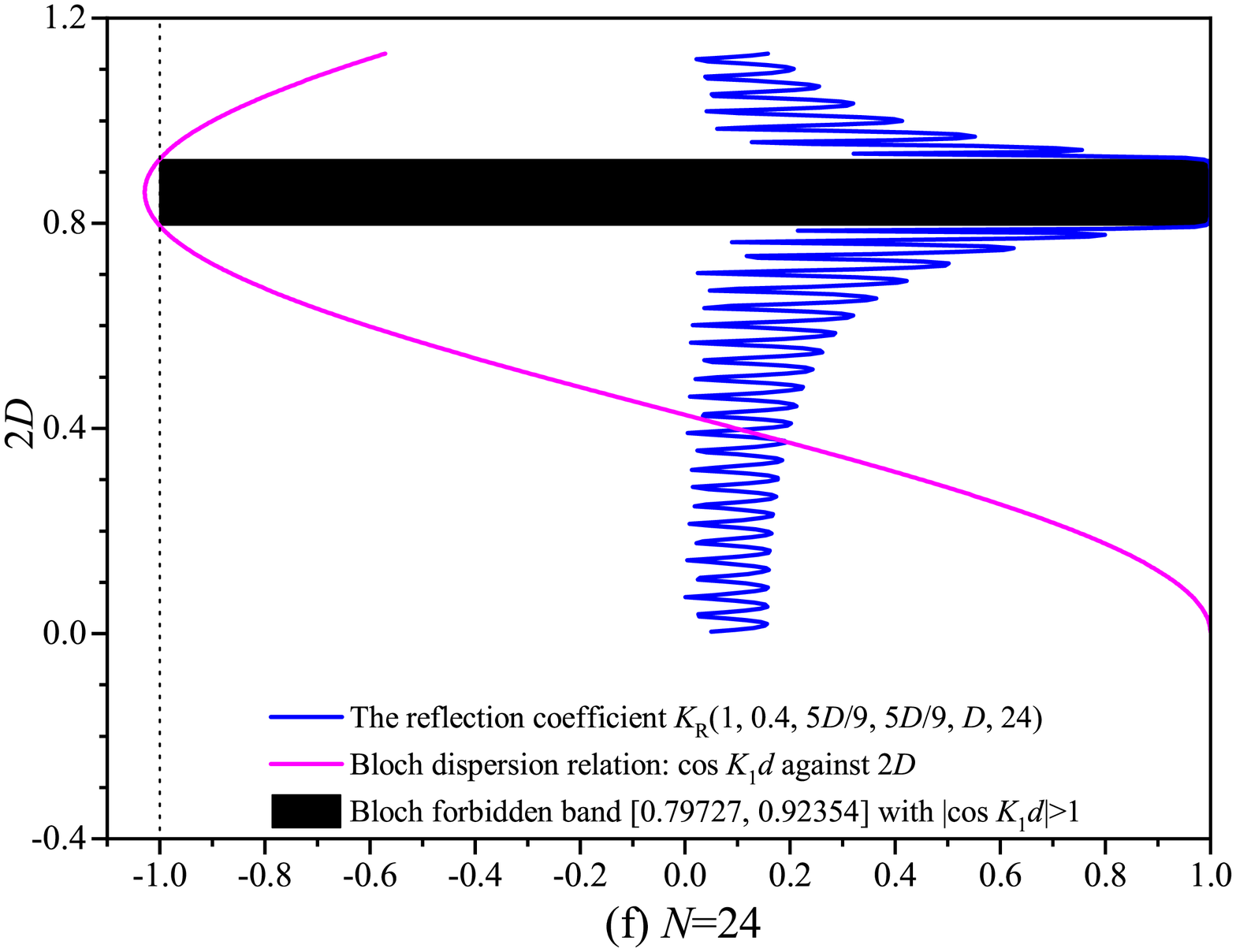}}
\vspace*{-10mm}
\caption[kdv]{\label{fig3} The correspondence between a Bragg resonance band caused by FPA-1 and a Bloch forbidden band caused by IPA-1, where $H=0.4$, $d=3.6\pi h_0$, and $W=\frac{5}{9}D$. (a) $N=4$; (b) $N=8$; (c) $N=12$; (d) $N=16$; (e) $N=20$; (f) $N=24$.}
\end{figure}

First, for a finite periodic array of rectangular bars, FPA-$1$, let $H$ and $W$ to be fixed as $H=0.4$ (i.e., $h_1=0.6h_0$), $W=\frac{5}{9}D$ (i.e., $w=\frac{5}{9}d$), and let $N$ take six values,  $N=4,8,12,16,20,24$. For any given $2D$, the wave reflection coefficient $K_R(1,0.4,\frac{5D}{9},\frac{5D}{9},D,N)$ against $2D$ can be calculated by using Eq. (\ref{Ref-Rectangle}). It is noted that $k_0h_0=2\pi D\frac{h_0}{d}$, in order to ensure that the calculation range is limited to the linear long-wave range, $2D$ must satisfy the condition $0<2D\le \frac{d}{10h_0}$, i.e., $0<k_0h_0\le \frac{\pi}{10}$. For any $h_0$, if we take $d=3.6\pi h_0$, then the long-wave condition is $0<2D \le 0.36\pi$. For $0.0036<2D\le 0.36\pi$, both the reflection coefficient $K_R(1,0.4,\frac{5D}{9},\frac{5D}{9},D,N)$ and Bloch dispersion $\cos K_1d$ defined in Eq. (\ref{rectangle}) against $2D$ are plotted and the results are shown in Figure (\ref{fig3})(a)-(f). It is easy to know that the forbidden band of Bloch long waves over IPA-$1$ is $[0.79727, 0.92354]$. As we can see in Figure (\ref{fig3})(a)-(f), no matter what $N$ is, the peak phase of the first order Bragg resonance coincides well with the center of the first Bloch forbidden band, i.e., $2D=0.860405$. Further, as $N$ increases, the Bragg resonance band approaches the Bloch forbidden band. By the way, in this case only the first order Bragg resonance occurs. If we fix $h_0$ and increase the value of $D$ to $20h_0$, then the second order Bragg resonance will occur.


\begin{figure}
\vspace*{-20mm}
\centerline{\hspace*{0mm}\epsfxsize=2.8in \epsffile{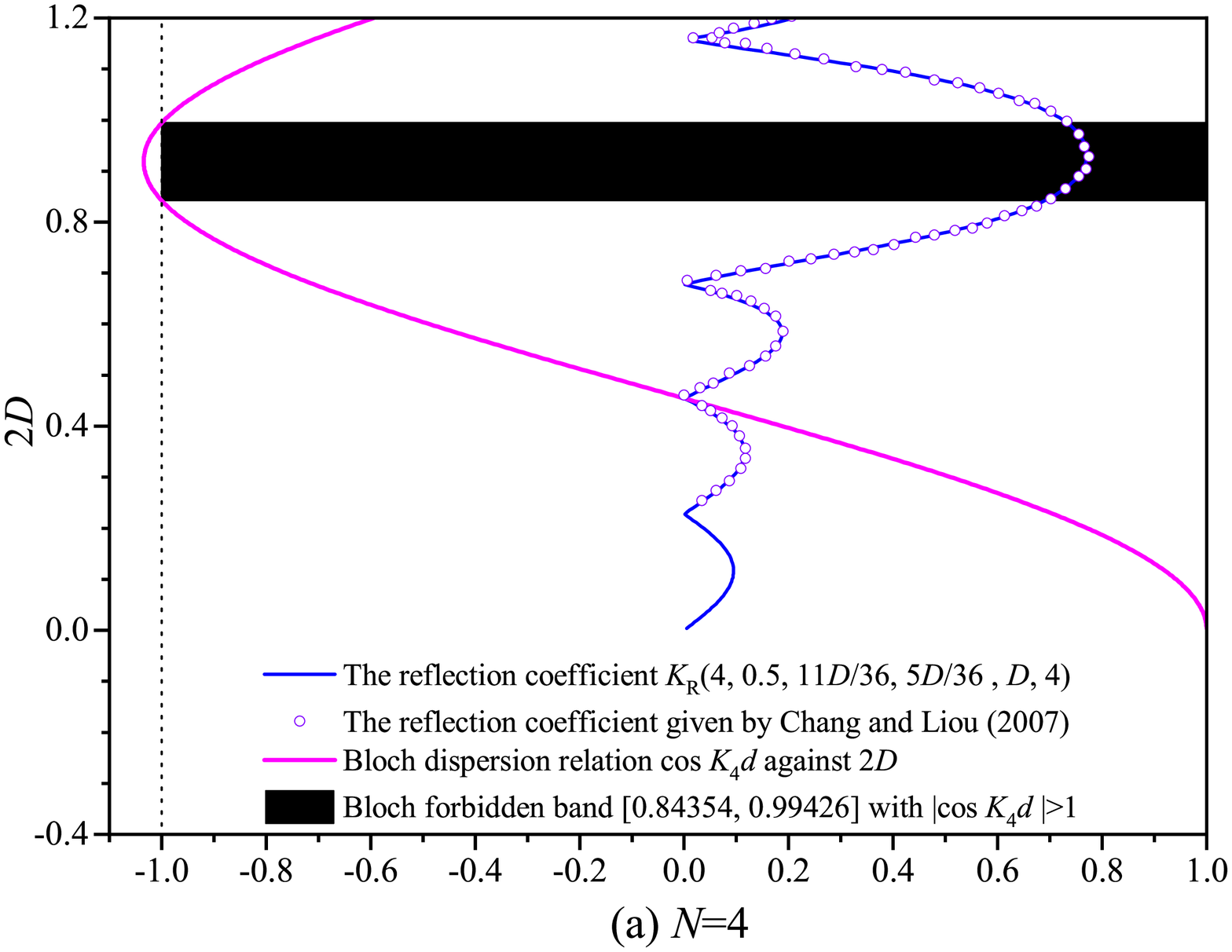}\hspace{-15mm}\epsfxsize=2.8in \epsffile{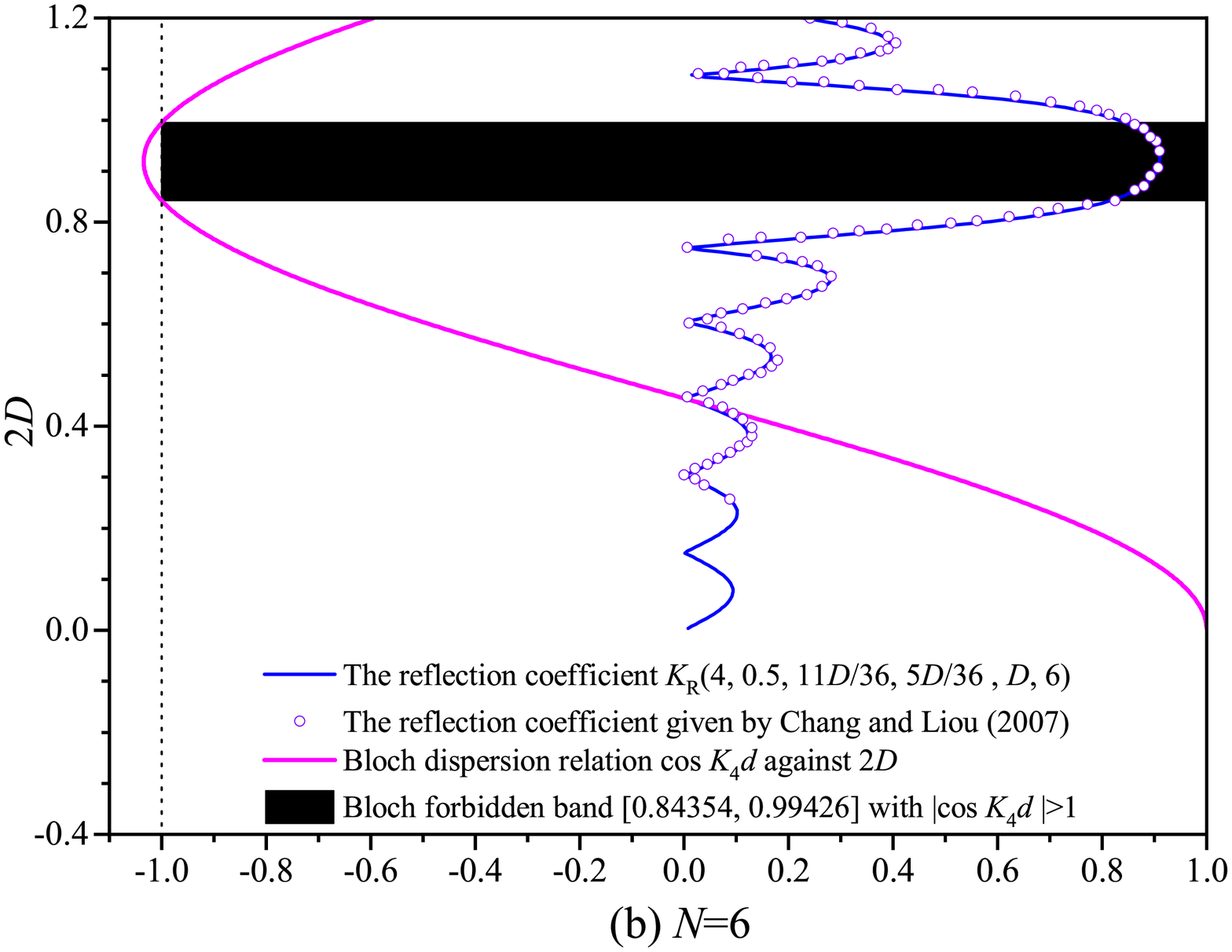}\hspace{-15mm}\epsfxsize=2.8in \epsffile{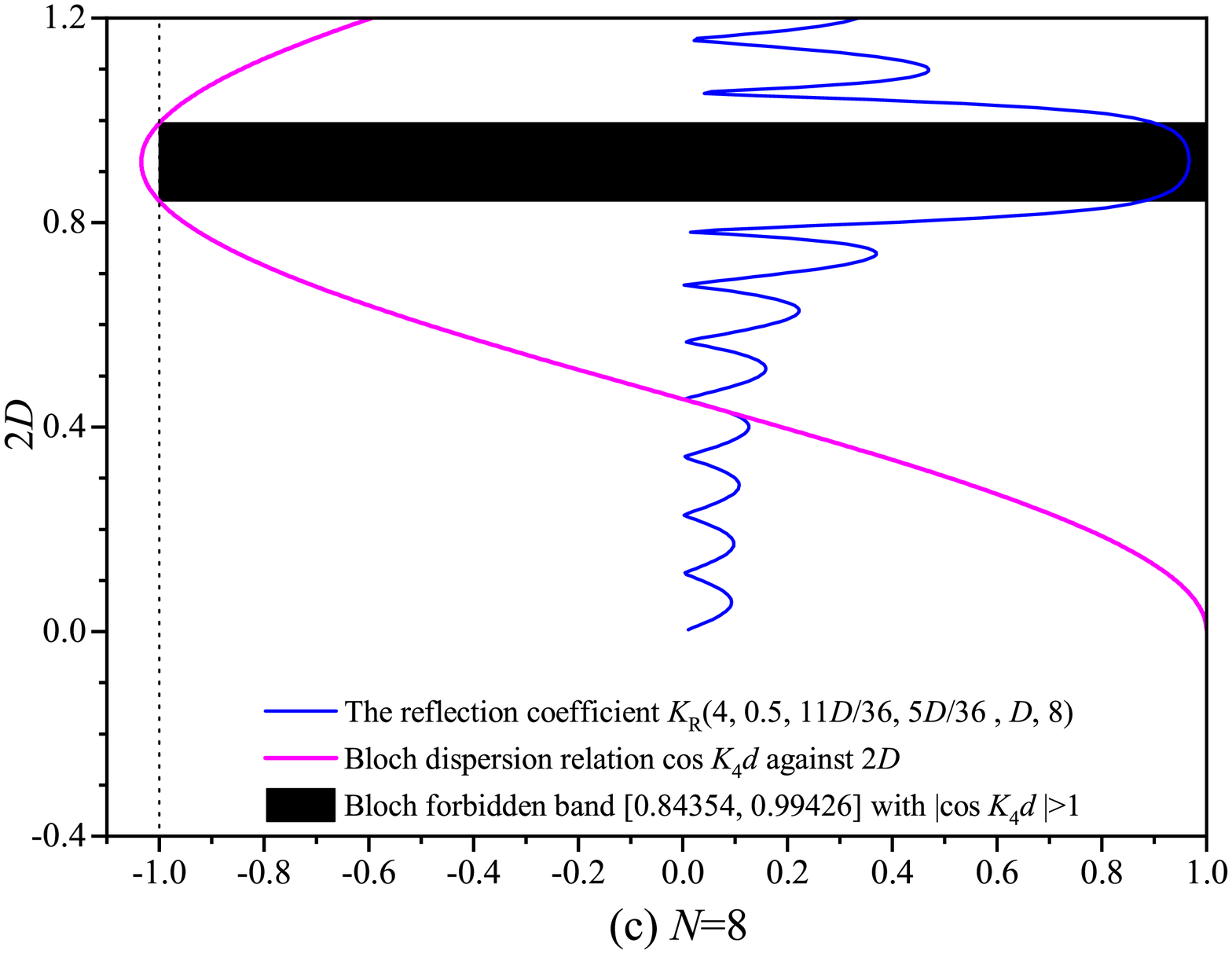}}
\vspace*{-6mm}
\centerline{\hspace*{0mm}\epsfxsize=2.8in \epsffile{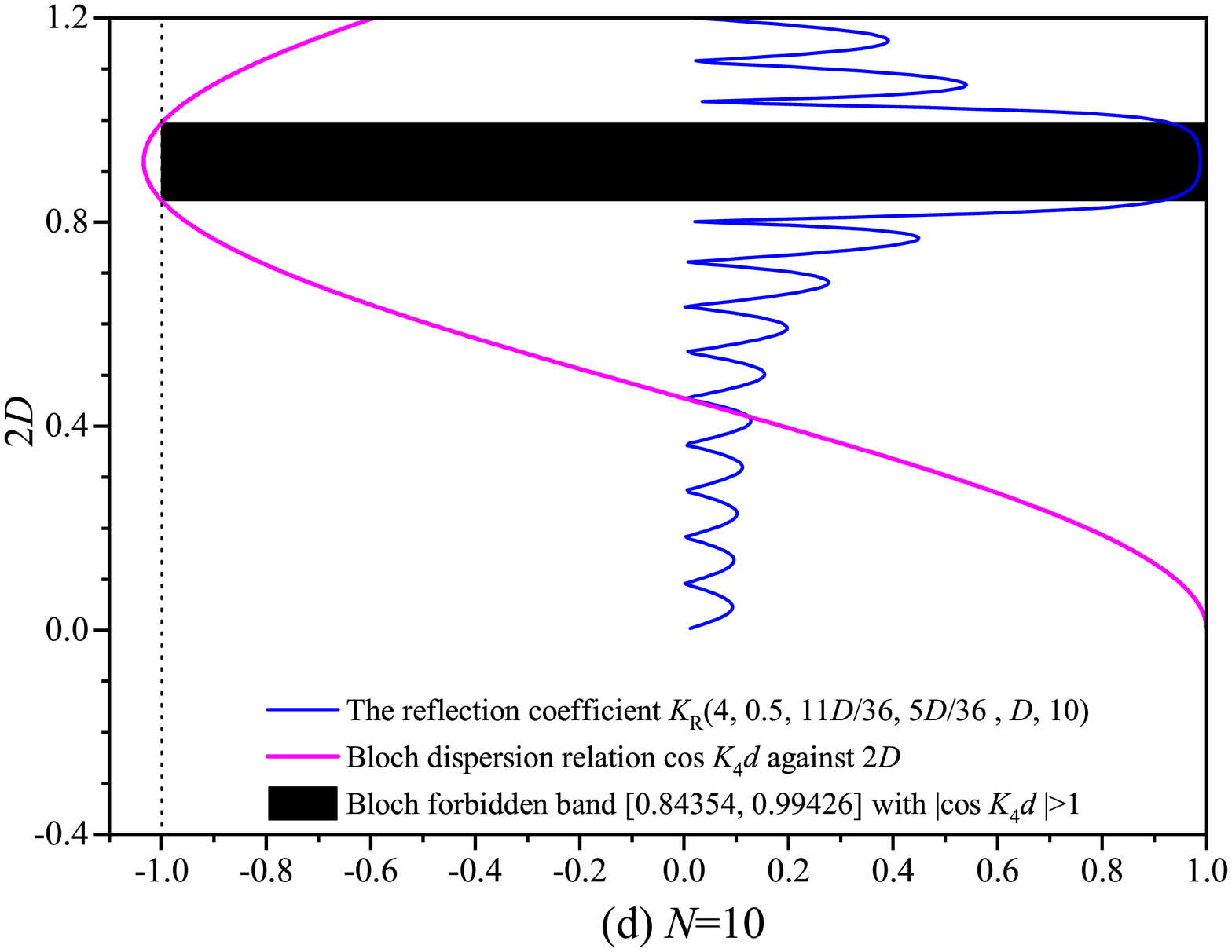}\hspace{-15mm}\epsfxsize=2.8in \epsffile{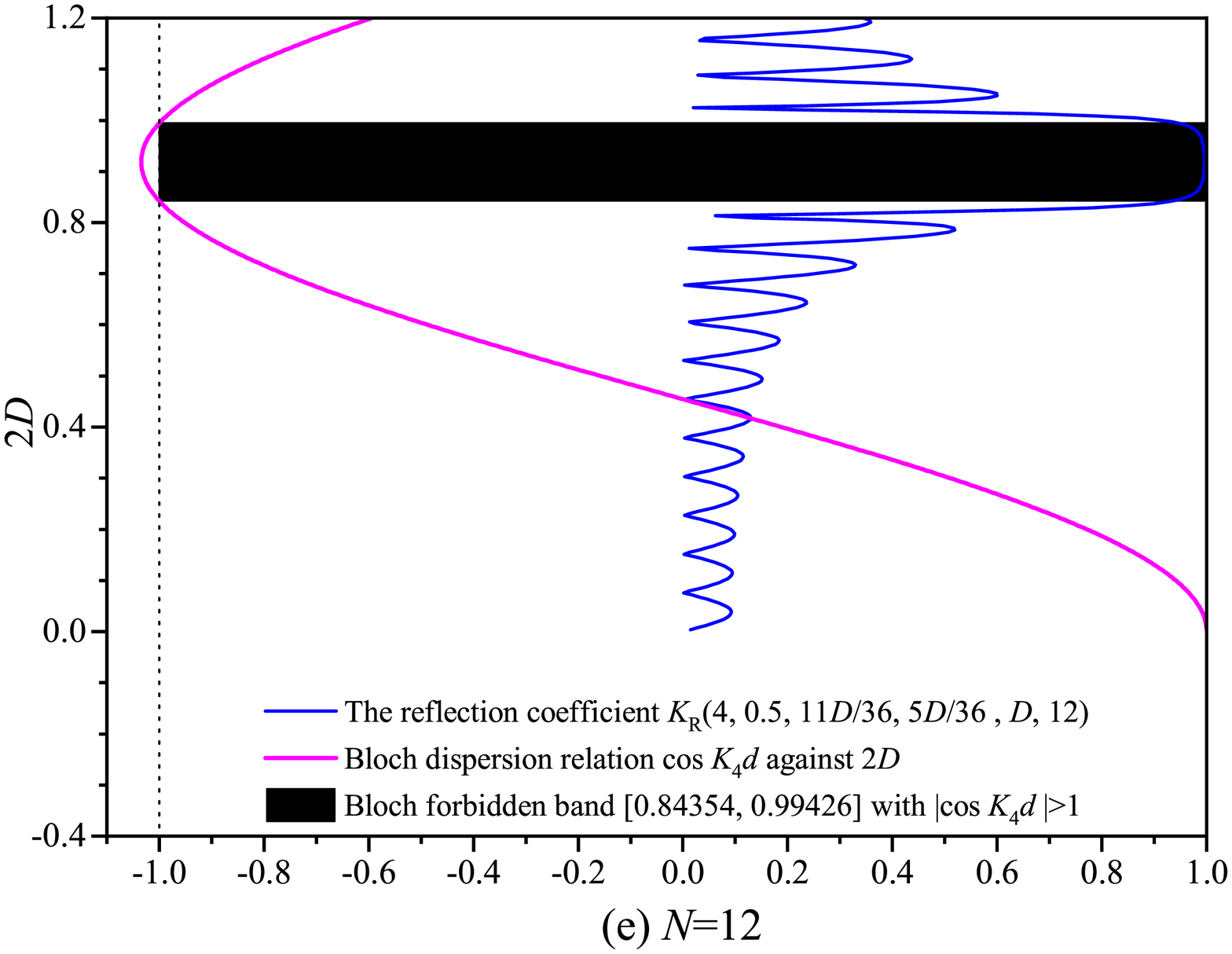}\hspace{-15mm}\epsfxsize=2.8in \epsffile{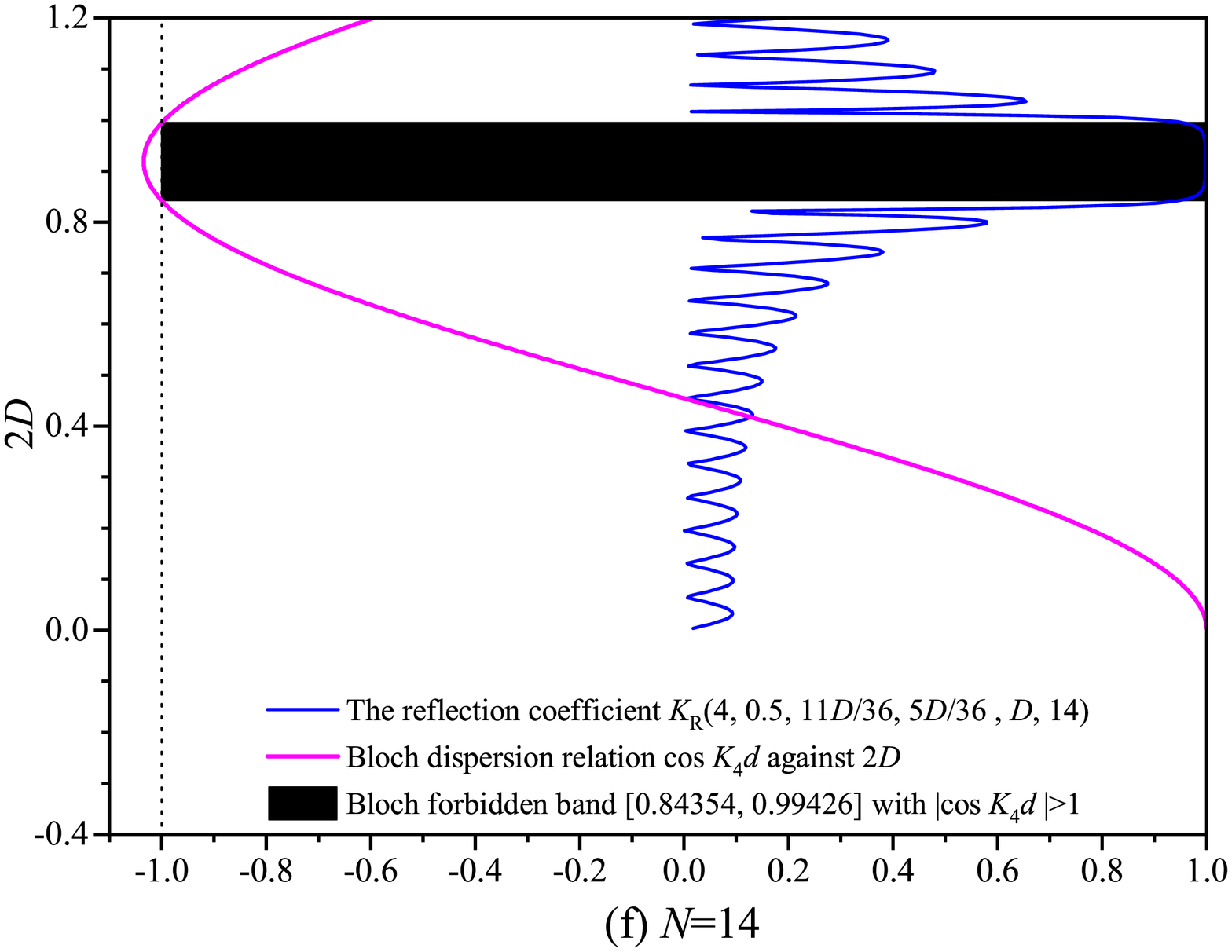}}
\vspace*{-10mm}
\caption[kdv]{\label{fig4} The correspondence between a Bragg resonance band caused by FPA-4 and a Bloch forbidden band caused by IPA-4, where $H$=0.5, $d=12h_0$, $W=\frac{11}{36}D$, and $W_t=\frac{5}{36}D$. (a) $N=4$; (b) $N=6$; (c) $N=8$; (d) $N=10$; (e) $N=12$; (f) $N=14$.}
\end{figure}

Second, for a finite periodic array of trapezoidal bars, FPA-$4$, let $H$, $W$, and $W_t$ to be fixed as $H=0.5$, $W=\frac{11}{36}D$, and $W_t=\frac{5}{36}D$, and let $N$ take six values, $N=4,6,8,10,12,14$. For any given $2D$, the wave reflection coefficients $K_R(4,0.5,\frac{11D}{36},\frac{5D}{36},D,N)$ against $2D$ can be calculated by using Eq. (\ref{Ref-trapezoid}). For any $h_0$, if we take $d=12h_0$, then the long-wave condition is $0<2D \le 1.2$, i.e., $0<k_0h_0\le \frac{\pi}{10}$. For $0.00382<2D\le 1.2$, both the reflection coefficient $K_R(4,0.5,\frac{11D}{36},\frac{5D}{36},D,N)$ and Bloch dispersion $\cos K_1d$ defined in (\ref{Bandgap-Trapezoid}) against $2D$ are plotted and the results are shown in Figure (\ref{fig4})(a)-(f), in which the two analytical solutions of the reflection coefficient given by Chang and Liou (2007) for the two cases of $N=4$ and $N=6$ are also presented in Figure \ref{fig4}(a) and \ref{fig4}(b), respectively. It is easy to check that the forbidden band of Bloch long waves over IPA-$4$ is $[0.84354,0.99426]$. As we can see in Figure (\ref{fig4})(a)-(f), no matter what $N$ is, the peak phase of the first order Bragg resonance coincides well with the center of the Bloch forbidden band, i.e., $2D=0.9189$. Further, as $N$ increases, the Bragg resonance band approaches the Bloch forbidden band $[0.84354,0.99426]$.

%
%

\begin{figure}
\vspace*{-5mm}
\centerline{\hspace*{0mm}\epsfxsize=2.8in \epsffile{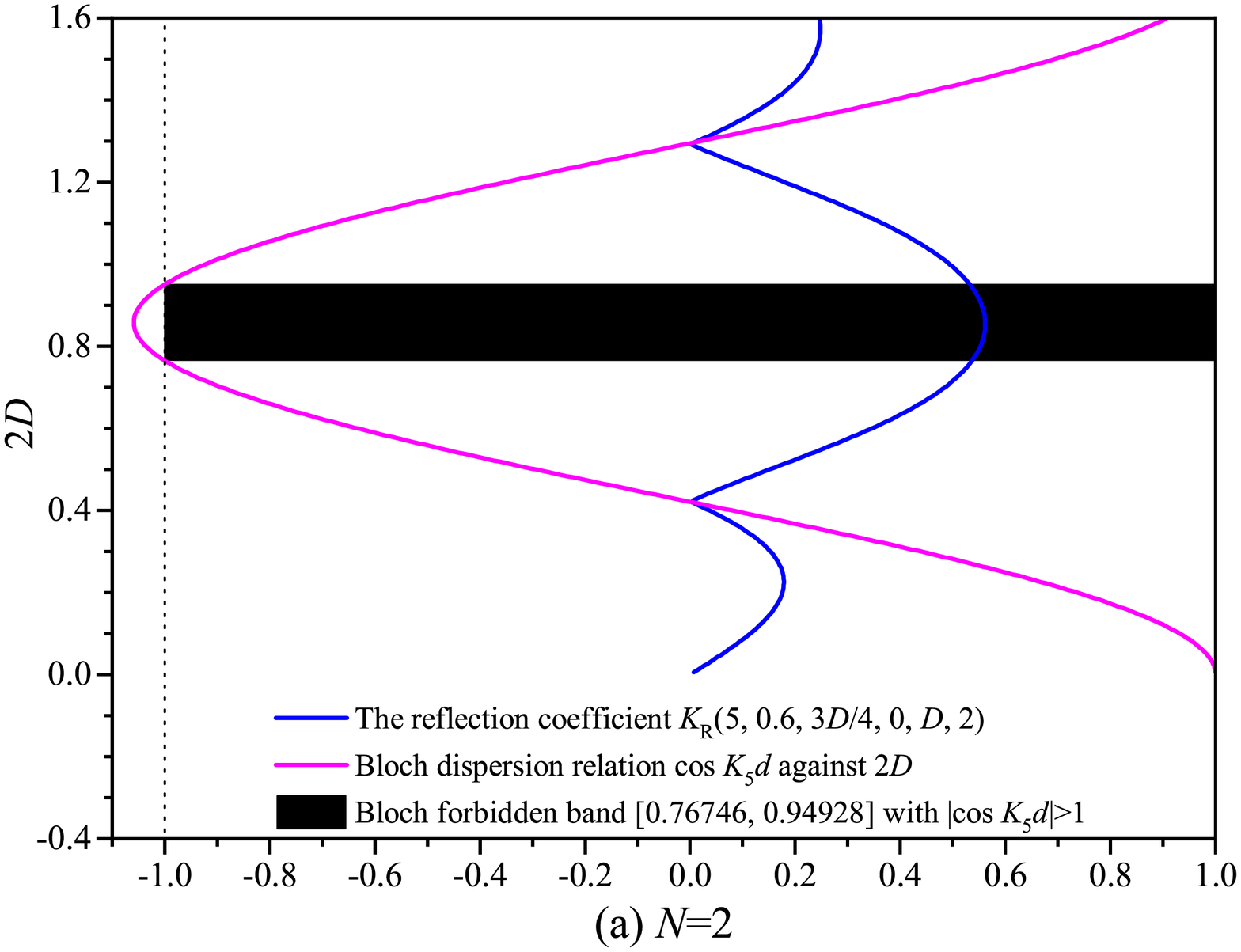}\hspace{-15mm}\epsfxsize=2.8in \epsffile{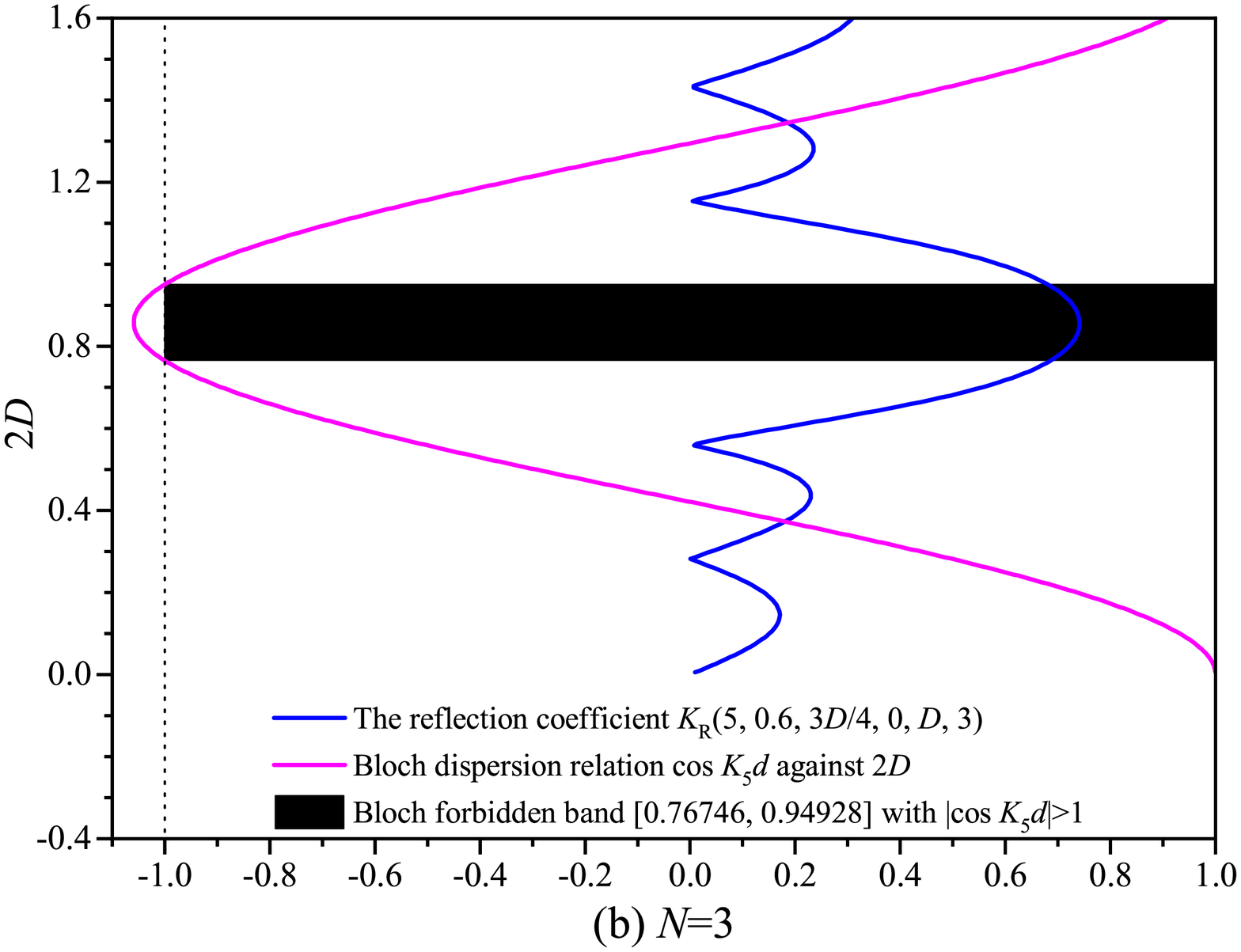}\hspace{-15mm}\epsfxsize=2.8in \epsffile{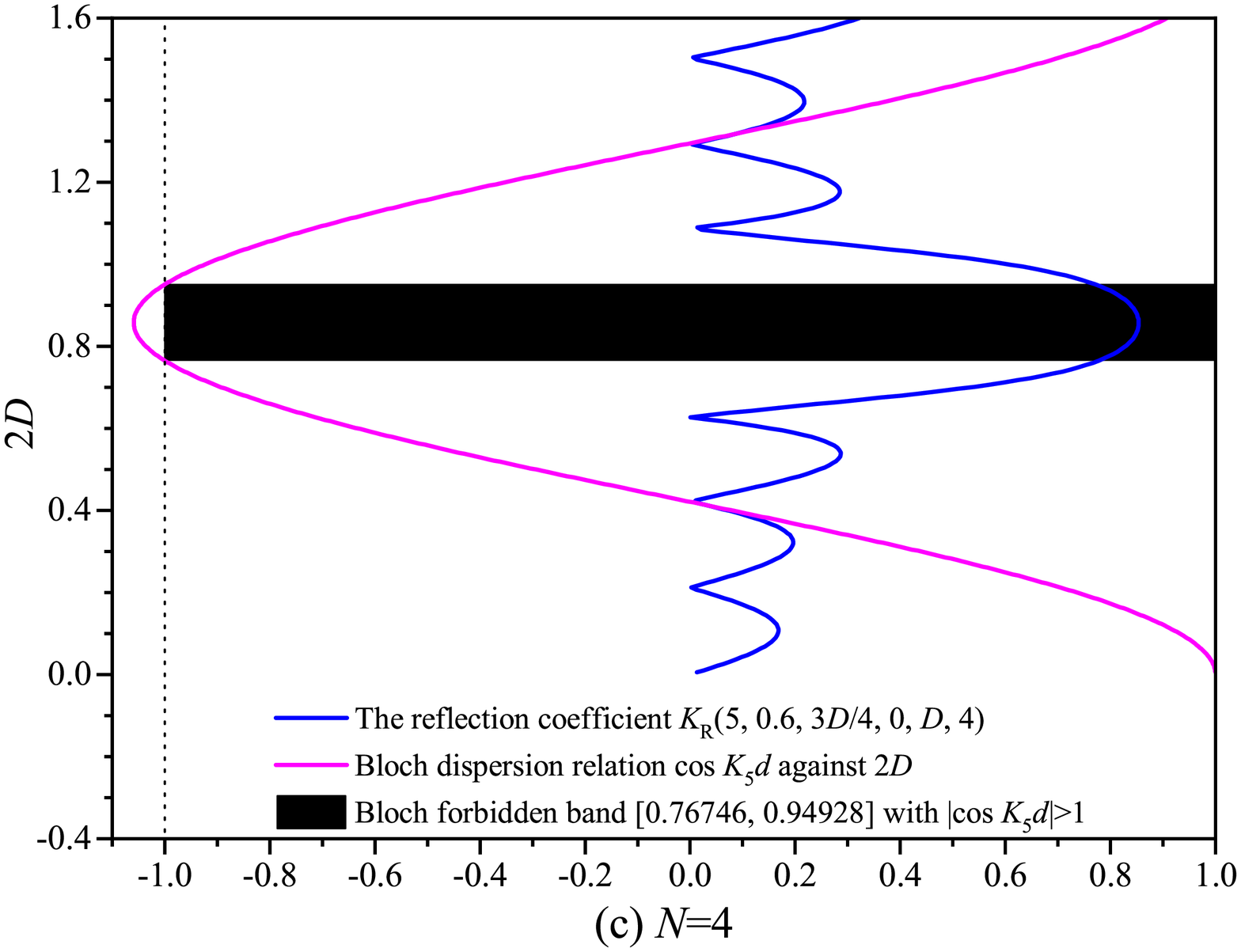}}
\vspace*{-6mm}
\centerline{\hspace*{0mm}\epsfxsize=2.8in \epsffile{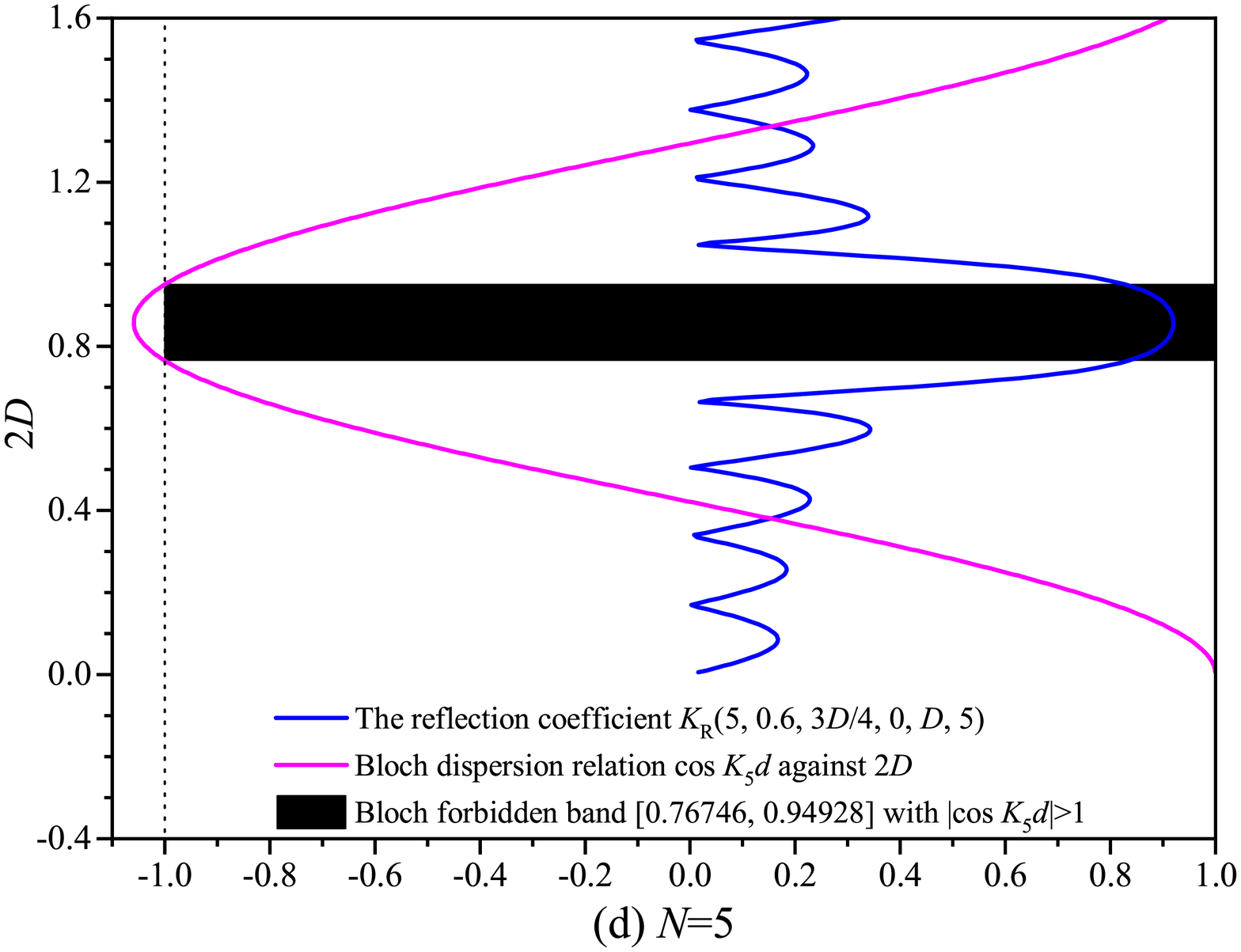}\hspace{-15mm}\epsfxsize=2.8in \epsffile{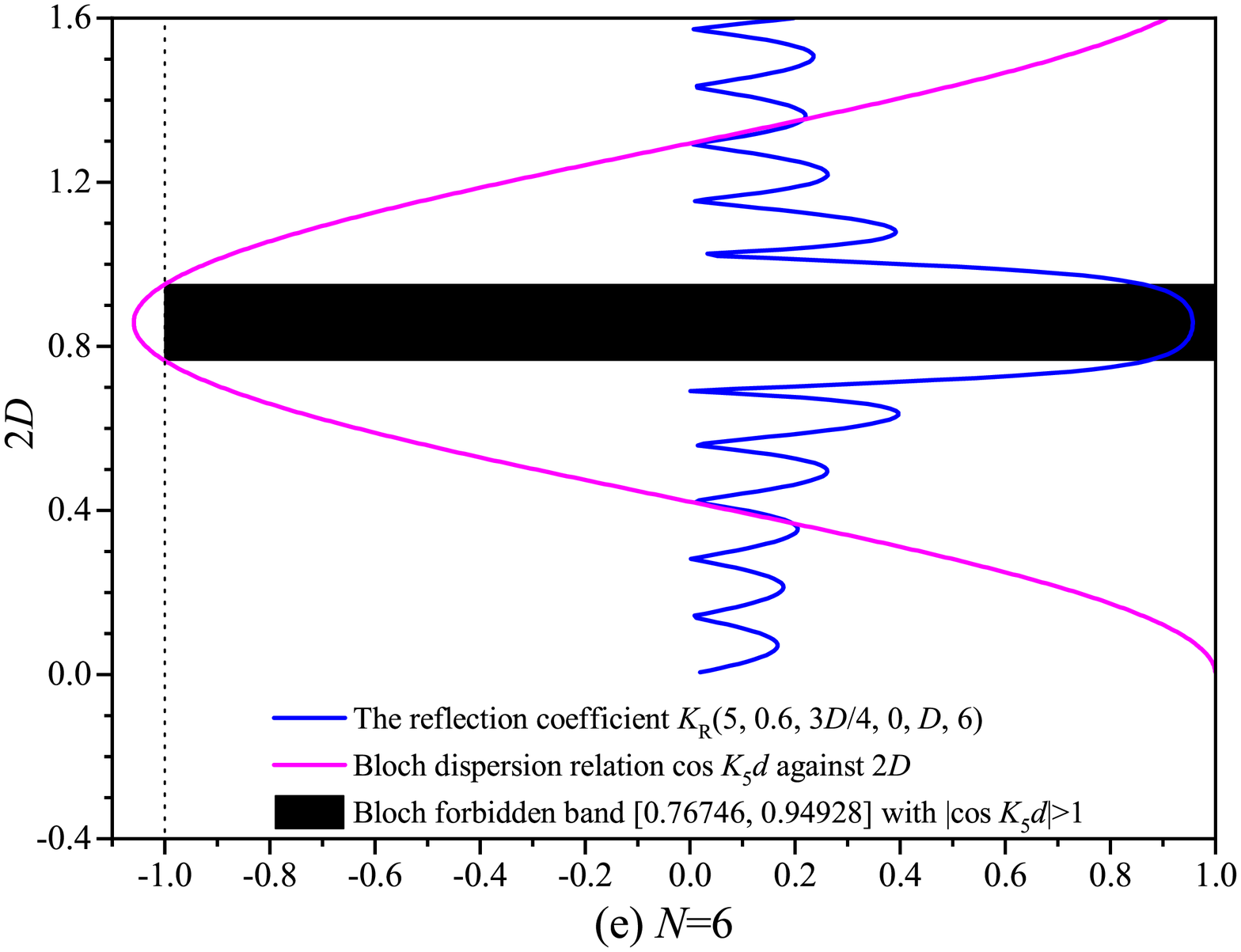}\hspace{-15mm}\epsfxsize=2.8in \epsffile{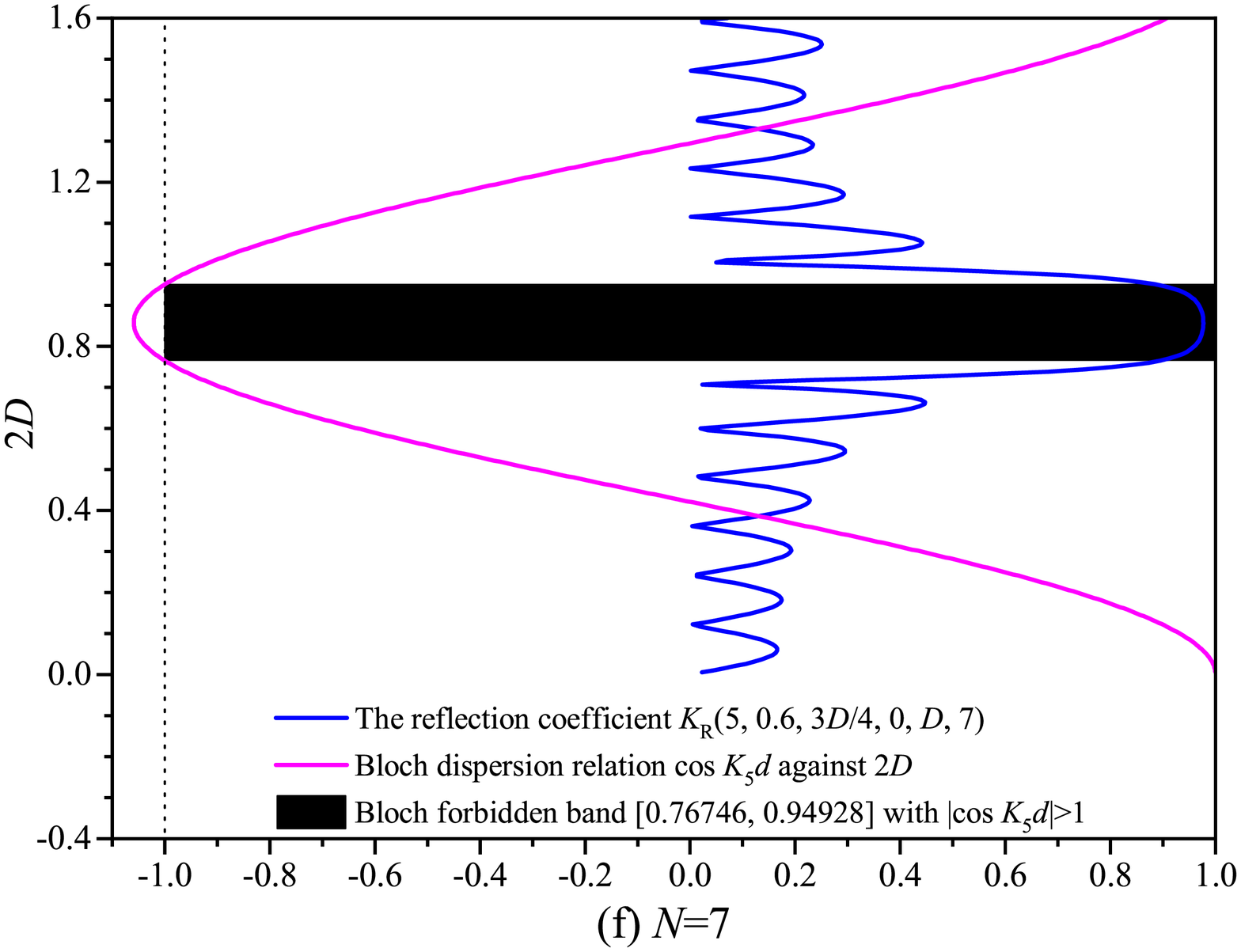}}
\vspace*{-10mm}
\caption[kdv]{\label{fig5} The correspondence between a Bragg resonance band caused by FPA-5 and a Bloch forbidden band caused by IPA-5, where $H=0.6$, $d=16h_0$, and $W=\frac{3}{4}D$. (a) $N=2$; (b) $N=3$; (c) $N=4$; (d) $N=5$; (e) $N=6$; (f) $N=7$.}
\end{figure}

Third, for a finite periodic array of triangular bars, FPA-$5$, let $H$ and $W$ to be fixed as $H=0.6$, $W=\frac{3}{4}D$, and $W_t=0$, and let $N$ take six values,  $N=2,3,4,5,6,7$. For any given $2D$, the wave reflection coefficient $K_R(5,0.6,\frac{3D}{4},0,D,N)$ against $2D$ can be calculated by using Eq. (\ref{Ref-triangular}). For any $h_0$, if we take $d=16h_0$, then the long-wave condition is $0<2D \le 1.6$. The wave reflection coefficient $K_R(5,0.6,\frac{3D}{4},0,D,N)$ against $2D$ for $0.00509\le 2D \le 1.6$ is calculated by using Eq. (\ref{Ref-triangular}). At the same time, $\cos K_5d$ against $2D$ for $0.00509\le 2D \le 1.6$ is also calculated by using Eq. (\ref{Bandgap-Triangle}). Both results are shown in Figure (\ref{fig5})(a)-(f). It is easy to check that the forbidden band of Bloch long waves over IPA-$5$ is $[0.76746, 0.94928]$. As we can see in Figure (\ref{fig5})(a)-(f), no matter what $N$ is, the peak phase of the first order Bragg resonance and the center of the Bloch forbidden band, i.e., $2D=0.85837$, match up quite good with each other. Further, as $N$ increases, the Bragg resonance band approaches the Bloch forbidden band $[0.76746, 0.94928]$.

\section{The modified Bragg's law}

In the previous section, all the comparison results tell us that Bragg resonance phase coincides with the center of Bloch forbidden band. Hence, to determine Bragg resonance phase, we need only to position the center of Bloch forbidden band, which is actually the stagnation point of the band expression, i.e., $\frac{\mbox{d}\left(\cos K_jd\right)}{\mbox{d}D}=0$.

By differentiating both sides of Eqs. (\ref{rectangle}), (\ref{Paraboloid}), (\ref{Rectified-band}), (\ref{Bandgap-Trapezoid}) and (\ref{Bandgap-Triangle}) with respect to $D$, and setting $\frac{\mbox{d}\left(\cos K_jd\right)}{\mbox{d}D}=0$, we infer that Bragg resonances occurs if and only if $D$ satisfies the following condition:
%
%
%
%
\begin{equation}
U_j\sin \left[2\pi(D-W)\right]+V_j\cos \left[2\pi(D-W)\right]=0, \quad j=1,2,3,4,5, \label{Modified-Bragg-Law-1}
\end{equation}
i.e.,
%
%
%
\begin{equation}
\cot(2\pi D)=\frac{U_j\cos(2\pi W)+V_j\sin(2\pi W)}
{U_j\sin(2\pi W)-V_j\cos(2\pi W)},\quad j=1,2,3,4,5, \label{Modified-Bragg-Law-2}
\end{equation}
where for those Bragg resonances excited by rectangular bars, parabolic bars, rectified cosinoidal bars, trapezoidal bars and triangular bars, the subscript $j$ takes 1, 2, 3, 4, 5 respectively.

It is noted that $D>0$, and the principal interval of the inverse cotangent function, $\mbox{arccot}\; x$, is $(0,\pi)$, by solving Eq. (\ref{Modified-Bragg-Law-2}), we obtain the modified Bragg's law as follows
\begin{equation}
D=\frac{n}{2}-\frac{1}{2\pi}\mbox{arccot}\frac{U_j\cos(2\pi W)+V_j\sin(2\pi W)}
{V_j\cos(2\pi W)-U_j\sin(2\pi W)},\;\;(2D,2W)\in\Omega_n,\;n=1,2,...
\label{Modified-Bragg-Law-3}
\end{equation}
where $\Omega_n=\left\{(2D,2W): n-1<2D\le n, D\ge W\right\}$, and the integer $n$ corresponds to the $n$th order Bragg resonance.

For convenience of statement, we denote
\begin{eqnarray}
\hspace*{-5mm}&&D^{\mbox{Bragg}}_n=\frac{n}{2}, \\
\hspace*{-5mm}&&D^{\mbox{Liu}}_n(j,H,W,W_t)=\frac{n}{2}-\frac{1}{2\pi}\mbox{arccot}\frac{U_j\cos(2\pi W)+V_j\sin(2\pi W)}
{V_j\cos(2\pi W)-U_j\sin(2\pi W)},\; j=1,2,3,4,5.
%
%
%
\end{eqnarray}
Then the traditional Bragg's law of the $n$th Bragg resonance is $D=D^{\mbox{Bragg}}_n$,
%
%
and the modified Bragg's law of the $n$th Bragg resonance (\ref{Modified-Bragg-Law-3}) can be rewritten as
\begin{equation}
D=D^{\mbox{Liu}}_n(j,H,W,W_t),\;j=1,2,3,4,5. \label{modified-law}
\end{equation}
It is clear that
\begin{equation}
D^{\mbox{Liu}}_n(j,H,W,W_t)-D^{\mbox{Bragg}}_n=-\frac{1}{2\pi}\mbox{arccot}\frac{U_j\cos(2\pi W)+V_j\sin(2\pi W)}
{V_j\cos(2\pi W)-U_j\sin(2\pi W)}.\label{modified-law-comparison}
\end{equation}
%
%
%

\begin{figure}
\vspace*{-40mm}
\centerline{\hspace*{15mm}\epsfxsize=3.4in \epsffile{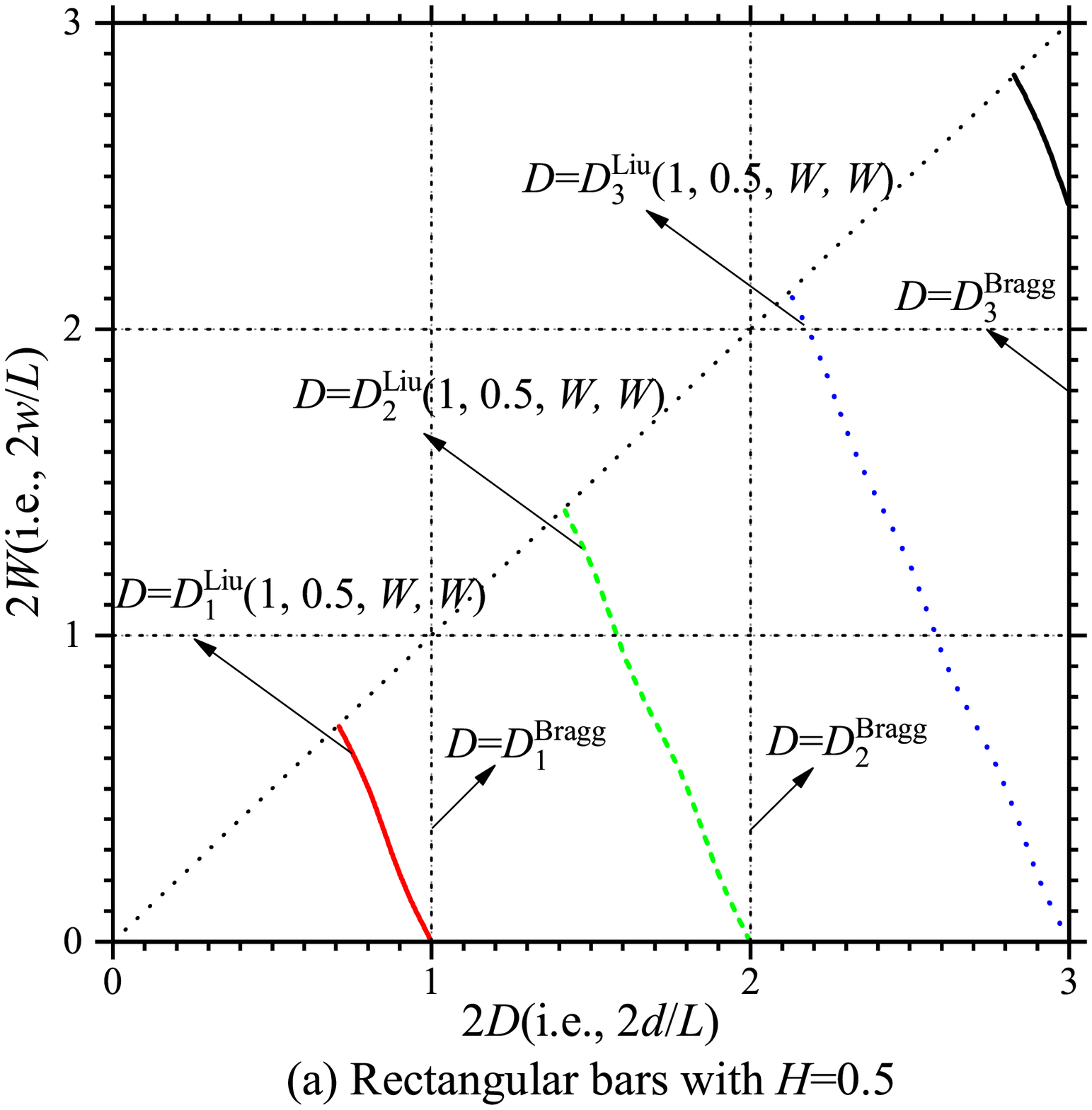}\hspace{-20mm}\epsfxsize=3.4in \epsffile{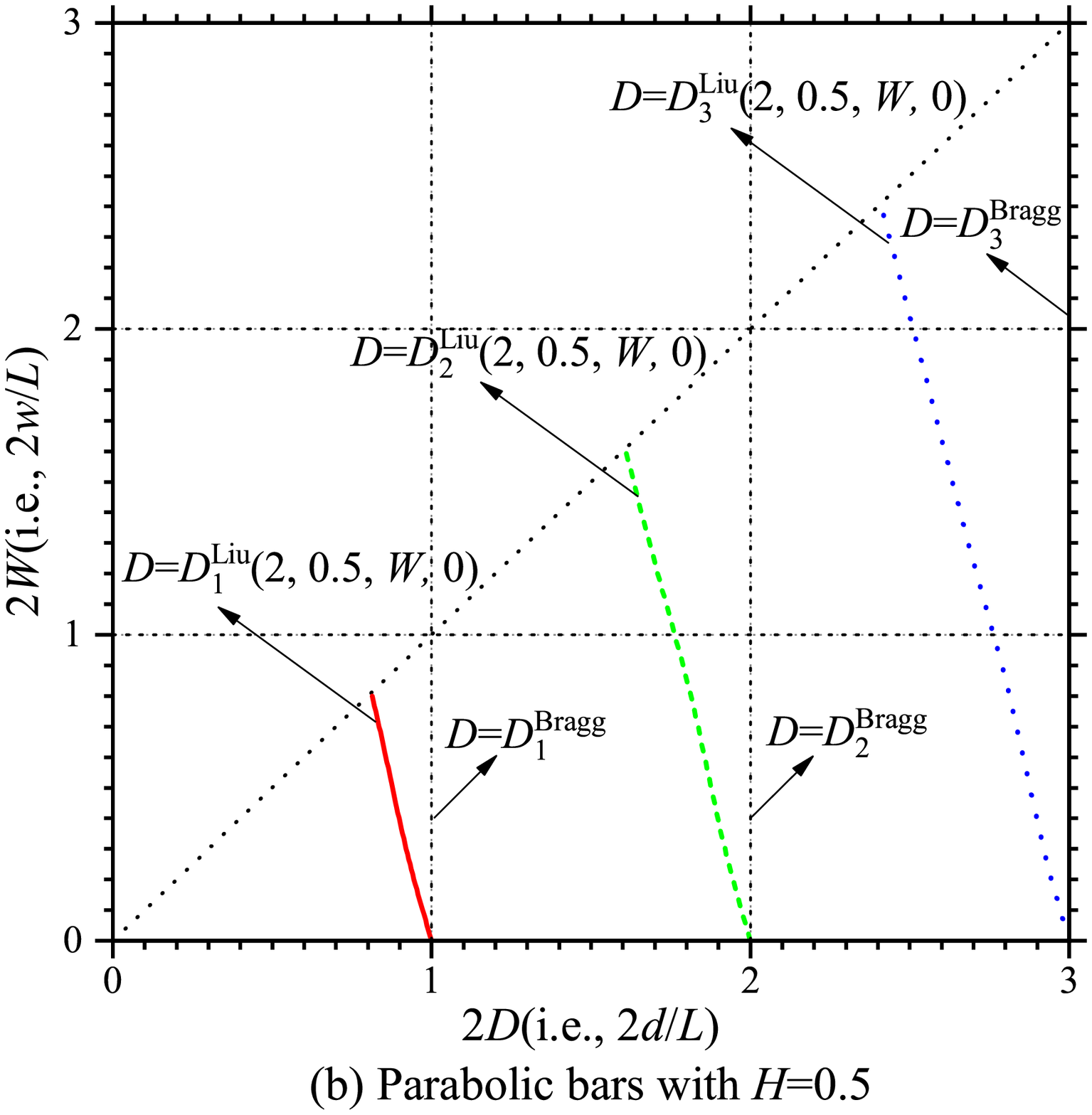}}
\vspace*{-5mm}
\centerline{\hspace*{15mm}\epsfxsize=3.4in \epsffile{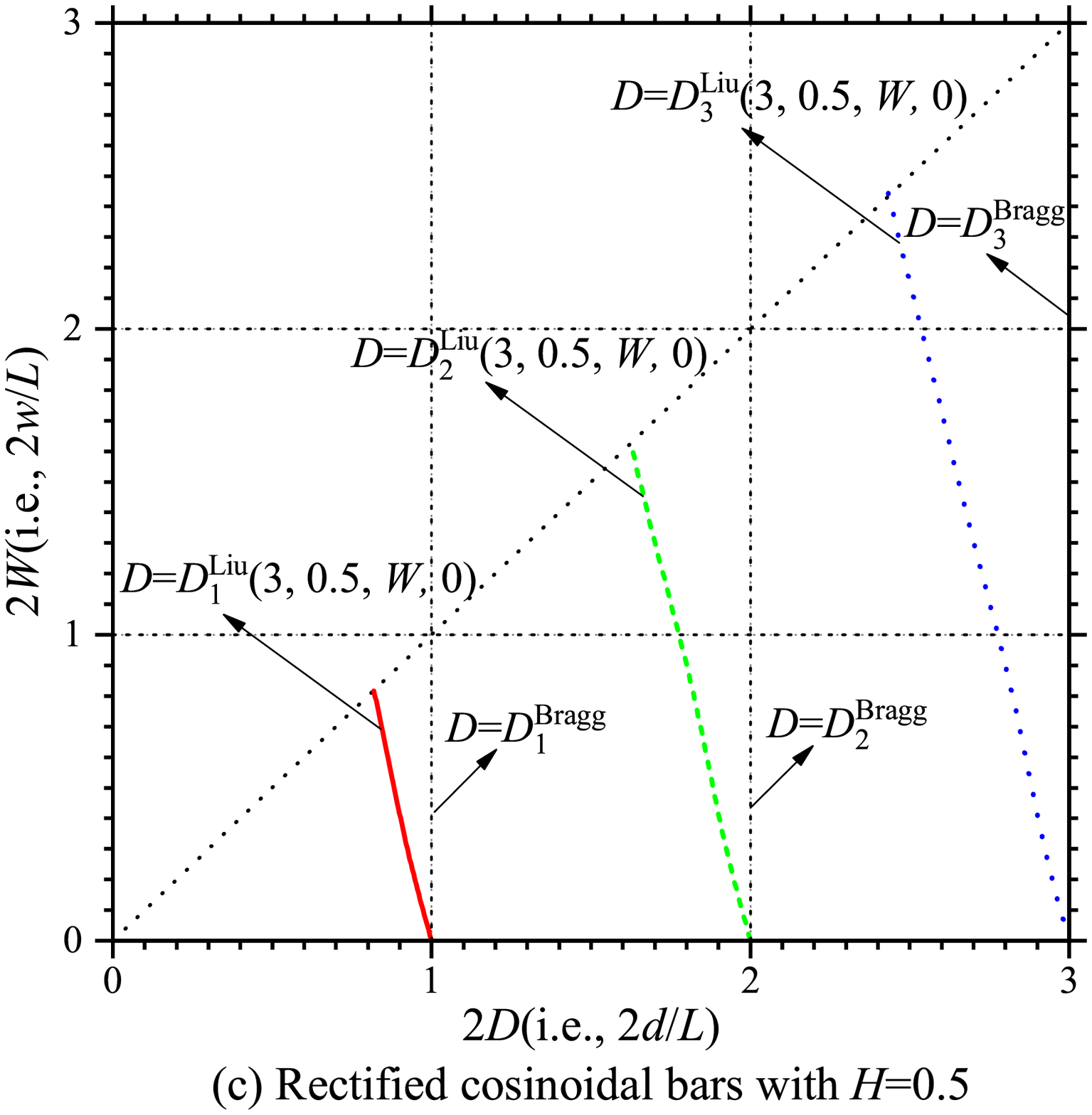}\hspace{-28mm}\epsfxsize=3.4in  \epsffile{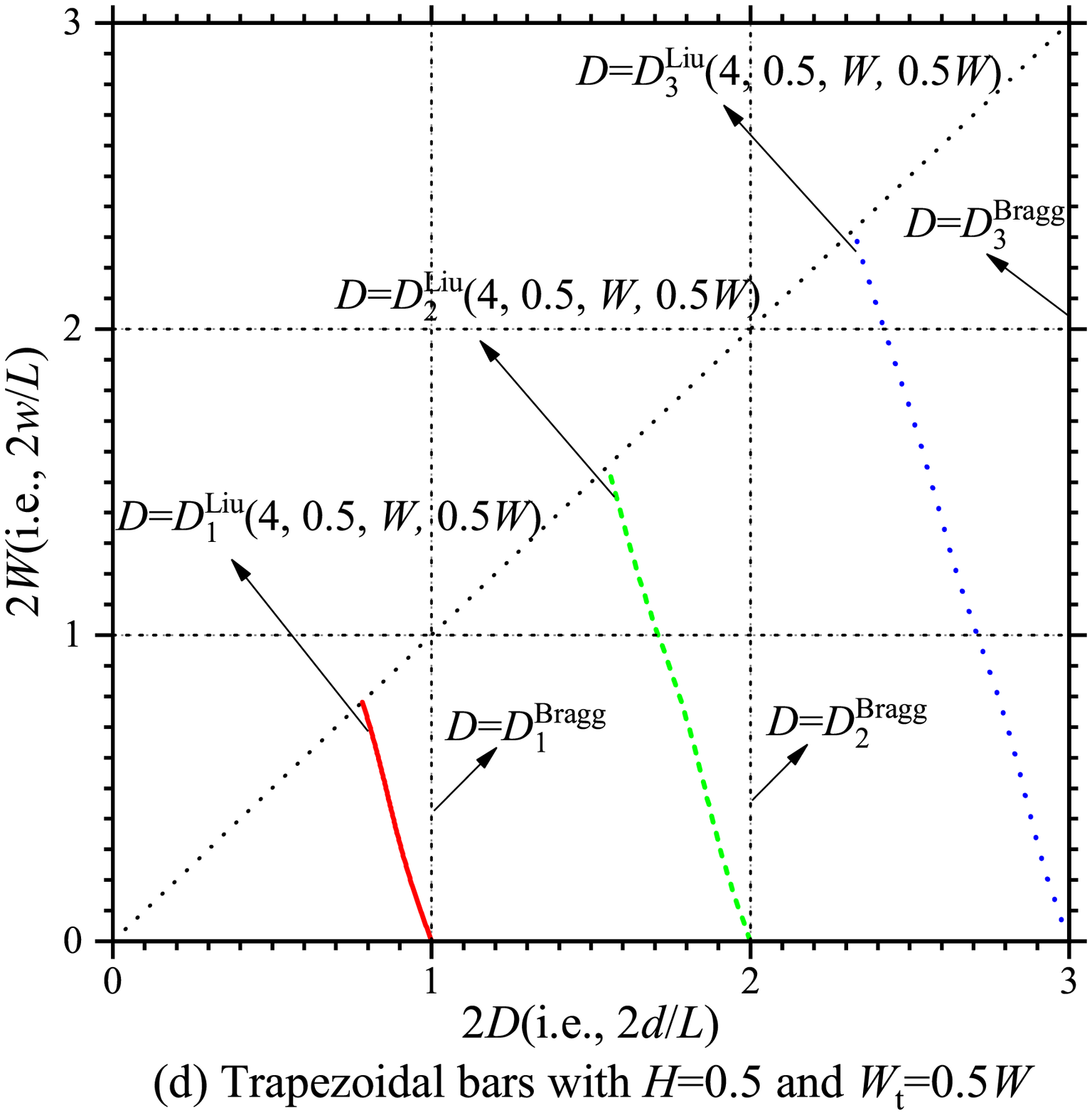}\hspace{-28mm}\epsfxsize=3.4in  \epsffile{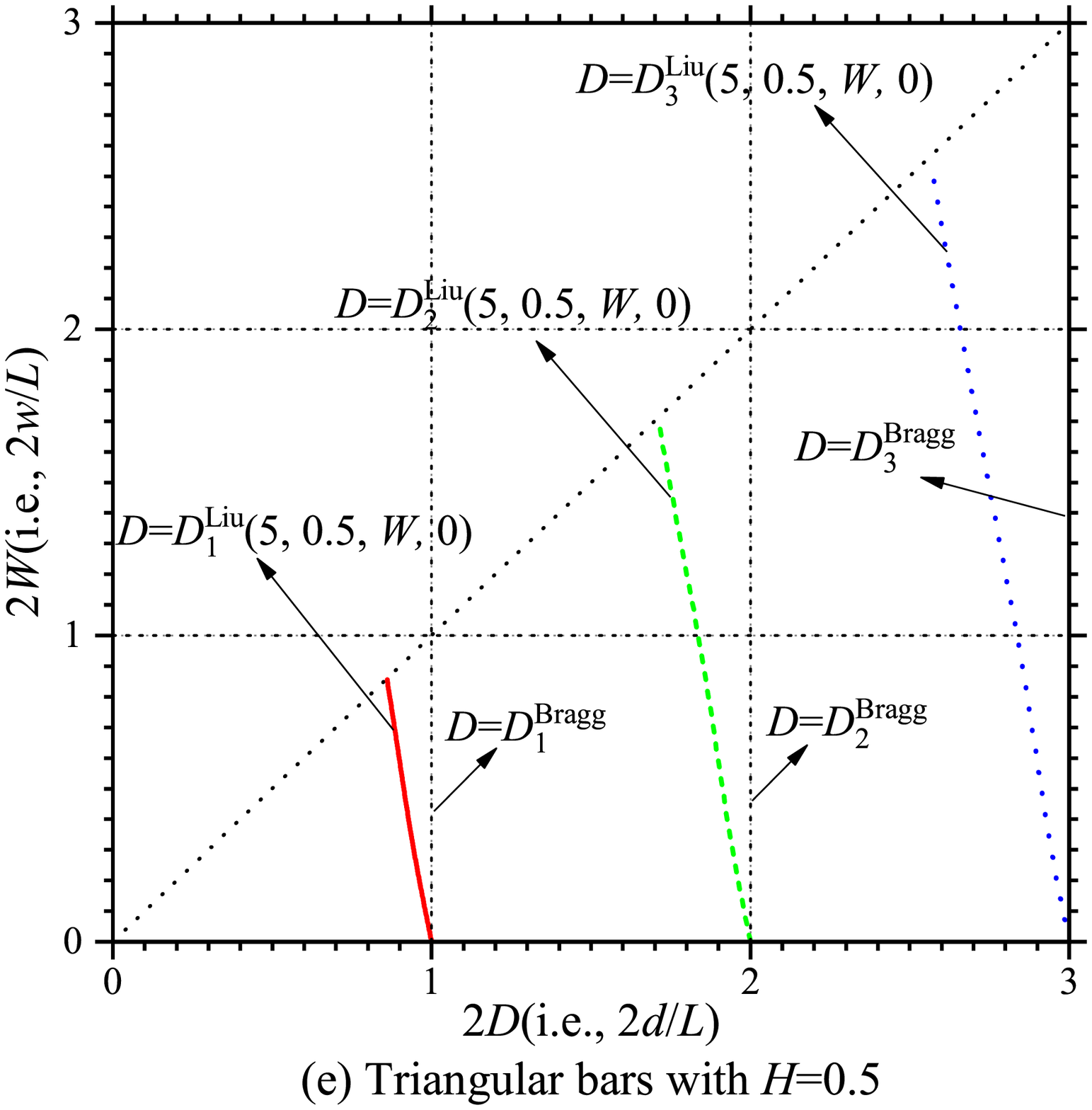}}
\vspace*{-5mm}
\caption[kdv]{\label{fig6} The modified Bragg's laws of Bragg resonances and the corresponding gap maps of Bloch waves over five types of artificial bars with $H$=0.5. (a) Rectangular bars; (b) Parabolic bars; (c) Rectified cosinoidal bars; (d) Isosceles trapezoidal bars with $W_t=0.5W$; (e) Isosceles triangular bars.}
\end{figure}

In Figure \ref{fig6}, all the curves representing the modified Bragg's laws of the 1st-, 2nd- and 3rd-order Bragg resonances excited by  FPA-$j$, $j$=1,...,5, are plotted. For a better understanding to the modified Bragg's law (\ref{Modified-Bragg-Law-3}) or (\ref{modified-law}), the gap maps of Bloch waves modulated by the corresponding IPA-$j$, $j$=1,...,5, are also plotted, where those red domains are Bloch forbidden bands with $\left|\cos K_jd\right|>1$. As we can see, each red subdomain looks like a leaf and each black curve representing the modified Bragg's law looks like the main vein of the corresponding leaf.

\begin{figure}
\vspace*{-35mm}
\centerline{\hspace*{5mm}\epsfxsize=3.0in \epsffile{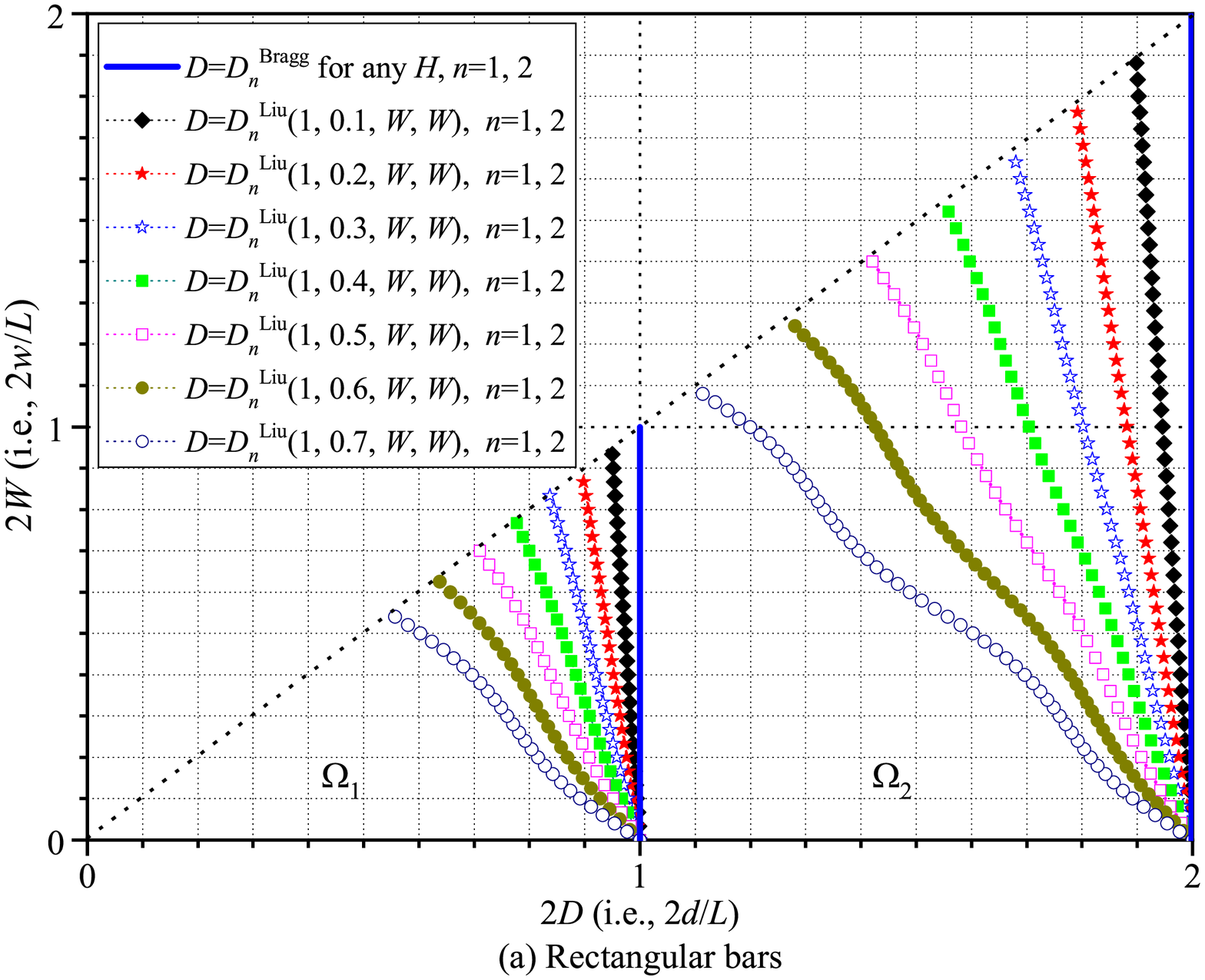}\hspace{-8mm}\epsfxsize=3.0in \epsffile{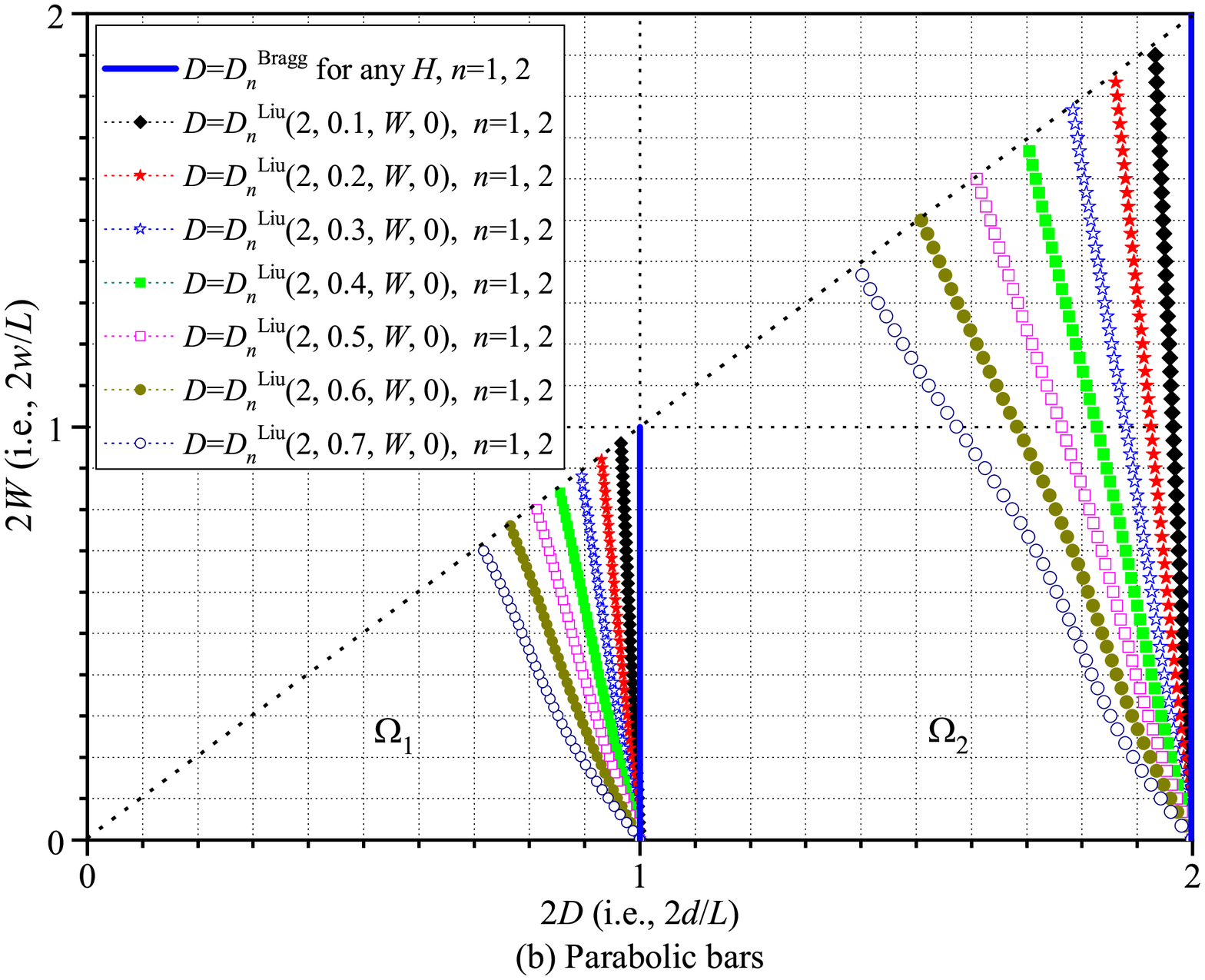}}
\vspace*{-5mm}
\centerline{\hspace*{0mm}\epsfxsize=3.0in \epsffile{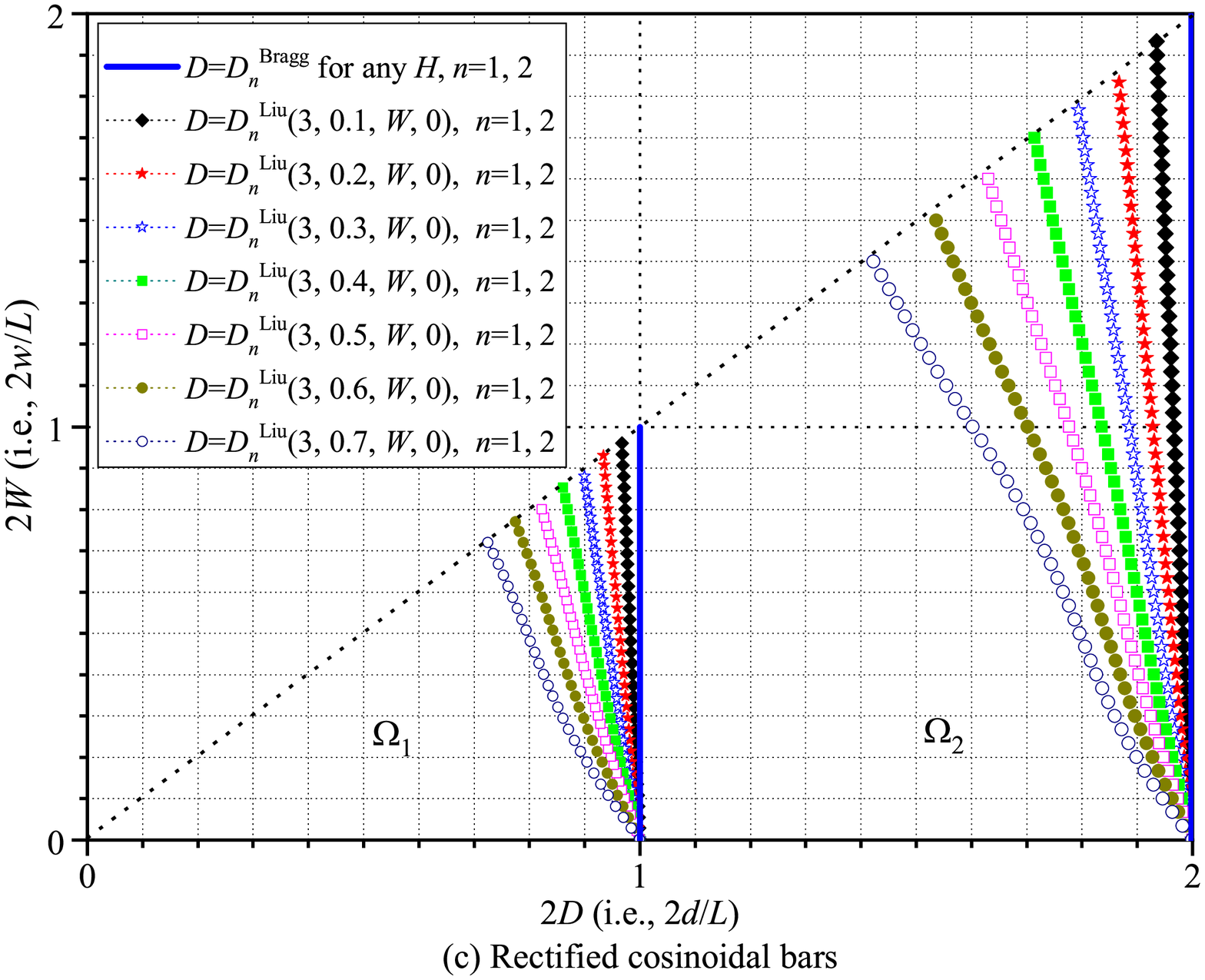}\hspace{-18mm}\epsfxsize=3.0in  \epsffile{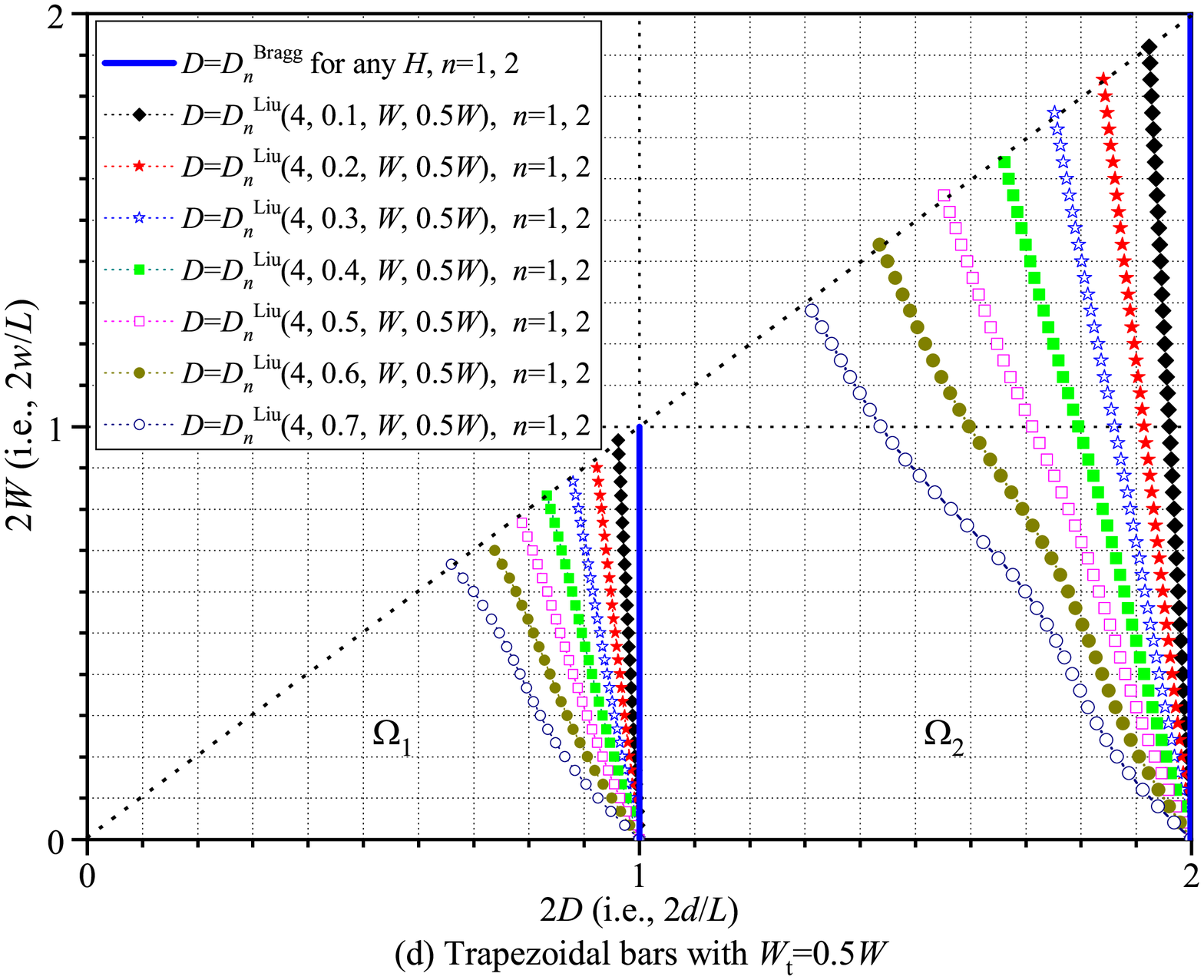}\hspace{-18mm}\epsfxsize=3.0in  \epsffile{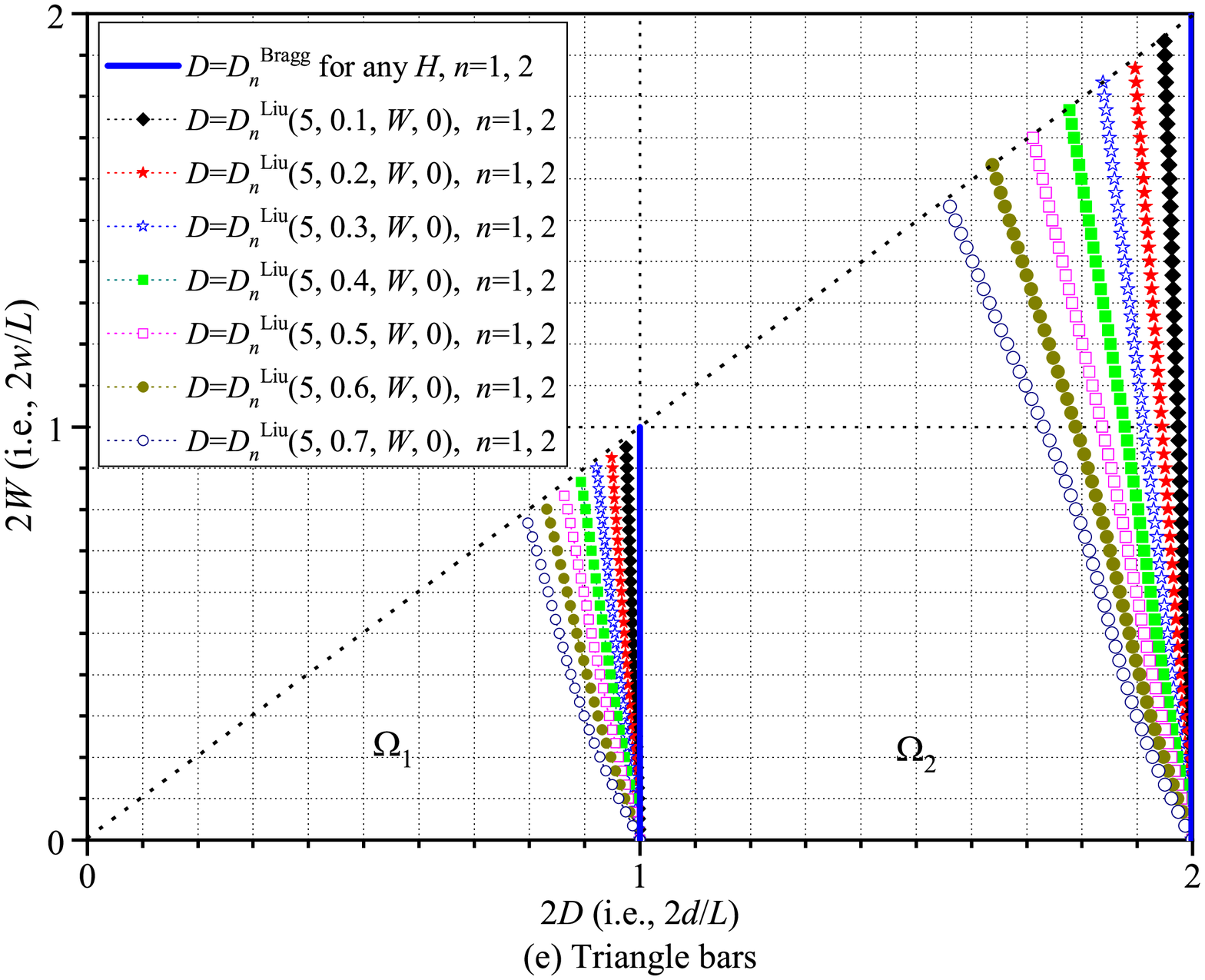}}
\vspace*{-8mm}
\caption[kdv]{\label{fig7} The modified Bragg's law of the 1st and 2nd order Bragg resonances excited by FPA-$j$, $j$=1,2,3,4,5, with $H$=0.1,...,0.7. (a) Rectangular bars; (b) Parabolic bars; (c) Rectified cosinoidal bars; (d) Isosceles trapezoidal bars with $W_t=0.5W$; (e) Isosceles triangular bars.}
\end{figure}

Further, for the five types of bars with the relative bar height, $H$, taking seven values, 0.1, ..., 0.7, the modified Bragg's laws of the 1st and 2nd order Bragg resonances are shown in Figure \ref{fig7}(a)-(e), and for convenience of comparison, the traditional Bragg's laws of the 1st and 2nd order Bragg resonances are also drawn in Figure \ref{fig7}(a)-(e).

%

From the modified Bragg's law (\ref{modified-law}) and Figures \ref{fig6}-\ref{fig7}, the following properties of Bragg resonances can be clarified.

1) The dimensionless phase, $D$, of the $n$th order Bragg resonance given by the modified Bragg's law varies with the bar shape, dimensionless bar height, and dimensionless bar width. This seems reasonable since water waves are mechanical waves, their phase velocities depend on the configuration and geometrical parameters of artificial bars deployed in the bar field. In comparison, X-rays are non-mechanical waves and their velocities do not depend on the propagation medium, so that the relative phase, $D$, of the $n$th order Bragg resonance given by the original Bragg's law in X-ray crystallography is a constant.

2) No matter what type of artificial bars, no matter what the height and width of the bars, and no matter what the order of Bragg resonance is,
the relative phase, $D$, of the $n$th order Bragg resonance given by the modified Bragg's law is always smaller than the phase given by the traditional Bragg's law, the latter comes from the mechanism of the nonlinear wave-wave interaction over a flat bottom (Phillips, 1960) for Class I Bragg resonances excited by artificial bars with the bar height
being quite small. In other words, for any $j$ ($j=1,2,3,4,5$), $H>0$, $W>0$, $W_t\ge0$ and any integer $n$, we always have $D^{\mbox{Liu}}_n(j,H,W,W_t)<D^{\mbox{Bragg}}_n$, since the value of $\mbox{arccot}\frac{U_j\cos(2\pi W)+V_j\sin(2\pi W)}{V_j\cos(2\pi W)-U_j\sin(2\pi W)}$ is positive. Hence, the well-known phenomenon of the phase downshift of Bragg resonances (Kirby and Anton, 1990; Chang and Liou, 2007; Linton, 2011; Liu et al., 2015a; Liu et al., 2016; Liu et al., 2019b; Liu et al., 2020; Xie, 2022) can be well explained here by the modified Bragg's law, see also Figures \ref{fig6}-\ref{fig7}.

3) As we can see in Figure \ref{fig7}, if $W$ and $W_t$ are fixed and let $H$ approach zero, then the modified Bragg's law degenerates into the traditional Bragg's law, which comes from the mechanism of the nonlinear wave-wave interaction over a flat seabed (Phillips, 1960) for Class I Bragg resonances. This is not strange since Phillips' (1960) mechanism was derived by using the perturbation method where $H$ is also assumed to be small. On the other hand, if $H$ is fixed and let $W$ approach zero (clearly, $W_t<W$ also approach zero), then as we can see in Figures \ref{fig6}-\ref{fig7}, the modified Bragg's law also degenerates into the traditional Bragg's law, this means that, even if $H$ is not small, when linear long waves are reflected by a finite periodic array of submerged thin plates, the $n$th order Bragg resonance will occur approximately at $D=D^{\mbox{Bragg}}_n$. This conclusion cannot be inferred from Phillips' (1960) mechanism since $H$ is not so small that the perturbation method cannot be employed. To the author's knowledge, Bragg resonances excited by a finite periodic array of thin plates are rarely studied, only a similar problem was studied by Losada et al. (1993).

Based on the conclusions on the above two aspects, we can further conclude that, for Class I Bragg resonances excited by each of five types of artificial bars, FPA-$j$ ($j$=1,...,5), if $H$ or $W$ is zero, i.e., if the area of the bar cross-section is zero, then the traditional Bragg's law holds. However, it is noted that no bar can be zero in height or width, hence, for Class I Bragg resonances excited by a finite periodic array of artificial bars, the traditional Bragg's law cannot hold although it has long been regarded as a standard for Class I Bragg resonances.

%
\begin{figure}
\vspace*{-35mm}
\centerline{\hspace*{0mm}\epsfxsize=3.4in \epsffile{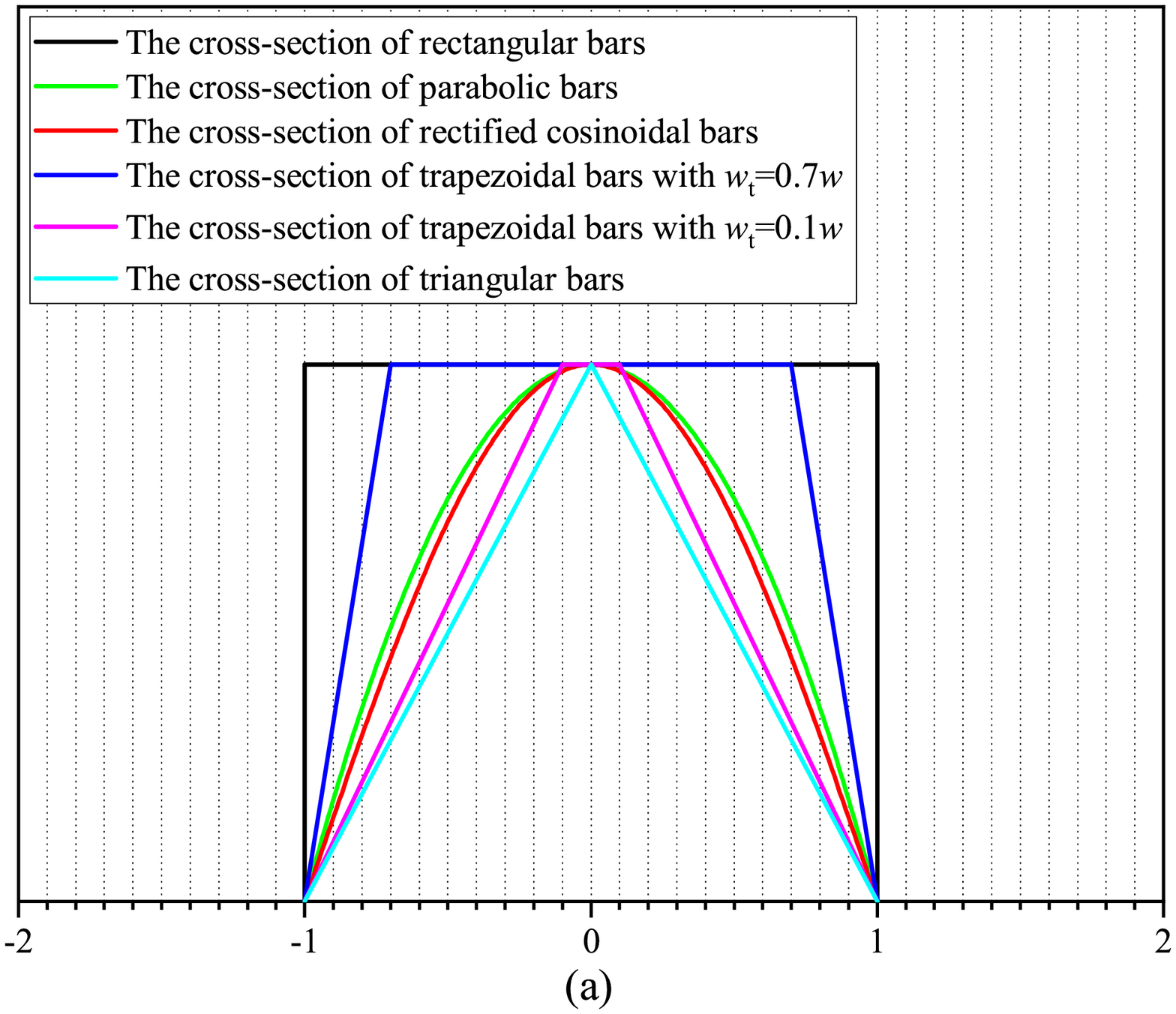}\hspace{-20mm}\epsfxsize=3.4in \epsffile{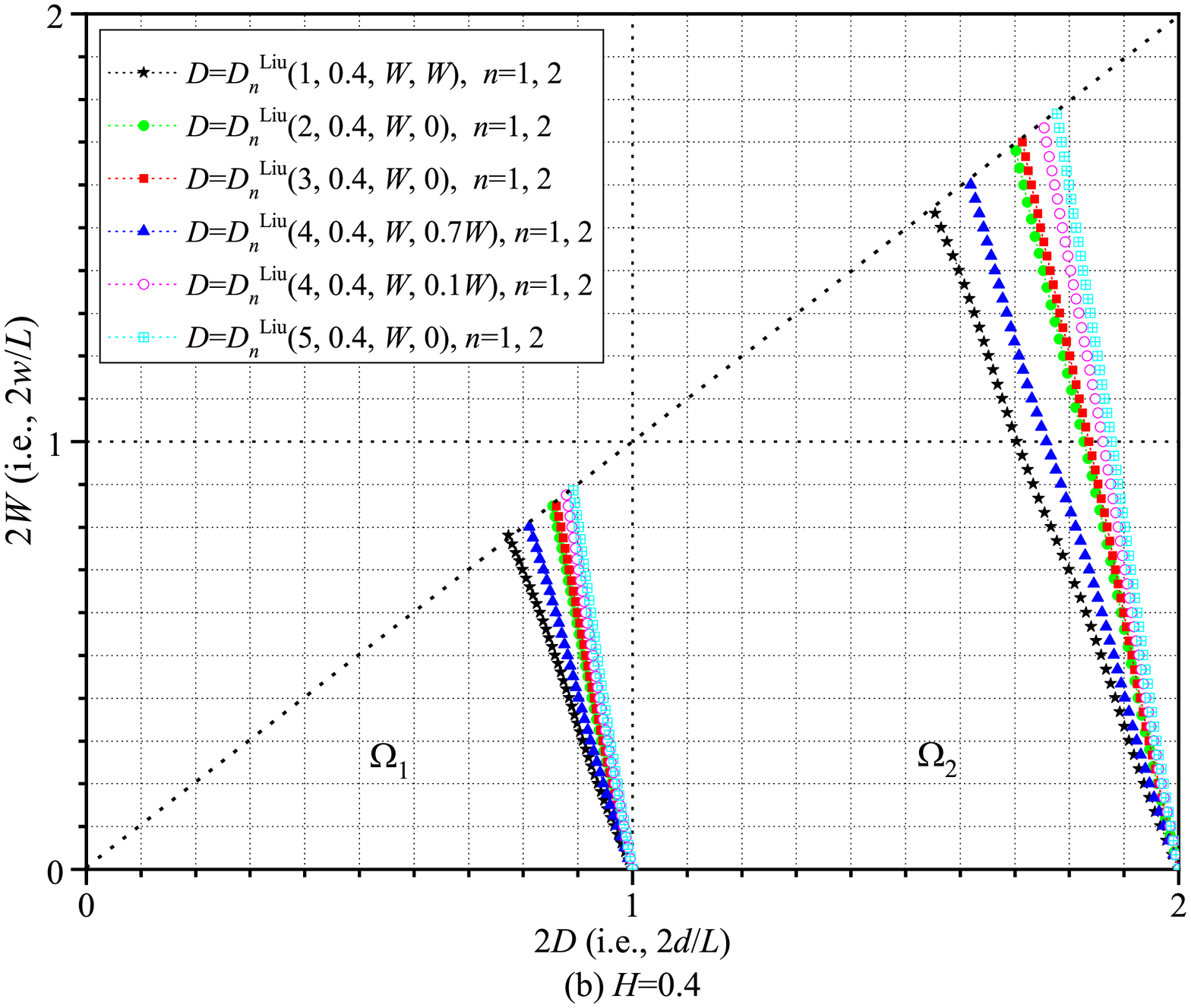}}
\vspace*{-8mm}
\centerline{\hspace*{0mm}\epsfxsize=3.4in \epsffile{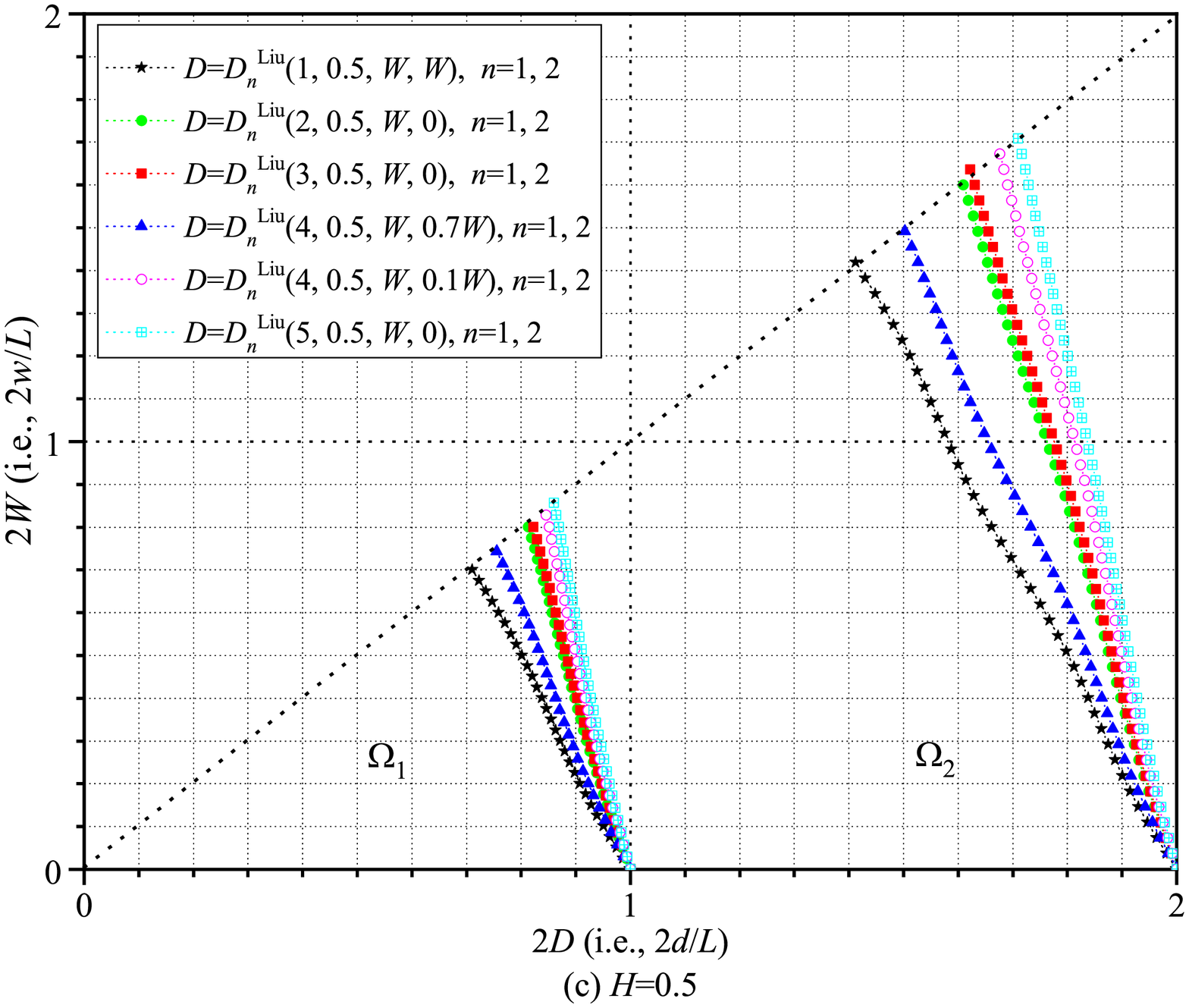}\hspace{-20mm}\epsfxsize=3.4in \epsffile{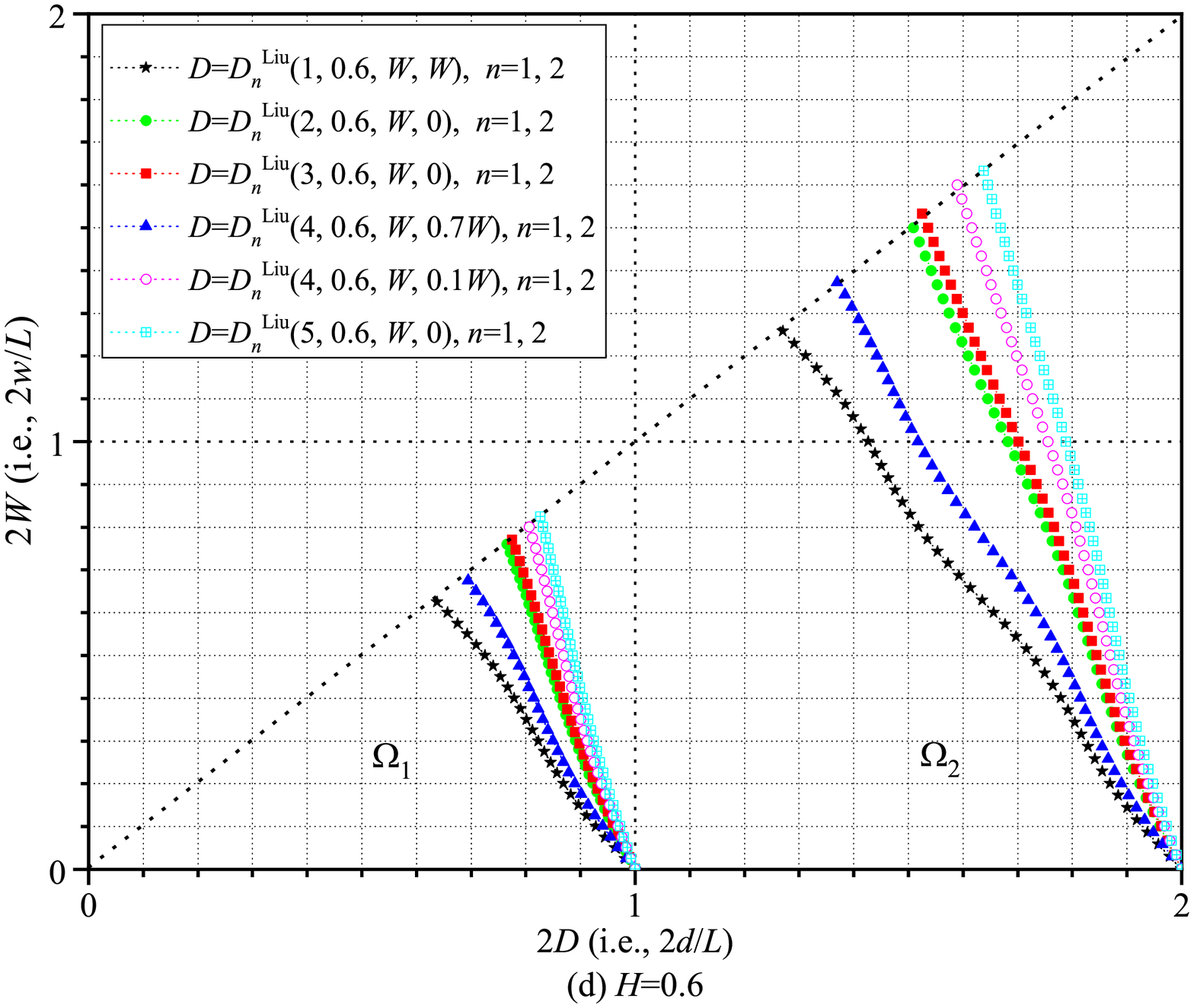}}
\vspace*{-10mm}
\caption[kdv]{\label{fig8} The influence of bar shape and area on the phase of Bragg resonance. (a) Different bars; (b)-(d) The modified Bragg's laws for different bars with $H$=0.4, 0.5, 0.6, respectively.}
\end{figure}

4) For all the five types of bars, if both $W$ and $W_t$ are fixed, then, as shown in Figure \ref{fig7}, the phase downshift of each order Bragg resonance increases with increasing the dimensionless bar height, $H$. This phenomenon has long been observed and reported by many scholars, for example, see Liu et al. (2020, Fig. 11), Guo et al. (2021, Fig. 7(a)-(b)), and Xie (2022, Fig. 6).

On the other hand, if the dimensionless bar height, $H$, and the ratio, $W_t/W$, are fixed, i.e., $H=H_0$ and $W_t/W=s$, such as $H_0=0.5$, $s=1$ in Figure \ref{fig6}(a), $H_0=0.5$, $s=0$ in Figure \ref{fig6}(b)-(c) and in Figure \ref{fig6}(e), and $H_0=0.5$, $s=0.5$ in Figure \ref{fig6}(d), then the phase downshift of each order Bragg resonance increases with increasing the dimensionless bar width, $W$. This feature coincides with the observations of Liu et al. (2020, Fig. 9) and Guo et al. (2021, Fig. 8(a)-(b)) in the whole wave range from deep water to shallow water. This implies that, the function $D^{\mbox{Liu}}_n(j, H_0, W, sW)$ is a monotonically decreasing function of $W$ though we cannot prove this in mathematics at the moment.

By combining the conclusions on the two aspects, we can further conclude that, for Bragg resonances excited by each of five types of artificial bars, i.e., FPA-$j$, $j$=1,...,5, the phase downshift of each order Bragg resonance increases with increasing the value of $H*W$.

5) To see the influence of the bar shape on the phases of Bragg resonance, six arrays of artificial bars with the same height and bottom width but with different cross-sections are plotted in Figure \ref{fig8}(a), and the corresponding modified Bragg's laws of the 1st and 2nd order resonances excited by these different bar arrays with $H=0.4$, 0.5 and 0.6 are plotted in Figure \ref{fig8}(b)-(d), respectively. It is clearly revealed by these results that, the larger the cross-sectional area of artificial bars, the more significant the phase downshift they cause.

\begin{figure}
\vspace*{-30mm}
\centerline{\hspace*{2mm}\epsfxsize=3.0in \epsffile{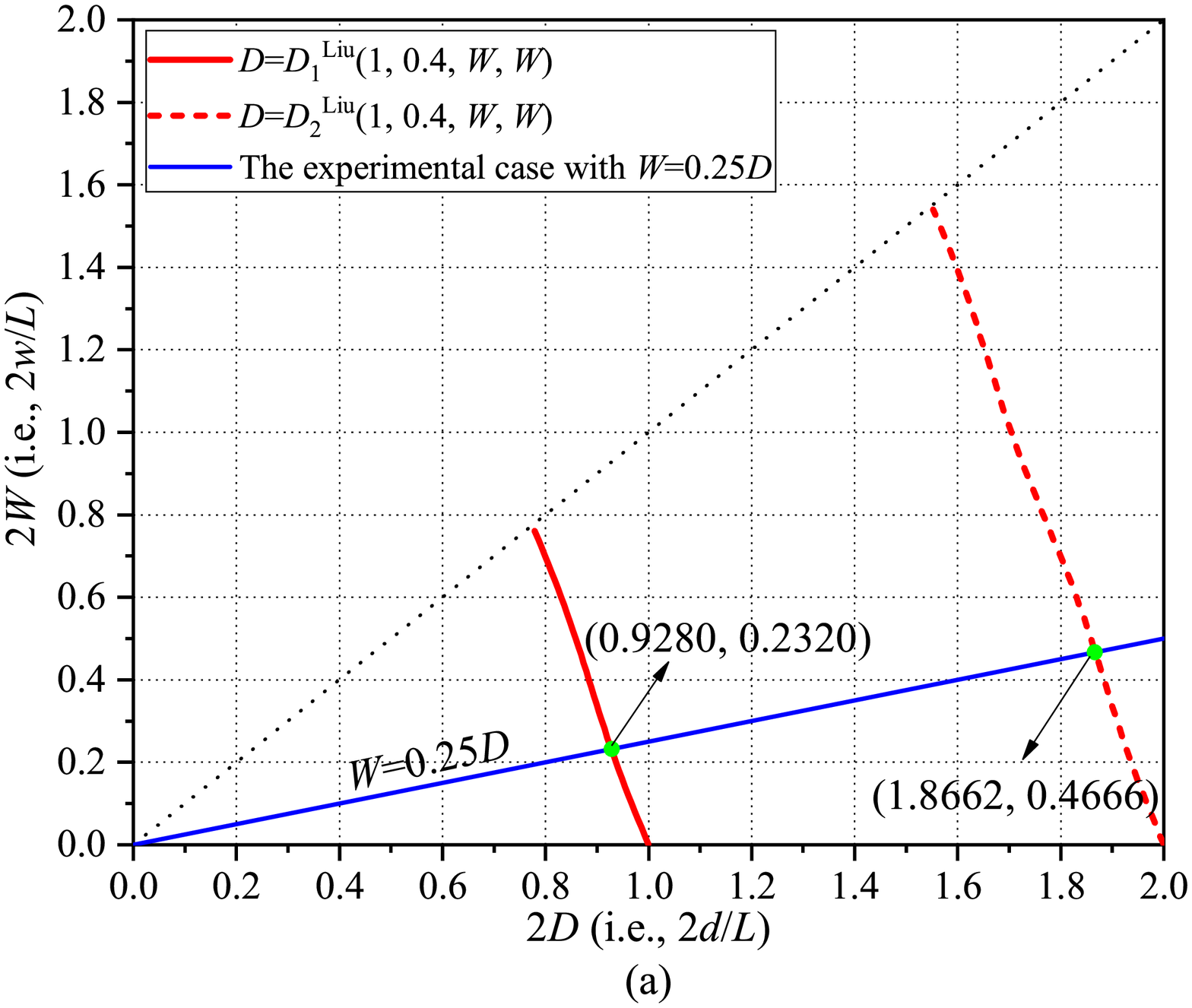}\hspace{-10mm}\epsfxsize=3.0in \epsffile{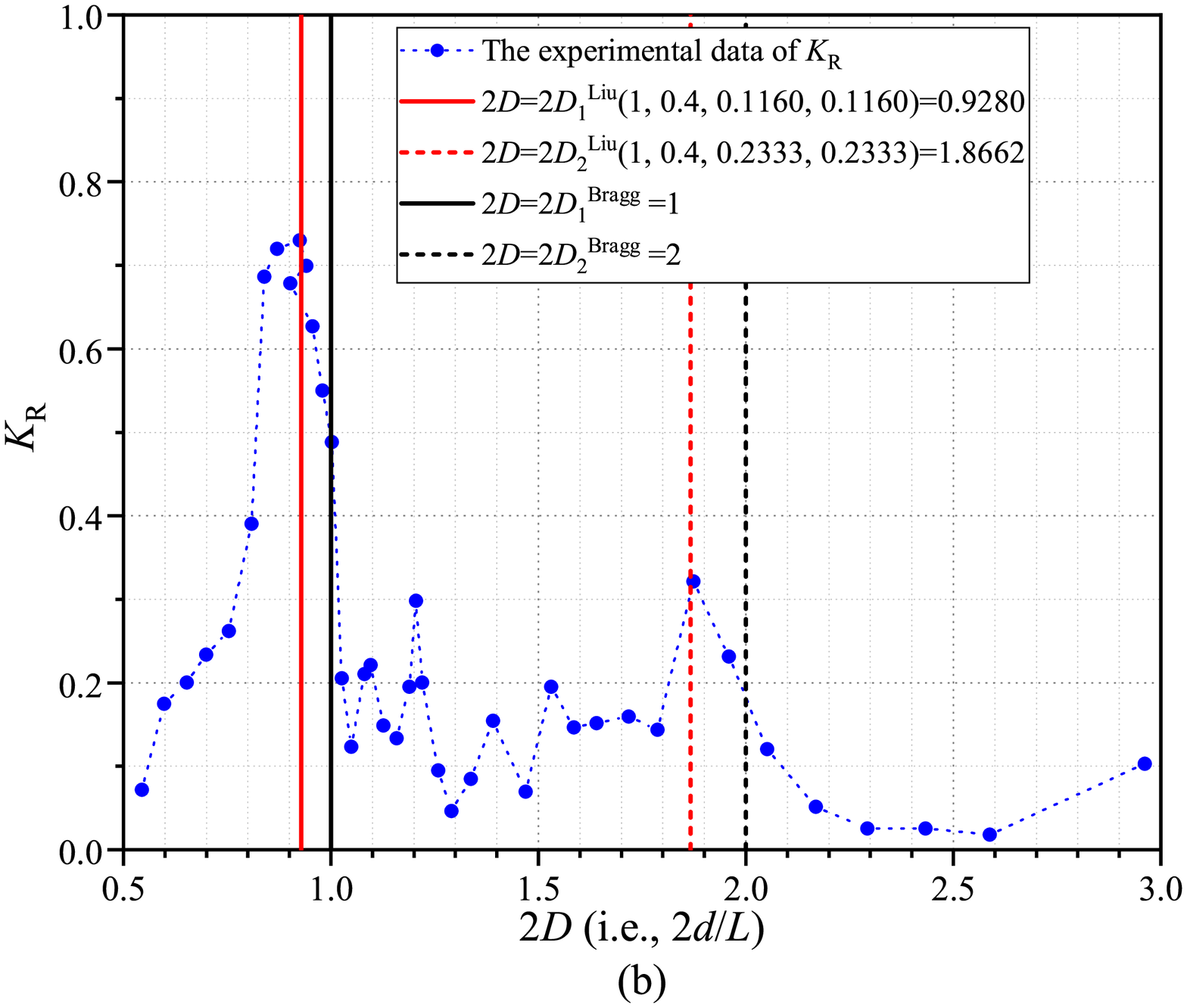}}
\vspace*{-9mm}
\caption[kdv]{\label{fig9} The predictions of the 1st- and 2nd-order resonances by the modified Bragg's law; (b) Comparison between the predictions and the experimental data.}
\end{figure}

\section{Validation of the modified Bragg's law}

In this section, to validate the modified Bragg's law, we consider a few cases which have been studied before by various methods including physical, numerical and analytical modellings.

\subsection{Bragg resonances excited by a FPA-1}

First, we consider wave reflection by a FPA-1. The parameters of the rectangular bars are as follows: $N=8$, $h_0$=0.60 m, $h_1$=0.36 m (i.e., $H$=0.4), $d$=2.40 m, $w$=0.60 m ($W=0.25D$), and the wave period ranges from 1.03 s to 4 s, i.e., $2D\in[0.5077, 2.9571]$, or $k_0h_0\in[0.3987, 2.3225]$. Clearly this goes beyond the long-wave range. This case was studied experimentally by Hsu et al. (2003), see their Fig. 7 and Wang et al. (2006), see their Fig. 5.

We now present the prediction results of the phases of the 1st- and 2nd-order Bragg resonances occurred in $[0.5077, 2.9571]$ by using the the modified Bragg's law, though recognizing that it is valid for long waves only. The modified Bragg's laws of the 1st- and 2nd-order Bragg resonances with $H=0.4$ are plotted in Figure \ref{fig9}(a), which are the red solid line and the red dashed line, respectively. The line $W=0.25D$ is also plotted in Figure \ref{fig9}(a), which together with the condition $H$=0.4 characterizes the bar field. The point where the line $W=0.25D$ intersects the curve representing the 1st-order modified Bragg's law is (0.9280, 0.2320), means that the 1st-order Bragg resonance will occur at $2D=0.9280$. And the point where the line $W=0.25D$ intersects the curve representing the 2nd-order modified Bragg's law is ((1.8662, 0.4666), means that the 2st-order Bragg resonance will occur at $2D=1.8662$. As we can see in Figure \ref{fig9}(b), both the predictions of the phases of the 1st- and 2nd-order Bragg resonances given by the modified Bragg's law are quite accurate.

\begin{figure}
\vspace*{-10mm}
\centerline{\hspace*{0mm}\epsfxsize=3.0in \epsffile{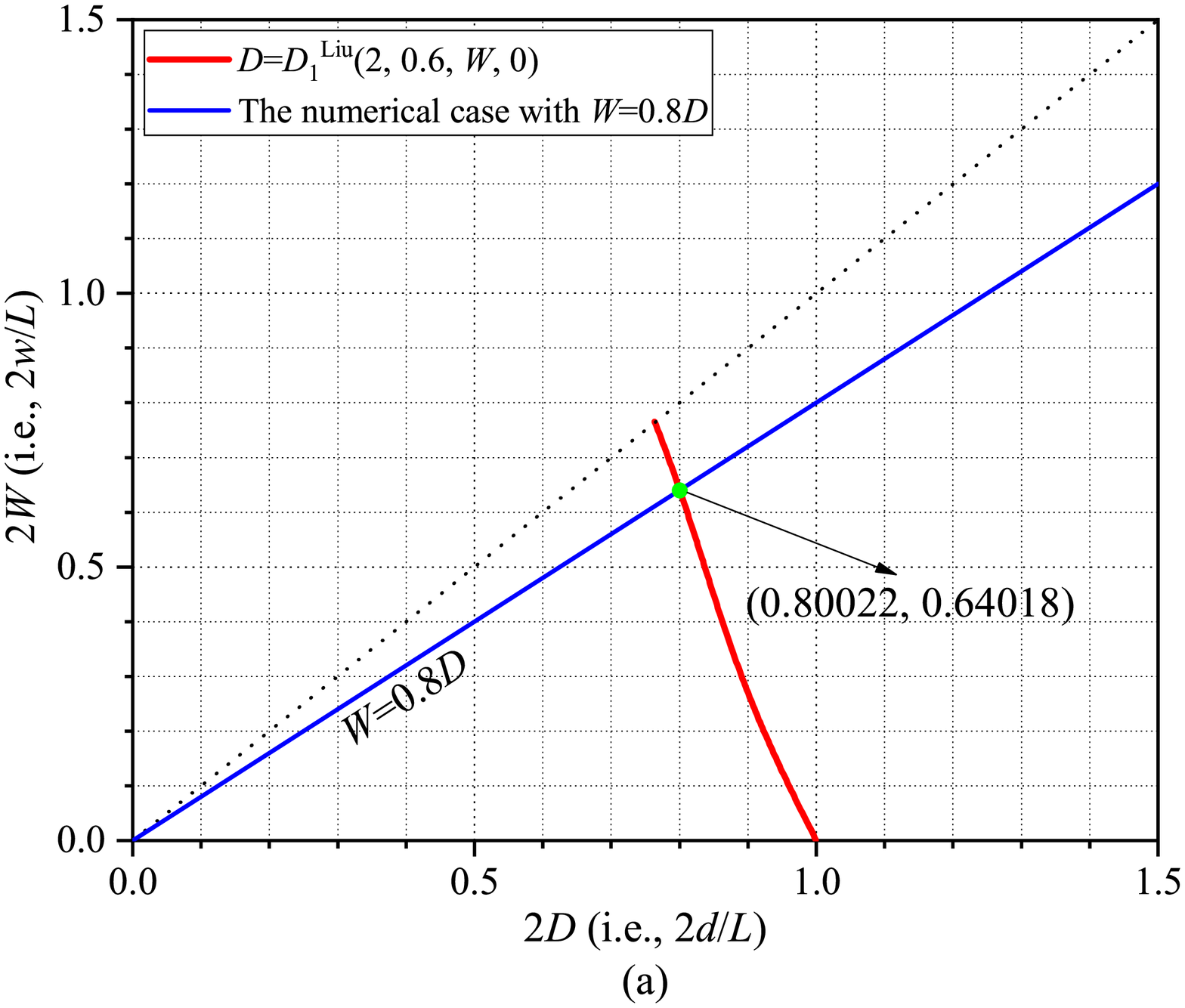}\hspace{-17mm}\epsfxsize=3.0in  \epsffile{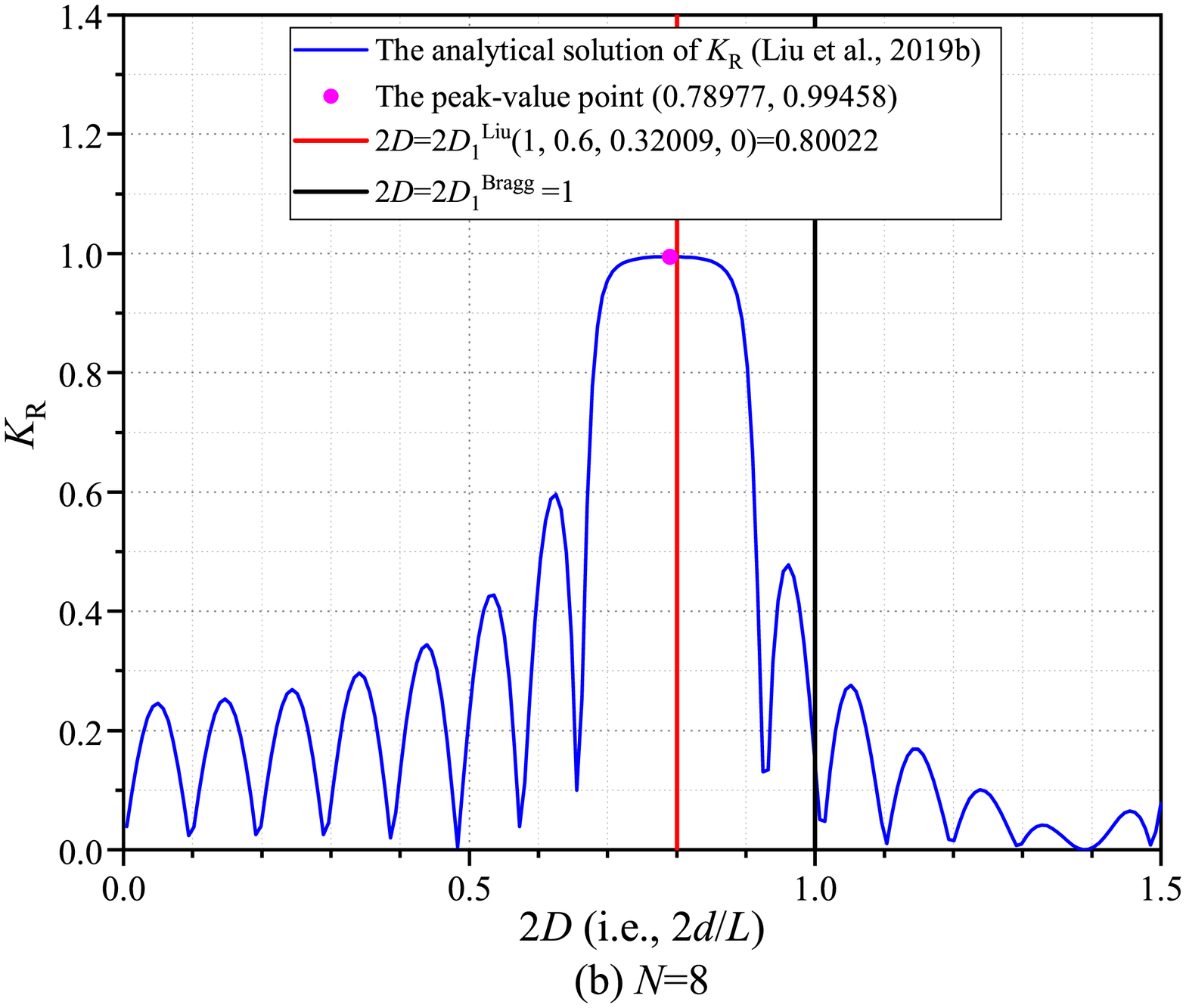}\hspace{-17mm}\epsfxsize=3.0in  \epsffile{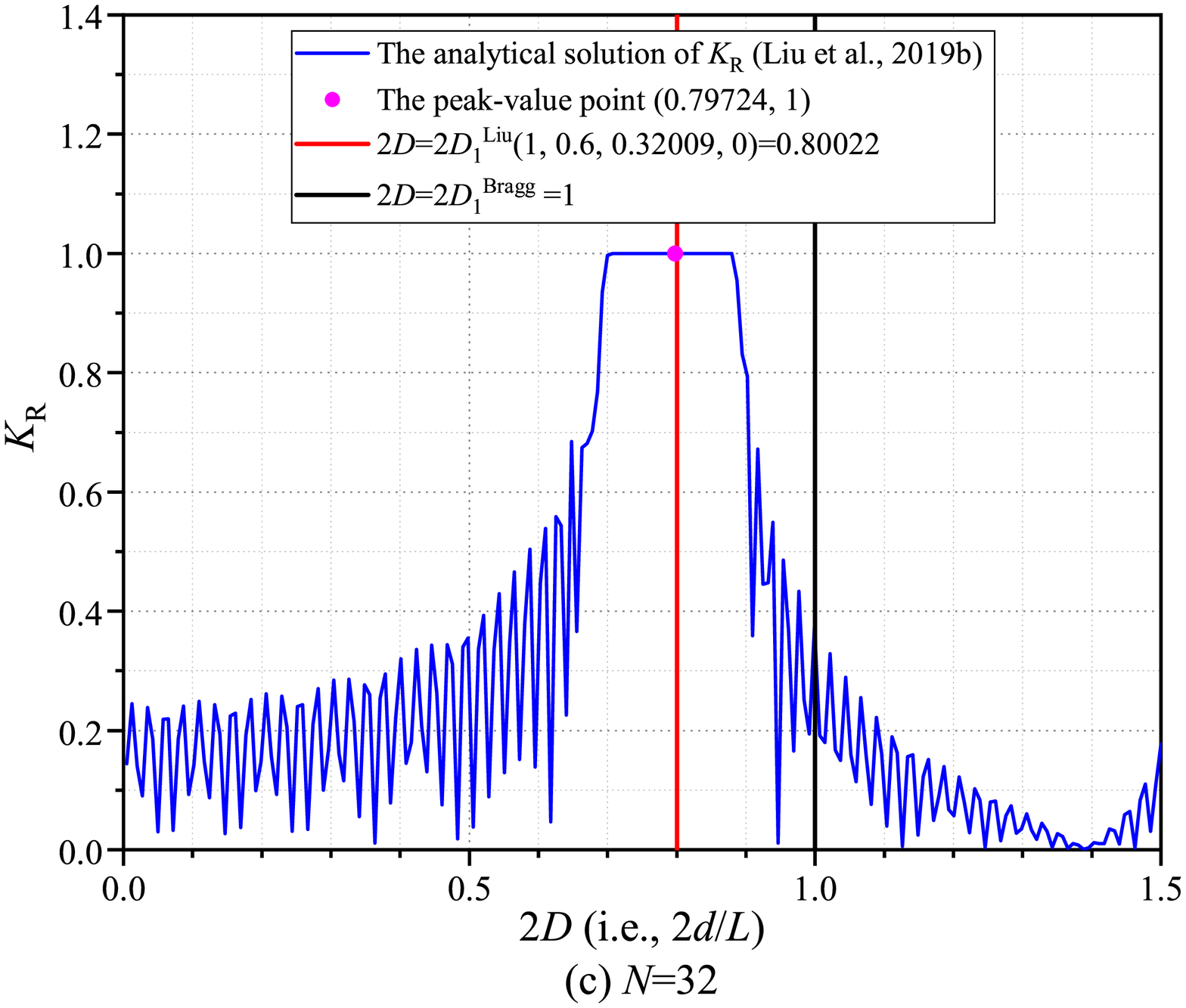}}
\vspace*{-7mm}
\caption[kdv]{\label{fig10} (a) The prediction of the phase of the 1st-order Bragg resonance; (b) Comparison between the prediction and the analytical solution with $N$=8; (c) Comparison between the prediction and the analytical solution with $N$=32.}
\end{figure}

\subsection{Bragg resonances excited by a FPA-2}

Next, we consider wave reflection by a FPA-2. The parameters of the parabolic bars are as follows: $H$=0.6, $d=15h_0$, $w=12h_0$ (i.e., $W=0.8D$), and the wave condition is $k_0h_0\in[0.001,\frac{\pi}{10}]$,  i.e., $2D\in [0.0048, 1.5]$. The analytical solutions of the reflection coefficient to the linear long-wave equation for three cases with $N$=8,16,32 were calculated by Liu et al. (2019b).

To predict the phase of the 1st-order Bragg resonance occurred in $[0.0048, 1.5]$, the modified Bragg's law of the 1st-order Bragg resonance with $H=0.6$ is plotted in Figure \ref{fig10}(a) and remarked by a red solid curve. The line $W=0.8D$ is also plotted in Figure \ref{fig10}(a), which together with the condition $H=0.6$ characterizes the bar field composed of parabolic bars. The point where the line $W=0.8D$ intersects the red solid curve is (0.8002, 0.6402), means that the 1st-order Bragg resonance predicted by the modified Bragg's law should occur at $2D=0.8002$. As we can see in Figure \ref{fig10}(b)-(c), the two 1st-order Bragg resonances occur at $2D=0.7898$ and $2D=0.7972$ when $N=8$ and $N=32$, respectively, both of them are very close to the prediction $2D=0.8002$. It is noted that the modified Bragg's law is derived based on the Bloch band theory, in which a FPA-$j$ is approximately replaced by the corresponding IPA-$j$, equivalently, the water surface waves over a FPA-$j$ are approximately regarded as Bloch waves over the corresponding IPA-$j$. Therefore, the accuracy of the prediction given by the modified Bragg's law increases with increasing the bar number $N$.

\begin{figure}
\vspace*{-10mm}
\centerline{\hspace*{0mm}\epsfxsize=3.0in \epsffile{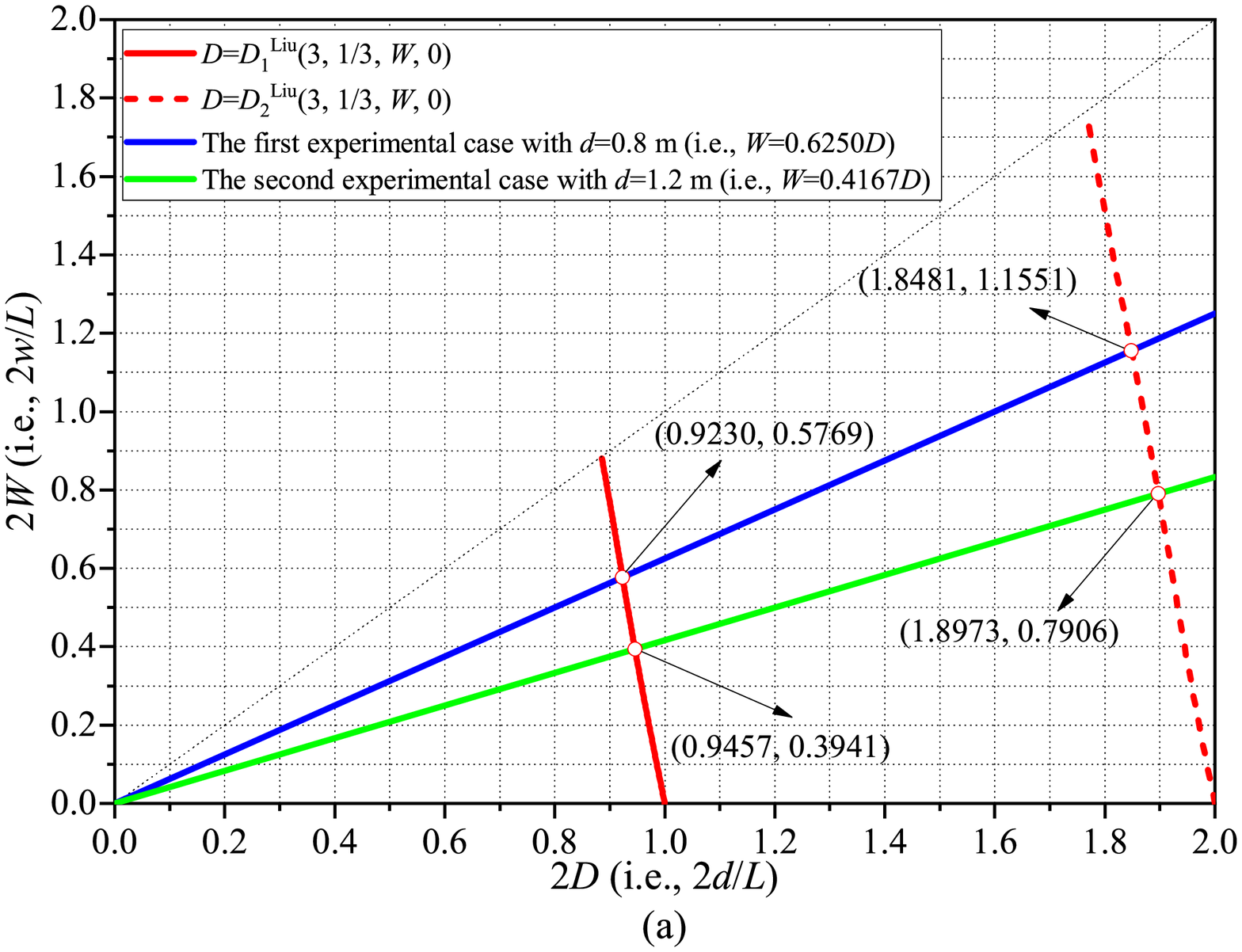}\hspace{-17mm}\epsfxsize=3.0in  \epsffile{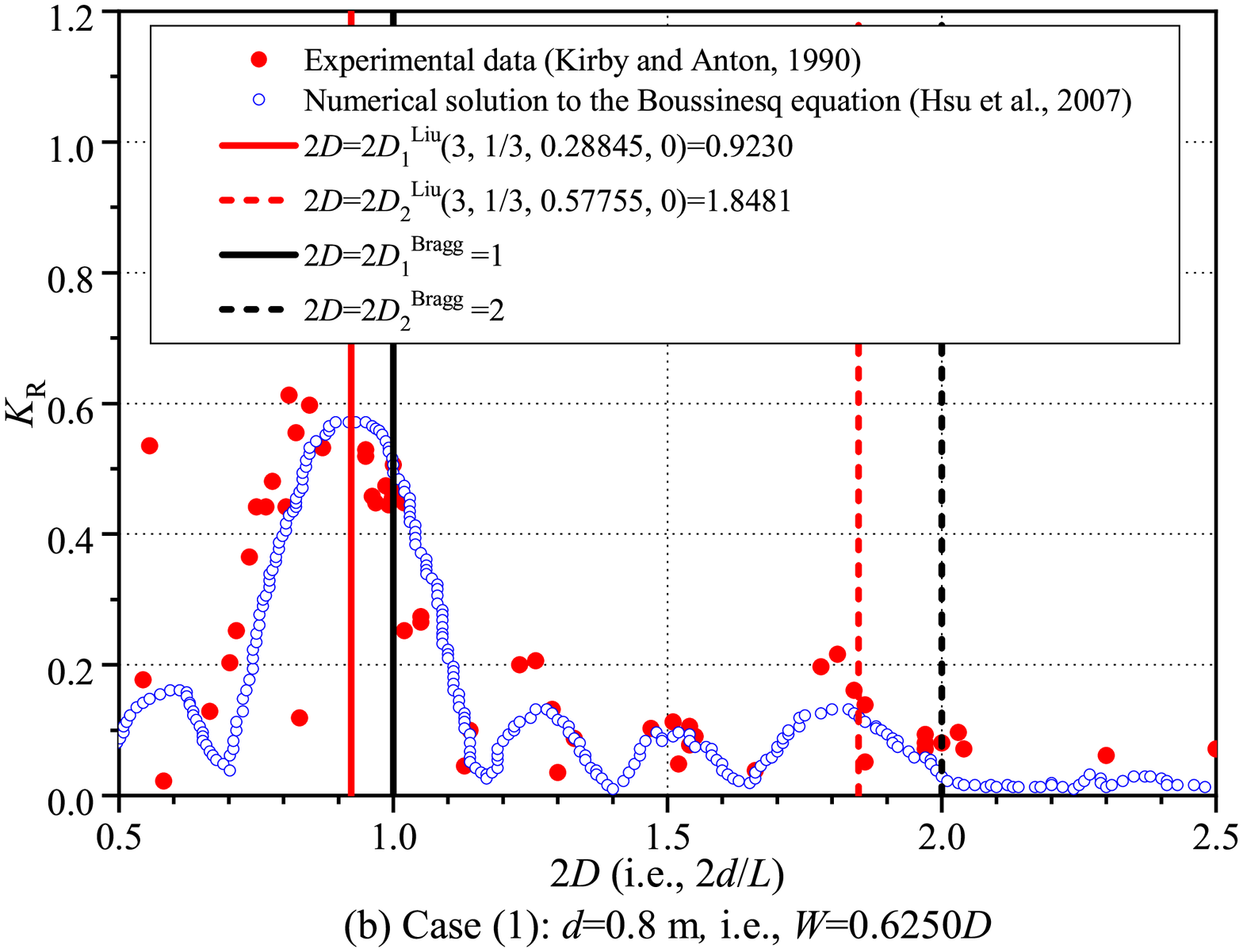}\hspace{-17mm}\epsfxsize=3.0in  \epsffile{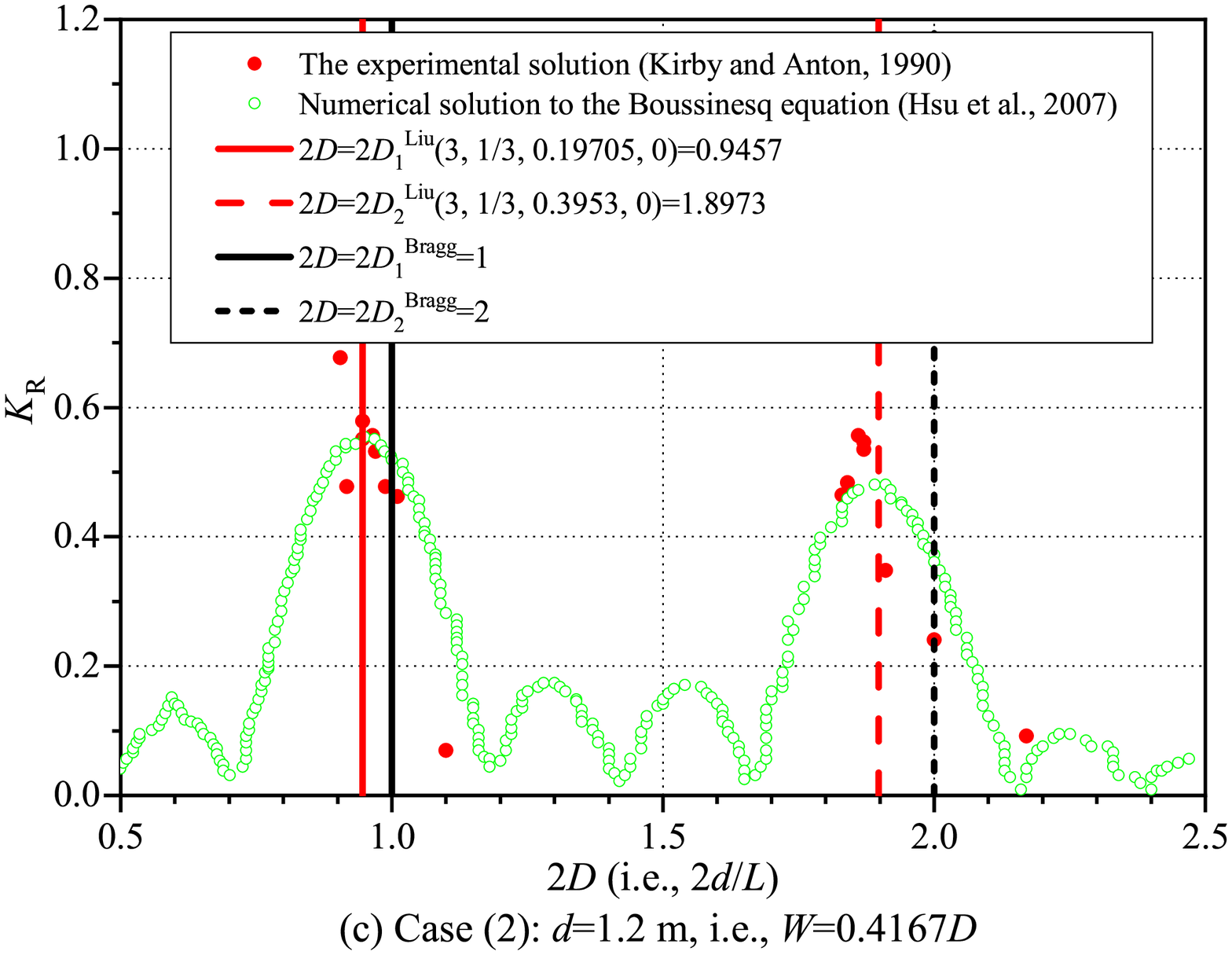}}
\vspace*{-7mm}
\caption[kdv]{\label{fig11} (a) The predictions of the phases of the 1st- and 2nd order Bragg resonances; (b) Comparison between the predictions and the experimental data with $d$=0.8 m; (c) Comparison between the predictions and the experimental data with $d$=1.2 m.}
\end{figure}

\subsection{Bragg resonances excited by a FPA-3}

Third, we consider wave reflection by a FPA-3. The parameters of the rectified cosinoidal bars are as follows: $h_0=0.15$ m, $H=\frac{1}{3}$, $w=0.5$ m, $d=0.8$ m (i.e., $W=0.6250D$) or $d=1.2$ m (i.e., $W=0.4167D$), and the wave condition is $2D\in [0.5, 2.5]$, i.e., $k_0h_0\in[0.2945,1.4726]$ for the case $d=0.8$ m, and $k_0h_0\in[0.1964,0.9818]$ for the case $d=1.2$ m. Clearly this goes beyond the long-wave range. For these two cases, experimental data were given by Kirby and Anton (1990) and numerical solutions to the Boussinesq equation were given by Hsu et al. (2007).

Now the modified Bragg's law is used to predict the phases of the 1st- and 2nd-order Bragg resonances occurred in $[0.5, 2.5]$, though it is valid for long waves only. For $n$=1,2, the modified Bragg's law $D=D^{\mbox{Liu}}_n(3,\frac{1}{3},W,0)$ with $H=\frac{1}{3}$ are plotted in Figure \ref{fig11}(a). For the case with $W=0.6250D$, the two points where the line $W=0.6250D$ intersects the two curves representing the 1st- and 2nd-order modified Bragg's laws are (0.9230, 0.5769) and (1.8481, 1.1551), respectively, which mean that, according to the predictions of the modified Bragg's law, the 1st- and 2nd-order Bragg resonances will occur at $2D=0.9230$ and $2D=1.8481$, respectively. Similarly, for the case with $W=0.4167D$, the 1st- and 2nd-order Bragg resonances will occur at $2D=0.9457$ and $2D=1.8973$, respectively. As we can see in Figure \ref{fig11}(b)-(c), all our predicted phases in two cases coincide with experimental data  quite well. Especially, the two predicted phases of the 1st-order Bragg resonance in two cases are almost identical to the two phases of the numerical solution.

\begin{figure}
\vspace*{-10mm}
\centerline{\hspace*{0mm}\epsfxsize=3.0in \epsffile{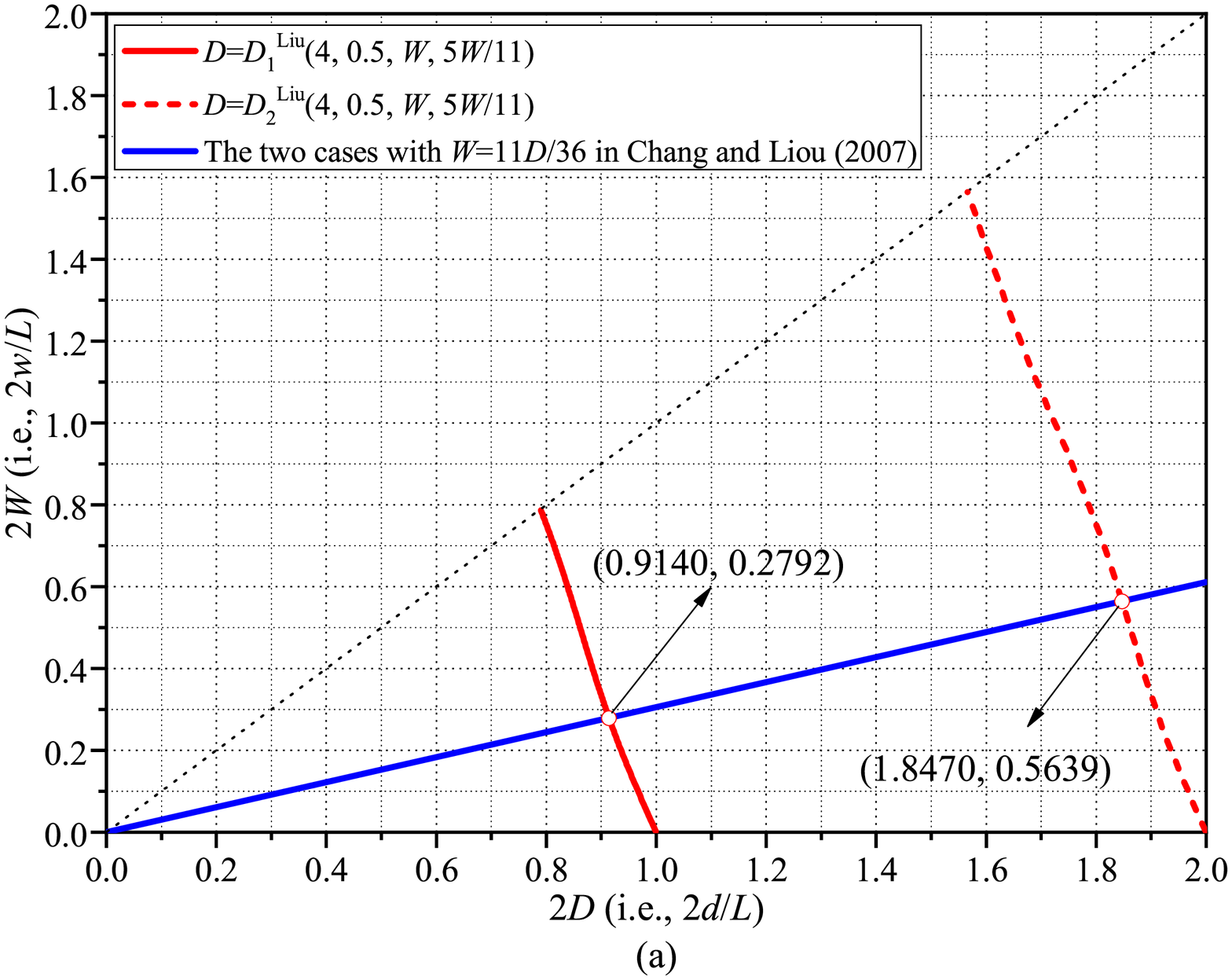}\hspace{-17mm}\epsfxsize=3.0in  \epsffile{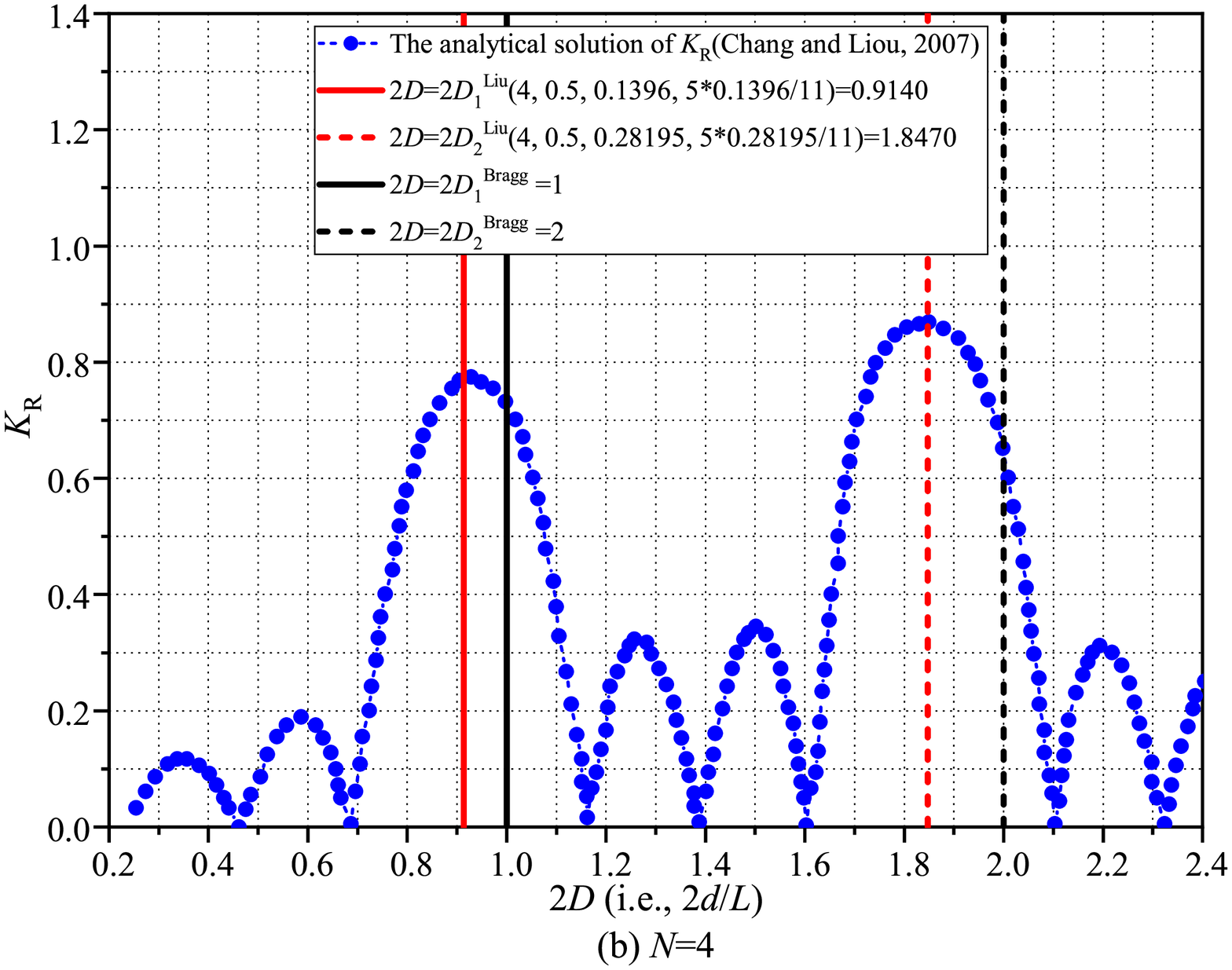}\hspace{-17mm}\epsfxsize=3.0in  \epsffile{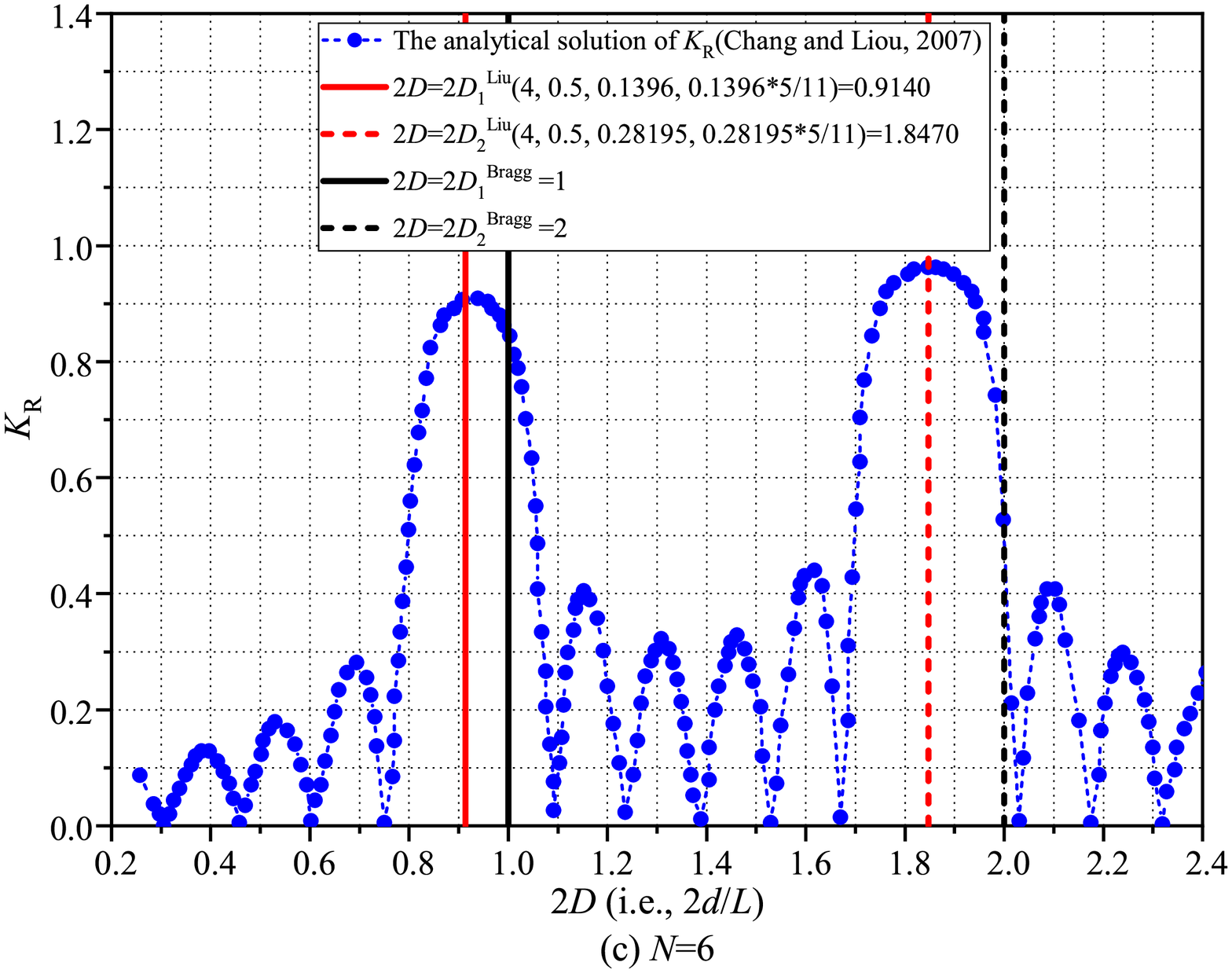}}
\vspace*{-9mm}
\caption[kdv]{\label{fig12} (a) The predictions of the phases of the 1st- and 2nd-order Bragg resonances; (b) Comparison between the predictions and the analytical solution with $N$=4; (c) Comparison between the predictions and the analytical solution with $N$=6.}
\end{figure}

\subsection{Bragg resonances excited by a FPA-4}

Fourth, we consider wave reflection by a FPA-4. The parameters of the trapezoidal bars are as follows: $h_0=2.4$ m, $H=0.5$, $d=28.8$ m, $w=8.8$ m (i.e., $W=\frac{11}{36}D$), $w_t=4$ m (i.e., $W_t=\frac{5}{11}W=\frac{5}{36}D$), and the wave condition is $2D\in [0.2547, 2.4044]$, i.e., $k_0h_0\in[0.0667,0.6295]$. Clearly, this still partially goes beyond the long-wave range: $2D\in (0, 1.2]$. For the two cases with $N$=4,6, analytical solutions to the long-wave equation were given by Chang and Liou (2007).

Now we employ the modified Bragg's law to predict the phases of the 1st- and 2nd-order Bragg resonances for $2D$ in $[0.2547, 2.4044]$, though recognizing that the modified law is valid for long waves only. The
modified Bragg's law $D=D^{\mbox{Liu}}_n(4,0.5,W,\frac{5}{11}W)$ with $n$=1,2 are plotted in Figure \ref{fig12}(a). The two points where the line $W=\frac{11}{36}D$ intersects the two curves representing the 1st- and 2nd-order modified Bragg's laws are (0.9140, 0.2792) and (1.8470, 0.5639), respectively, hence, the 1st- and 2nd-order Bragg resonances should occur at $2D=0.9140$ and $2D=1.8470$, respectively. It can be seen in Figure \ref{fig12}(b)-(c) that, for the case with $N=4$, the phases of 1st- and 2nd-order Bragg resonances of the analytical solution are $2D=0.9287$ and $2D=1.8493$, respectively. And for the case with $N=6$, the phases of 1st- and 2nd-order Bragg resonances of the analytical solution are $2D=0.9390$ and $2D=1.8539$, respectively. Obviously, the predicted phases in two cases given by the modified Bragg's law coincide with analytical solutions quite well and are much more accurate than the two phases $2D=1$ and $2D=2$ predicted by the traditional Bragg's law.

\begin{figure}
\vspace*{-10mm}
\centerline{\hspace*{0mm}\epsfxsize=3.0in \epsffile{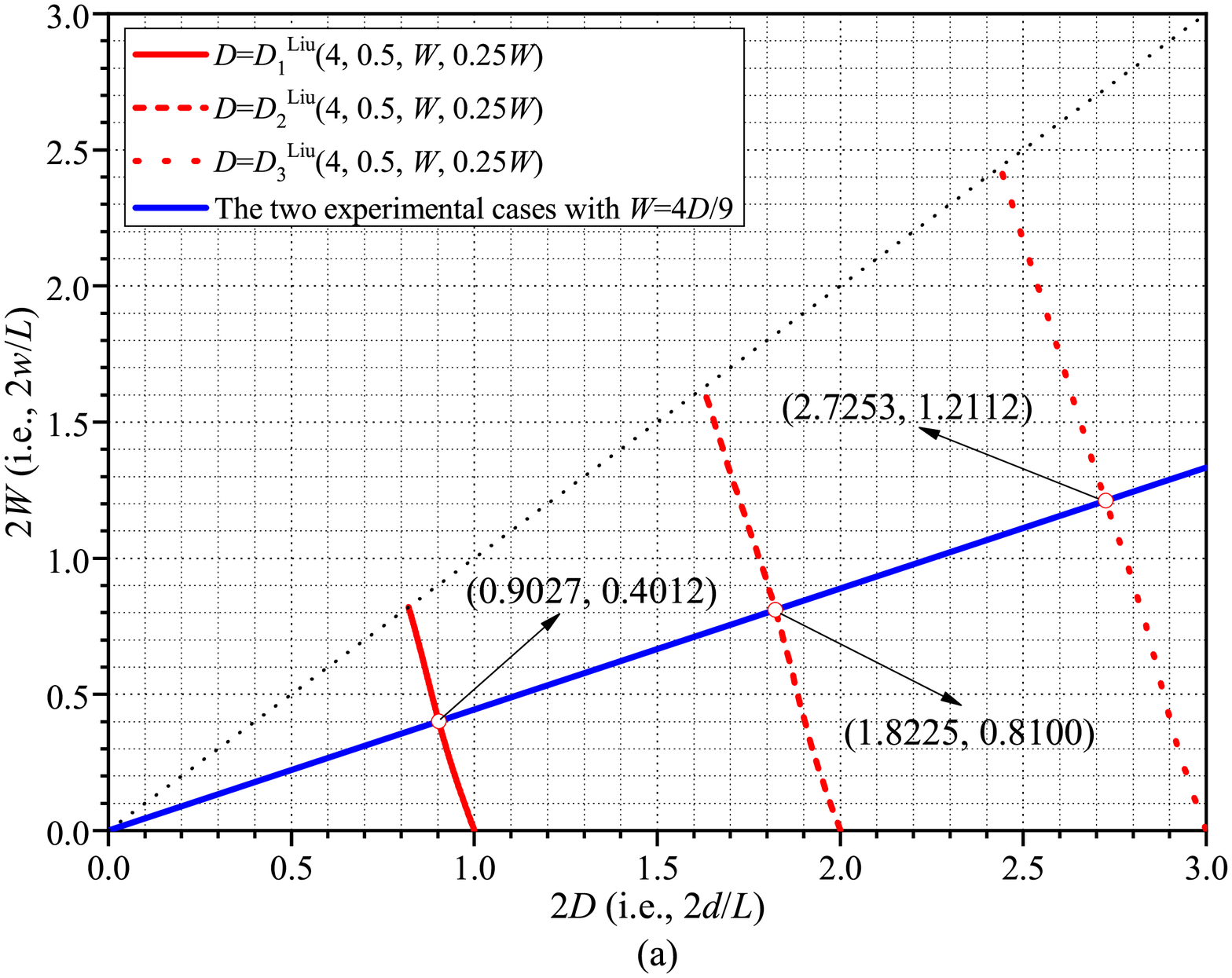}\hspace{-18mm}\epsfxsize=3.0in  \epsffile{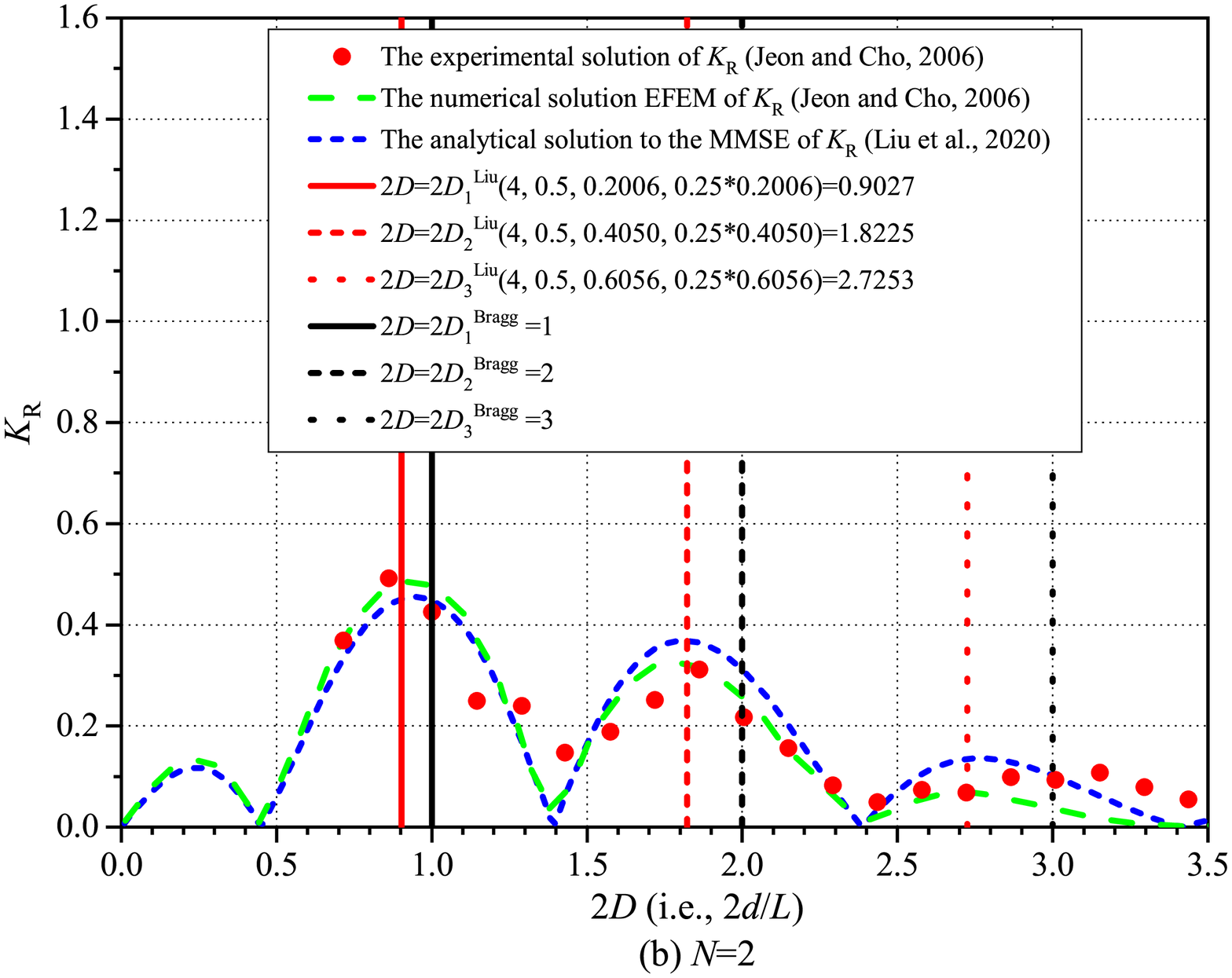}\hspace{-18mm}\epsfxsize=3.0in  \epsffile{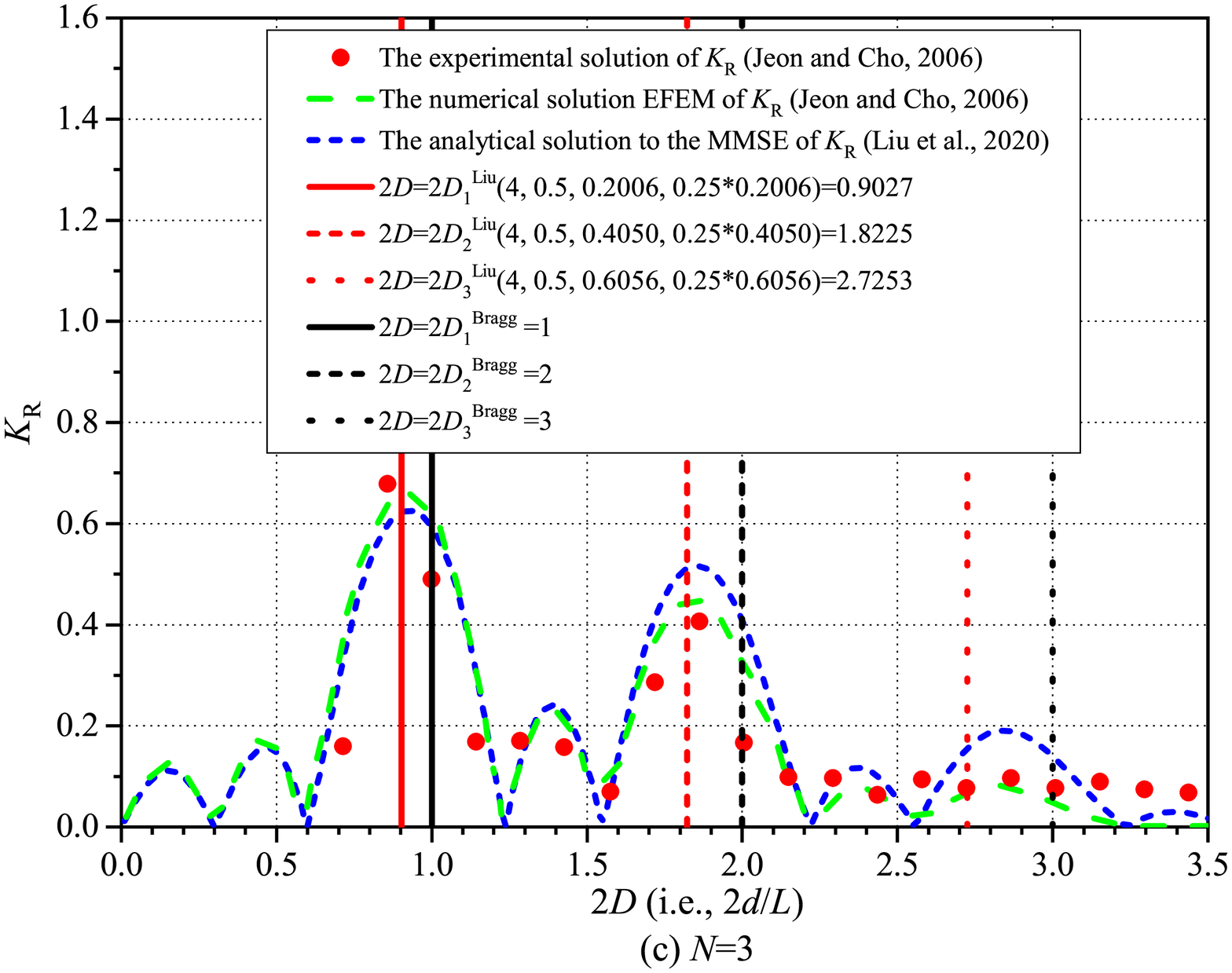}}
\vspace*{-9mm}
\caption[kdv]{\label{fig13} (a) The predictions of the phases of the 1st-, 2nd-, and 3rd-order Bragg resonances; (b) Comparison between the predictions and the analytical solution with $N$=2; (c) Comparison between the predictions and the analytical solution with $N$=3.}
\end{figure}

Now we further consider wave reflection by a FPA-4. The parameters of the trapezoidal bars are as follows: $h_0=0.8$ m, $H=0.5$, $d=3.6$ m, $w=1.6$ m (i.e., $W=\frac{4}{9}D$), $w_t=0.4$ m (i.e., $W_t=\frac{W}{4}=\frac{D}{9}$), and the calculating range of water waves is $2D\in (0,3.5]$, i.e., $k_0h_0\in(0,2.4435]$, which is beyond the long-wave range. For the two cases with $N$=2,3, both experimental modelling and numerical modelling were conducted by Jeon and Cho (2006) and an analytical modelling based on the modified mild-slope equation was given by Liu et al. (2020). According to the modified Bragg's law for trapezoidal bars, i.e., $D=D^{\mbox{Liu}}_n(4,0.5,W,\frac{W}{4})$, the 1st-, 2nd- and 3rd-order (i.e., $n$=1,2,3) Bragg resonances should occur at $2D$=0.9027, 1.8225, and 2.7253, respectively, see Figure \ref{fig13}(a). Actually, compared with all experimental, numerical and analytical solutions, our predictions of the phases of the 1st-, 2nd-, and 3rd-order Bragg resonances are quite accurate regardless of $N=2$ or $N=3$, see Figure \ref{fig13}(b)-(c).

\subsection{Bragg resonances excited by a FPA-5}

Finally, we consider wave reflection by a FPA-5. The parameters of the triangular bars are as follows: $H=0.6$, $d=16h_0$, $w=12h_0$ (i.e., $W=0.75D$), and the calculating range of water waves is restricted in the long-wave range $k_0h_0\in(0,\frac{\pi}{10}]$, i.e., $2D\in (0,1.6]$. For the five cases with $N$=2,3,4,5,6, the analytical solution (\ref{Ref-triangular}) to the long-wave equation (Liu et al., 2015a) is employed to calculate the reflection coefficient $K_R$.

\begin{figure}
\vspace*{-10mm}
\centerline{\hspace*{2mm}\epsfxsize=3.0in \epsffile{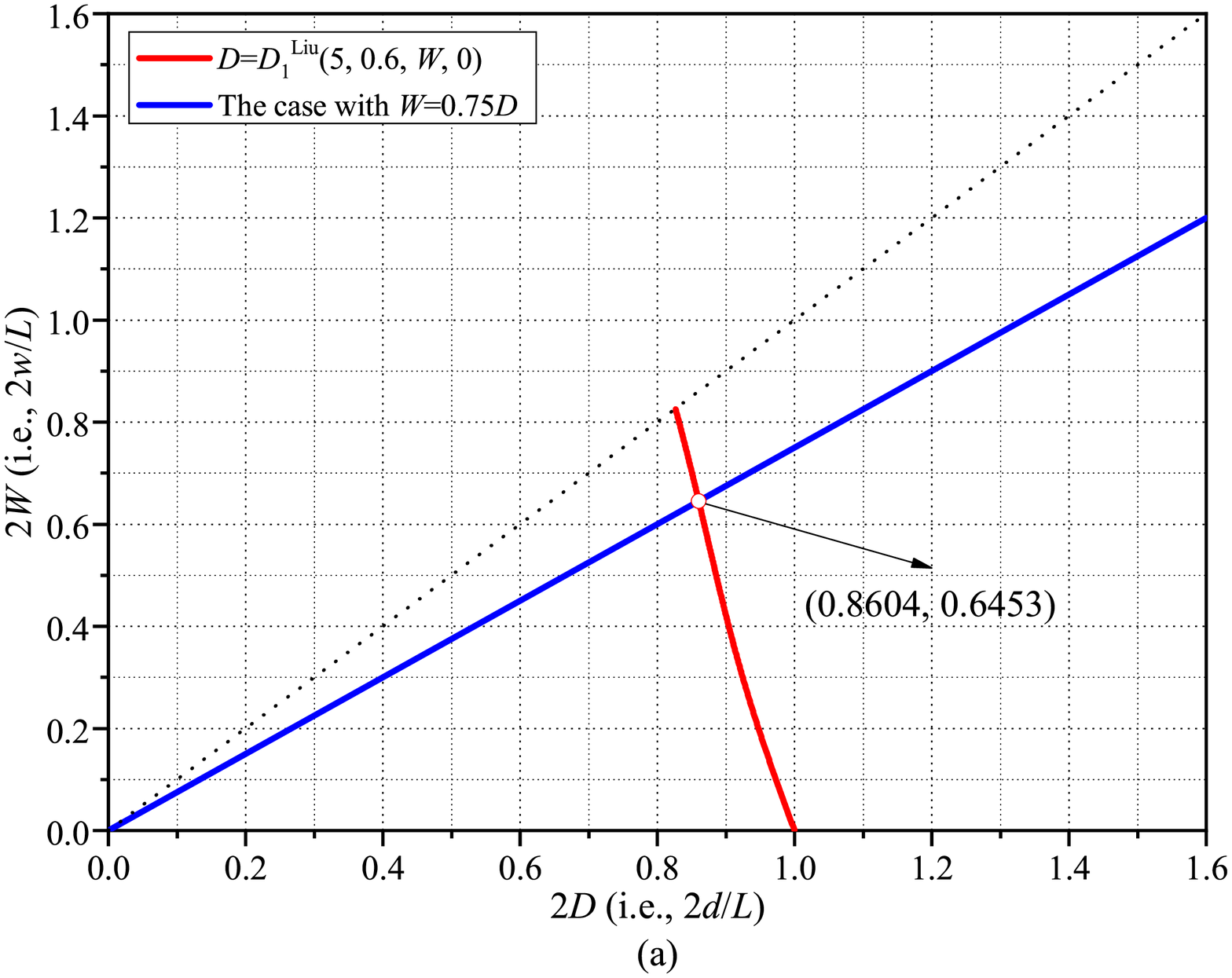}\hspace{-10mm}\epsfxsize=3.0in \epsffile{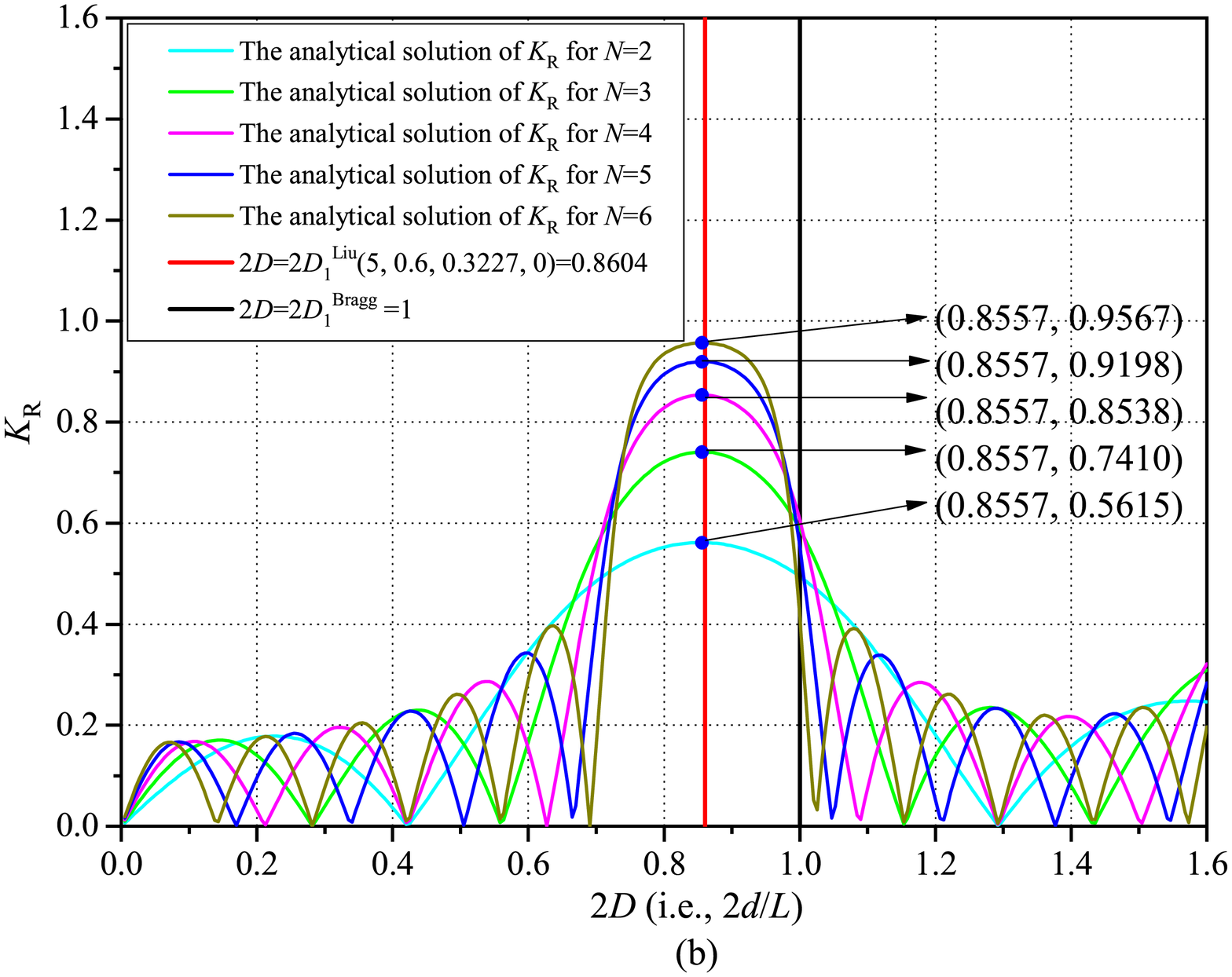}}
\vspace*{-9mm}
\caption[kdv]{\label{fig14} The prediction of the 1st-order resonance by the modified Bragg's law; (b) Comparison between the prediction and analytical solutions with $N$=2,3,4,5,6.}
\end{figure}

According to the modified Bragg's law for triangular bars, $D=D^{\mbox{Liu}}_n(5,0.6,W,0)$, the 1st-order (i.e., $n=1$) Bragg resonance should occur at $2D$=0.8604, see Figure \ref{fig14}(a). Then as we can see in Figure \ref{fig14}(b), all the phases of the five 1st-order Bragg resonances with $N$=2,3,4,5,6 are the same, i.e., $2D=0.8557$, very closed to the prediction $2D=0.8604$, which is much more accurate than the prediction of the traditional Bragg's law.

\section{Conclusion}

Based on Bloch band theory, a modified Bragg's law is derived to determine the phase of the $n$-th order Bragg resonance of linear long waves excited by a finite periodic array of artificial bars. The cross-section of artificial bars can be a rectangle, paraboloid, rectified cosine, isosceles trapezoid, and isosceles triangle.

For Bragg resonances excited by trapezoidal bars, the modified Bragg's law is described by a function of four variables, including the bar shape, dimensionless bar height, dimensionless bar width and bar top width; for Bragg resonances excited by other kinds of bars, the modified Bragg's law is described by a function of three variables due to the zero top width.

It is shown that Bragg's law is just a degenerated result of the modified Bragg's law when the bar height or width approaches zero.
However, the height and width of any artificial bar cannot be zero, hence, the Bragg's law in X-ray crystallography cannot be simply applied to Class I Bragg resonances excited by any finite periodic array of artificial bars. Second, the phase of each order Bragg resonance given by the modified Bragg's law is always smaller than that given by Bragg's law. Hence, the downshift of the phase of each order Bragg resonance can be well explained and accurately predicted by using the modified Bragg's law. Third, it is revealed at first time that the phase downshift of any order Bragg resonance excited by each type of bars increases with increasing the cross-sectional area of artificial bars.\\

{\bf Declaration of competing interest}\\

The author declares that he has no known competing financial interests or personal relationships that could have appeared to influence the work reported in this paper.\\

{\bf Acknowledgment}\\

This work is supported by the Natural Science Foundation of China (11572092, 51879237), the Qianjiang Scholar Project in Zhejiang Province, and Zhejiang Ocean University (Q1607).
\\
\\

{\bf References}

Abramowitz, M., Stegun, I.A., 1972. Handbook of Mathematical Functions. National Bureau of Standards, Applied Mathematics Series,
Vol. 55, U.S. Government Printing Office, Washington, DC.

Alam, M.R., Liu, Y., Yue, D.K., 2010. Oblique sub-and super-harmonic Bragg resonance of surface waves by bottom ripples. J. Fluid Mech. 643, 437-447.

An, Z., Ye, Z., 2004. Band gaps and localization of water waves over one-dimensional topographical bottoms. Appl. Phys. Lett. 84, 2952-2954.

Ardhuin, F., Magne, R., 2007. Scattering of surface gravity waves by bottom topography with a current. J. Fluid Mech. 576, 235-264.

Ascroft, N.W., Mermin, N.D., 1976. Solid State Physics, Saunders College, Philadelphia.

Bailard, J.A., DeVries, J.W., Kirby, J.T., Guza, R.T., 1990. Bragg reflection breakwater: a new shore protection method? Proc. 22nd Int. Conf. on Coastal Eng. ASCE, New York, 1990, pp. 1702-1715.

Bailard, J.A., DeVries, J.W., Kirby, J.T., 1992. Considerations in using Bragg reflection for storm erosion protection. J Waterw Port Coast Ocean Eng 118(1), 62-74.

Belzons, M., Rey, V., Guazzelli, E., 1991. Subharmonic Bragg resonance for surface water waves. Europhys. Lett. 16, 189-194.


Bragg, W.H., Bragg, W.L., 1913. The reflection of X-rays by crystal. Proc. Royal Soc. A: Math. Phys. Eng. Sci. 88, 428-438.

Chang, H.-K., Liou, J.-C., 2007. Long wave reflection from submerged trapezoidal breakwaters. Ocean Eng. 34(1), 185-191.

Chen, L.-S., Kuo, C.-H., Ye, Z., Sun X., 2004. Band gaps in the propagation and scattering of surface water waves over cylindrical steps. Phys. Rev. E 69, 066308.

Cho, Y.-S., Lee, J.-I., Kim, Y.-T., 2004. Experimental study of strong reflection of regular water waves over submerged breakwaters in tandem. Ocean Eng. 31(10), 1325-1335.

Couston, L.-A., Jalali, M.A., Alam, M.-R., 2017. Shore protection by oblique seabed bars. J. Fluid Mech. 815, 481-510.

Davies, A.G., 1982. The reflection of wave energy by undulations on the seabed. Dyn. Atmos. Oceans 6, 207-232.

Dalrymple, R.A., Kirby, J.T., 1986. Water waves over ripples. J. Waterw. Port, Coast. Ocean Eng. 112, 309-319.

Davies, G., Heathershaw, A.D., 1984. Surface-wave propagation over sinusoidally varying topography. J. Fluid Mech. 144, 419-443.

Davies, A.G., Guazzelli, E., Belzons, M., 1989. The propagation of long waves over an undulating bed. Phys. Fluids 1, 1331-1340.

Gao, J., Ma, X., Dong, G., Chen, H., Liu, Q., Zang, J., 2021. Investigation on the effects of Bragg reflection on harbor oscillations. Coast. Eng. 170, 103977.

Guazzelli, E., Rey, V., Belzons, M., 1992. Higher-order Bragg reflection of gravity surface waves by periodic beds. J. Fluid Mech. 245, 301-317.

Guo, F.-C., Liu, H.-W., Pan, J.-J., 2021. Phase downshift or upshift of Bragg resonance for water wave reflection by an array of cycloidal bars or trenches. Wave Motion 106, 102794.

Heathershaw, A.D., 1982. Seabed-wave resonance and sand bar growth. Nature 296, 343-345.

Hsu, T.-W., Tsai, L.-H., Huang, Y.-T., 2003. Bragg scattering of water waves by multiply composite artificial bars, Coast. Eng. J. 45, 235-253.

Jeon, C.-H., Cho, Y.-K., 2006. Bragg reflection of sinusoidal waves due to trapezoidal submerged breakwaters. Ocean Eng. 33, 2067-2082.

Joannopoulos, J.D., Meade, R.D., Winn, J.N., 1995. Photonic Crystals. Princeton University Press, Princeton.

Kirby, J.T., 1986. A general wave equation for waves over rippled beds. J. Fluid Mech. 162, 171-186.

Kirby, J.T., 1987. A Program for Calculating the Reflectivity of Beach Profiles, University of Florida, Gainesville, Florida.

Kirby, J.T., Anton, J.P., 1990. Bragg reflection of waves by artificial bars. In: Proc. the 22nd Int. Conf. Coast. Eng., pp. 757-768, New York.

Liao, S., Xu, D., Stiassnie, M., 2016. On the steady-state nearly
resonant waves. J. Fluid Mech. 794, 175-199.

Liang, B., Ge, H., Zhang, L., Liu, Y., 2020. Wave resonant scattering mechanism of sinusoidal seabed elucidated by Mathieu Instability theorem. Ocean Eng. 218, 108238.

Linton, C., 2011. Water waves over arrays of horizontal cylinders: band gaps and Bragg resonance. J. Fluid Mech. 670, 504-526.

Liu, H.-W., 2017. Band gaps for Bloch waves over an infinite array of trapezoidal bars and triangular bars in shallow water. Ocean Eng. 130, 72-82.

Liu, H.-W., Li, X.-F., Lin, P., 2019a. Analytical study of Bragg resonance by singly periodic sinusoidal ripples based on the modified mild-slope
equation. Coast. Eng. 150, 121-134.

Liu, H.-W., Liu, Y., Lin, P., 2019b. Bloch band gap of shallow-water waves over infinite arrays of parabolic bars and rectified cosinoidal bars and Bragg resonance over finite arrays of bars, Ocean Eng. 188, 106235.

Liu, H.-W., Luo, H., Zeng, H.-D., 2015a. Optimal collocation of three kinds of Bragg breakwaters for Bragg resonant reflection by long waves. J. Waterw. Port Coast. Ocean Eng. 141, 04014039.

Liu, H.-W., Shi, Y.-P., Cao, D.-Q., 2015b. Optimization of parabolic bars for maximum Bragg resonant reflection of long waves. J. Hydrodyn. 27, 840-847.

Liu, Z., Xu, D.L., Liao, S.J., 2018. Finite amplitude steady-state
wave groups with multiple near resonances in deep water. J. Fluid Mech., 835, 624-653.

Liu, Y., Yue, D.K.P., 1998. On generalized Bragg scattering of surface waves by bottom ripples. J. Fluid Mech. 356, 297-326.

Liu, H.-W., Zeng, H.-D., Huang H.-D., 2020. Bragg resonant reflection of surface waves from deep water to shallow water by a finite array of trapezoidal bars. Appl. Ocean Res., 94, 101976.

Liu, Y., Li, H.J., Zhu, L., 2016. Bragg reflection of water waves by multiple submerged semi-circular breakwaters. Appl. Ocean Res. 56, 67-78.

Losada, I.J., Losada, M.A., Baquerizo, A., 1993. An analytical method to evaluate the efficiency of porous screens as wave dampers. Appl. Ocean Res. 15, 207-215.

Madsen, P.A., Fuhrman, D.R., Wang, B., 2006. A Boussinesq-type method for fully nonlinear waves interacting with a rapidly varying bathymetry. Coast. Eng. 53, 487-504.

Massel, S.R., 1993. Extended refraction-diffraction equation for surface waves. Coast. Eng. 19, 97-126.

Mattioli, F., 1990. Resonant reflection of a series of submerged breakwaters. Il Nuovo Cimento 13, 823-833.

Mei, C.C., 1985. Resonant reflection of surface water waves by periodic sandbars. J. Fluid Mech. 152, 315-335.

Mei, C.C., 1989. The Applied Dynamics of Ocean Surface Waves. World Scientific, Singapore.

Mei, C.C., Hara, T., Naciri, M., 1988. Note on Bragg scattering of water waves by parallel bars on the seabed. J. Fluid Mech. 186, 147-162.

Miles, J., 1981. Oblique surface-wave diffraction by a cylindrical obstacle. Dyn. Atmos. Oceans, 6, 121-123.

Pan, J.-J., Liu, H.-W., Li, C.J., 2022. Bragg resonance and the phase upshift of linear water waves excited by a finite periodic array of parabolic trenches (in Chinese). Appl. Math. Mech. 43, 1-18.

Peng, J., Tao, A., Liu, Y., Zheng, J., Zhang, J., Wang, R., 2019. A laboratory study of class III Bragg resonance of gravity surface waves by periodic beds. Phys. of Fluids 31(6), 067110.

Peng, J., Tao, A.-F., Fan, J., Zheng, J.-H., Liu, Y.-M., 2022. On the downshift of wave frequency for Bragg resonance. China Ocean Eng. 36, 76-85.

Phillips, O.M. 1960. On the dynamics of unsteady gravity waves of finite amplitude, Part 1. J. Fluid Mech. 9, 193-217.

Rey, V., Guazzelli, E., Mei, C.C., 1996. Resonant reflection of surface gravity waves by one-dimensional doubly sinusoidal beds. Phys. Fluids 8, 1525-1530.

Torres, M., Adrados, J.P., Montero de Espinosa, F.R., 1999. Visualization of Bloch waves and domain walls. Nature 398, 114-115.

Tsai, C.-P., Kuo, K.-W., Yeh, P.-H., Chen, H.-B., 2017. Bragg reflection of waves by slope sandy rippled bed and its induced soil response. Ocean. Eng. 143, 186-197.

Tsai, L.-H., Kuo, Y.-S., Lan, Y.-J., Hsu, T.-W., Chen, W.-J., 2011. Investigation of multiply composite artificial bars for Bragg scattering of water waves. Coast. Eng. J. 53(4), 521-548.

Wang, S.-K., Hsu, T.-W., Tsai, L.-H., Chen, S.-H., 2006. An application of Miles' theory to Bragg scattering of water waves by doubly composite artificial bars. Ocean Eng. 33, 331-349.

Xie, J.-J., 2022. Long wave reflection by an array of submerged trapezoidal breakwaters on a sloping seabed. Ocean Eng. 252, 111138.

Xu, D., Lin, Z., Liao, S., 2015. Equilibrium states of class-I Bragg resonant wave system. Europ. J. Mech.-B/Fluids 50, 38-51.

Yoon, S.B., Liu, P.L.-F., 1987. Resonant reflection of shallow-water waves due to corrugated boundaries. J. Fluid Mech. 180, 451-469.

\end{document}